\global\def\draftcontrol{0}

   \def\versionno{On-shell YM \& the Grassmannian -- draft   }

\catcode`\@=11

\expandafter\ifx\csname draftcontrol\endcsname\relax\global\def\draftcontrol{0}
\fi

{\count255=\time\divide\count255 by 60
\xdef\hourmin{\number\count255}
\multiply\count255 by-60\advance\count255 by\time
\xdef\hourmin{\hourmin:\ifnum\count255<10 0\fi\the\count255}}
\def\draftdate{\number\month/\number\day/\number\year\ \ \ \hourmin }

\newcommand\makepapertitle{\par
  \begingroup
    \renewcommand\thefootnote{\@fnsymbol\c@footnote}%
    \def\@makefnmark{\rlap{\@textsuperscript{\normalfont\@thefnmark}}}%
    \long\def\@makefntext##1{\parindent 1em\noindent
            \hb@xt@1.8em{%
                \hss\@textsuperscript{\normalfont\@thefnmark}}##1}%
     \newpage
     \global\@topnum\z@   
     \@makepapertitle
     \thispagestyle{empty}\@thanks
  \endgroup
  \setcounter{footnote}{0}%
  \global\let\thanks\relax
  \global\let\makepapertitle\relax
  \global\let\@makepapertitle\relax
  \global\let\@thanks\@empty
  \global\let\@author\@empty
  \global\let\@date\@empty
  \global\let\@title\@empty
  \global\let\title\relax
  \global\let\author\relax
  \global\let\date\relax
  \global\let\and\relax
  \def\version{\let\version\@version\@gobble}
}
\def\@makepapertitle{%
  \newpage
   \ifnum\draftcontrol=1 {}
   \version\versionno
   \vskip 3em%
   \else
   \hfill\hbox to 3cm {\parbox{4cm}{\@pubnum}\hss}%
   \vskip 3em%
   \fi
   \begin{center}%
   \let \footnote \thanks
     {\LARGE {\@title}}%
     \vskip 1.5em%
     {\normalsize
       \lineskip .5em%
       \begin{tabular}[t]{c}%
         \@author
       \end{tabular}\par}%
     \vskip 1.5em%
     {\@bstract}%
     \end{center}%
     \vskip 1.5em
     \@date%
   \par
}

\gdef\@pubnum{}
\def\pubnum#1{%
  \gdef\@pubnum{#1}}

\gdef\@bstract{}
\def\Abstract#1{%
  \gdef\@bstract{%
   \parbox{\textwidth-0pc}{%
   \centerline{\bf Abstract}\penalty1000%
\kern.2cm%
\noindent
\renewcommand\baselinestretch{1.0}%
{#1}}}
}

\def\ps@paper{\let\@mkboth\@gobbletwo%
     \ifnum\draftcontrol=1
    \def\@oddfoot{\hbox to \textwidth{\tiny \versionno \hfil\tiny\draftdate}%
    \hskip -\textwidth \hbox to \textwidth{\hfil\rm\thepage\hfil}}%
     \else\def\@oddfoot{\hbox to \textwidth{\hfil\rm\thepage\hfil}}
     \fi
     \let\@evenfoot\@oddfoot
}

\def\body{\clearpage
          \pagestyle{paper}
    }

\def\@version#1{\ifnum\draftcontrol=1
\typeout{}\typeout{#1}\typeout{}
\vskip3mm\centerline{\hbox{\fbox{\normalsize{\tt DRAFT -- #1 -- }
                   {\draftdate}}}}\vskip3mm
\fi}
\let\version\@version
\long\def\eqlabel#1{\ifnum\draftcontrol=1
                    \tag@false  
                    \tag*{(\theequation) \hbox to -0.2cm{\hspace{0cm}\small{#1}\hss}}
                    \refstepcounter{equation}
                    \edef\@currentlabel{\theequation}
                    \ltx@label{#1}          
                    \else
                    \label{#1}
                    \fi
                    }
\let\st@bibitem\@bibitem
\let\st@lbibitem\@lbibitem
\ifnum\draftcontrol=1
  \def\@bibitem#1{%
    \st@bibitem{#1}\a@@label{#1}\ignorespaces}
  \def\@lbibitem[#1]#2{%
    \st@lbibitem[#1]{#2}\a@@label{#2}\ignorespaces}
  \def\a@@label#1{%
    \gdef\a@lab{\smash{\normalfont\small#1}}
    \ifvmode
      \if@inlabel
        \global\setbox\@labels\hbox{%
          \llap{\a@lab\let\a@lab\relax
                \kern\@totalleftmargin\kern\marginparsep}%
          \box\@labels}%
      \fi
    \fi}
\fi

\documentclass[11pt,A4paper]{article}

\usepackage{amsmath,amssymb,array,calc,epsfig}
\usepackage{xcolor}
\usepackage{cite}

\usepackage[
      colorlinks=true,
      linkcolor=red,  
      urlcolor=blue,    
      filecolor=red,     
      linkcolor=blue,
      citecolor = red,
      pdfstartview=FitV,
      pdftitle={},
        pdfauthor={Paolo Benincasa},
        pdfsubject={},
        pdfkeywords={},
        pdfpagemode=None,
        bookmarksopen=true    
      ]{hyperref}
      
\usepackage{graphicx}
\usepackage{epstopdf}
\usepackage{ccaption}

\ifnum\draftcontrol=1
\fi

\renewcommand\baselinestretch{1.25}
\renewcommand\section{\@startsection {section}{1}{\z@}%
                                   {-3.5ex \@plus -1ex \@minus -.2ex}%
                                   {2.3ex \@plus.2ex}%
                                   {\normalfont\large\bfseries}}
\renewcommand\subsection{\@startsection{subsection}{2}{\z@}%
                                   {-3.25ex\@plus -1ex \@minus -.2ex}%
                                   {1.5ex \@plus .2ex}%
                                   {\normalfont\normalsize\bfseries}}
\renewcommand\subsubsection{\@startsection{subsubsection}{3}{\z@}%
                                   {-3.25ex\@plus -1ex \@minus -.2ex}%
                                   {1.5ex \@plus .2ex}%
                                   {\normalfont\normalsize\it}}
\renewcommand\paragraph{\@startsection{paragraph}{4}{\z@}%
                                   {-3.25ex\@plus -1ex \@minus -.2ex}%
                                   {1.5ex \@plus .2ex}%
                                   {\normalfont\normalsize\bf}}

\numberwithin{equation}{section}

\def\revise#1       {\raisebox{-0em}{\rule{3pt}{1em}}%
                     \marginpar{\raisebox{.5em}{\vrule width3pt\
                     \vrule width0pt height 0pt depth0.5em
                     \hbox to 0cm{\hspace{0cm}{%
                     \parbox[t]{4em}{\raggedright\footnotesize{#1}}}\hss}}}}

\def\sqr#1#2{{\vcenter{\vbox{\hrule height.#2pt
 \hbox{\vrule width.#2pt height#1pt \kern#1pt
 \vrule width.#2pt}\hrule height.#2pt}}}}

\def\aa1{\phi}
\def\cc1{\psi}

\catcode`\@=12

\begin{document}


\title{\bf On-shell diagrams and the geometry of planar $\mathcal{N}\,<\,4$ SYM theories}
\pubnum{%
arXiv:xxxx.xxxxx}
\date{September 2016}

\author{
\scshape Paolo Benincasa, David Gordo \\[0.4cm]
\ttfamily ${}^{\dagger}$Instituto de F{\'i}sica Te{\'o}rica, \\
\ttfamily Universidad Aut{\'o}noma de Madrid / CSIC \\
\ttfamily Calle Nicolas Cabrera 13, Cantoblanco 28049, Madrid, Spain\\
\small \ttfamily paolo.benincasa@csic.es, david.gordo@csic.es \\[0.2cm]
}

\Abstract{We continue the discussion of the decorated on-shell diagrammatics for planar $\mathcal{N}\,<\,4$ Supersymmetric Yang-Mills theories started in \cite{Benincasa:2015zna}.  In particular, we focus on its relation
 with the structure of varieties on the Grassmannian. The decoration of the on-shell diagrams, which physically keeps tracks of the helicity of the coherent states propagating along their edges, defines new on-shell functions
 on the Grassmannian and can introduce novel higher-order singularities, which graphically are reflected into the presence of helicity loops in the diagrams. These new structures turn out to have similar features as in the non-planar 
 case: the related higher-codimension varieties are identified by either the vanishing of one (or more) Pl{\"u}cker coordinates involving at least two non-adjacent columns, or new relations among Pl{\"u}cker coordinates. A distinctive
 feature is that the functions living on these higher-codimenson varieties can be thought of distributionally as having support on derivative delta-functions. After a general discussion, we explore in some detail the
 structures of the on-shell functions on $Gr(2,4)$ and $Gr(3,6)$ on which the residue theorem allows to obtain a plethora of identities among them.}

\makepapertitle

\body

\version\versionno

\tableofcontents

\section{Introduction}\label{sec:Intro}

The direct analysis of observables provides a new perspective on particle physics and its underlying structures. In the last decade it became clear that the perturbative physics can be fully described in terms of gauge 
invariant (and thus physical) objects such as the scattering amplitudes, at least in the regime where asymptotic states can be defined and for a certain class of theories. Specifically, the first indication that this were possible
came from the discovery of on-shell recursion relations in Yang-Mills theory at tree level \cite{Britto:2004aa, Britto:2005aa}, which afterwards were extended also to general relativity \cite{Benincasa:2007qj} and to a larger
class of theories with particles having at most spin $2$ \cite{Cheung:2008dn, Benincasa:2011kn, Benincasa:2011pg} again at tree level, as well as to $\mathcal{N}\,\le\,4$ Supersymmetric Yang-Mills theories at loop level 
\cite{ArkaniHamed:2010kv, ArkaniHamed:2012nw, Benincasa:2015zna}. These recursion relations\footnote{For an overview on scattering amplitudes and recursion relations, see \cite{Elvang:2013cua, Benincasa:2013faa}.} allow to express 
scattering amplitudes in terms of lower point/lower level physical sub-amplitudes. Thus, iterating the algorithm, one finds  that a scattering amplitude with an arbitrary number of external states can be totally determined in terms of 
the smallest possible amplitude allowed by a given theory. 

For the cases of interest, which typically deal with massless  particles, such building blocks are provided by the three-particle amplitudes, which are fully determined by Poincar{\'e} invariance \cite{Benincasa:2007xk}. Suitably 
gluing these building blocks along one edge, which boils down to  integrate out the degrees of freedom along such an edge so that momentum conservation and the on-shell condition are satisfied, it is possible to construct more 
complicated  on-shell processes, whose peculiarity lies in the {\it on-shell-ness}  of all the states, both external and internal. This procedure generates objects which are always {\it physical} and gauge invariant, with no need
of introducing {\it virtual} particle and breaking gauge invariance as it happens in the individual Feynman diagrams. This new diagrammatics has been extensively studied in the case of planar $\mathcal{N}\,=\,4$ Supersymmetric Yang-Mills
(SYM) theory \cite{ArkaniHamed:2012nw} for which the construction of the on-shell processes preserves and makes manifest the infinite dimensional Yangian invariance \cite{Drummond:2009fd} (which is made of superconformal
and dual superconformal invariance \cite{Drummond:2006rz, Bern:2006ew})

Notably, the on-shell diagram formulation of $\mathcal{N}\,=\,4$ SYM is intimately related to novel mathematical structures such as the Grassmannian $Gr(k,n)$ \cite{Postnikov:2006kva, ArkaniHamed:2009dn, Mason:2009qx, 
ArkaniHamed:2009go, Kaplan:2009mh}, whose positivity preserving diffeomorphisms correspond to the Yangian symmetry of the theory \cite{ArkaniHamed:2012nw}, and the permutations, which define equivalence classes for the diagrams 
\cite{ArkaniHamed:2012nw, Franco:2013nwa}.

While these structures for {\it planar} $\mathcal{N}\,=\,4$ SYM have been extensively discussed, the non-planar sector has been object of studies just more recently \cite{Chen:2014ara, Arkani-Hamed:2014bca, Bern:2014kca, Franco:2015rma, 
Bourjaily:2016mnp, Frassek:2016wlg}. 
Even less is known outside the context of $\mathcal{N}\,=\,4$ SYM, with the notable exception of the three-dimensional ABJM theory \cite{ArkaniHamed:2012nw, Huang:2013owa, Kim:2014hva, Huang:2014xza, Elvang:2014fja}
and for $\mathcal{N}\,<\,4$ SYM theories\footnote{For $\mathcal{N}\,<\,4$ SYM theories, an on-shell treatment of the scattering amplitudes was discussed in \cite{Elvang:2011fx}.} \cite{Benincasa:2015zna}. 
In the latter case, the on-shell diagrammatics acquires new features: it is endowed with a {\it physical} decoration which encodes the helicities of the coherent
states propagating along the edges of each on-shell diagrams. Such a decoration, which is represented as incoming/outgoing arrows for negative/positive helicity coherent states, induces directed paths along the edges (named helicity 
flows) of the diagrams which beautifully encode the singularity structure of a given on-shell process. Furthermore, the equivalence relations are now codified in terms of such helicity flows, while the permutations represent Ward 
identities among different on-shell processes. The helicity flows can form loops in a diagram: this corresponds to singularities (higher-order poles) which are completely absent in the maximally supersymmetric case and are associated to
further structures in loop amplitudes, such as the UV divergences and the rational terms \cite{Benincasa:2015zna}.

In this paper, we continue the investigation of the on-shell diagrammatics and the related mathematical structures for less/no-supersymmetric Yang-Mills theories in the planar sector. In particular, we focus on the possibility of
associating an {\it auxiliary} Grassmannian to the on-shell processes along similar lines of what happens in the maximally supersymmetric case. A property of the Grassmannian integral is its invariance under $GL(k)$-transformations.
In the maximally supersymmetric case this is guaranteed by the fact that both the Grassmannian form and the kinematic support turn out to be separately $GL(k)$-invariant. This is no-longer true for $\mathcal{N}\,<\,4$. However,
the decoration induces a further function of the Pl{\"u}cker coordinates which complete the integrand, satisfying the required invariance. 
Concretely, also in this case, an on-shell diagram corresponds to a particular stratification of the Grassmannian, with integral representation given by:
\begin{equation}\eqlabel{eq:GenGrass}
 \mathcal{M}_{\mbox{\tiny $k,n$}}^{\mbox{\tiny $(OD)$}}\:=\:
 \int_{\mbox{\tiny $\Pi$}}\frac{d^{k\times n}C}{\mbox{Vol}\{GL(k)\}}\,\frac{\delta^{\left(4k|k\mathcal{N}\right)}\left(C\cdot\mathcal{W}\right)\:\Delta_{s_1\ldots s_k}^{4-\mathcal{N}}}{
   \Delta_{i_{1}\ldots i_{k}}\Delta_{i_2\ldots i_{k+1}}\ldots\Delta_{i_{n-k+1}\ldots i_{n}}\Delta_{i_{n-k+2}\ldots i_n i_1}}\,
  \left[\mathfrak{h}\left(\frac{\Delta_I}{\Delta_J}\right)\right]^{4-\mathcal{N}},
\end{equation}
where the kinematic data $\mathcal{W}$ are represented in twistor space, the $\Delta$'s are the maximal minors of $C$ (the indices $s_1,\,\ldots\,s_k$ represents the helicity sources, {\it i.e.} the columns of $C$ related to the 
coherent states with negative helicity), and $\mathfrak{h}$ is a rational function of the maximal minors. More precisely, $\mathfrak{h}$ turns out to be both invariant under little group transformations and $GL(k)$-invariant, given
that it is a function of the ratios of the Pl{\"u}cker coordinates.

The structure in \eqref{eq:GenGrass} can be easily obtained by gluing the trivalent nodes, with such a procedure determining also the functional form of $\mathfrak{h}(\Delta_I/\Delta_J)$. The presence of a helicity loop in the
on-shell diagrams is reflected in the structure of $\mathfrak{h}$ through the presence of a non-planar pole. The on-shell function related to the codimension-$1$ variety identified by such a pole can be seen as having support on
a derivative delta-function.

We will discuss this new structures in relation to both the top-varieties and the higher-codimension ones, and discuss new relations among the on-shell functions that beautifully the Grassmann representation encodes, with
specific examples for $Gr(2,4)$ and $Gr(3,6)$.

The paper is organised as follows. In Section \ref{sec:DOSrev} we highlight the features of the decorated on-shell diagrammatics for $\mathcal{N}\,<\,4$ theories, along the lines of \cite{Benincasa:2015zna}, with particular attention
on the equivalence relations and the helicity flows which encode the singularity structure of the on-shell processes. Section \ref{sec:GrassDOS} is devoted to a quick review of the Grassmannian and to a general discussion of the
features of the Grassmannian integrals for on-shell processes in the cases of interest. Section \ref{sec:SingGrass} contains a detailed analysis of the new structures emerging in our context, with a particular focus on
$Gr(2,4)$. Interestingly, the residue theorem on $Gr(2,4)$ returns two classes of identities, one of which corresponds to the equivalence between two different BCFW representations while the other one provides an equivalence
relation between the on-shell diagram with a helicity loop and a {\it non-planar} diagram. In Section \ref{sec:Gr36} we discuss the structure on-shell functions on $Gr(3,6)$. Notably, new poles appear, whose location imposes
a relation among the Pl{\"u}cker coordinates keeping them all non-zero. In momentum space such a relation turns out to have two solutions, both of which need to be considered (see Appendix \ref{app:NPpMomSp}).
Finally, Section \ref{sec:Concl} is devoted to the conclusion and further discussion of the results.


\section{On-shell Diagrams for $\mathcal{N}\,<\,4$ SYM}\label{sec:DOSrev}

The building blocks of the construction are the three-particle amplitudes which, for massless particles, are fixed (up to a coupling constant) by (super)-Poincar{\'e} invariance \cite{Benincasa:2007xk}. More precisely, while for real
momenta, invariance under space-time translations forces these objects to vanish, for complexified momenta there are two non-trivial solutions with support on the momentum-conserving sheet\footnote{The three-particle amplitudes are
also non-zero if we consider our space-time in $(2,2)$-signature. In this case, the Lorentz group is isomorphic to $SL(2,\mathbb{R})\times SL(2,\mathbb{R})$ and the two spinors in which a four-dimensional massless momentum can
be decomposed transform under a different copy of $SL(2\,\mathbb{R})$ each.}. Taking the complexified Lorentz group as
$SL(2,\mathbb{C})\times SL(2,\mathbb{C})$, the kinematic data can be encoded into spinorial variables $p_{a\dot{a}}\,=\,\lambda_a\tilde{\lambda}_{\dot{a}}$, with the two spinors transforming under different copies of $SL(2,\mathbb{C})$,
and the two solutions identified by having either all the $\lambda$'s or all the $\tilde{\lambda}$'s proportional to each other. The functional expression of each of these two solutions is fixed by Little group covariance, through
the requirement that the amplitudes are eigenfunctions of the helicity operator:

\begin{equation}\eqlabel{eq:3ptAmpl}
 \begin{split}
  &\mathcal{M}_3^{\mbox{\tiny $(\mathfrak{1})$}}\:=\:
    \raisebox{-1cm}{\scalebox{.30}{\includegraphics{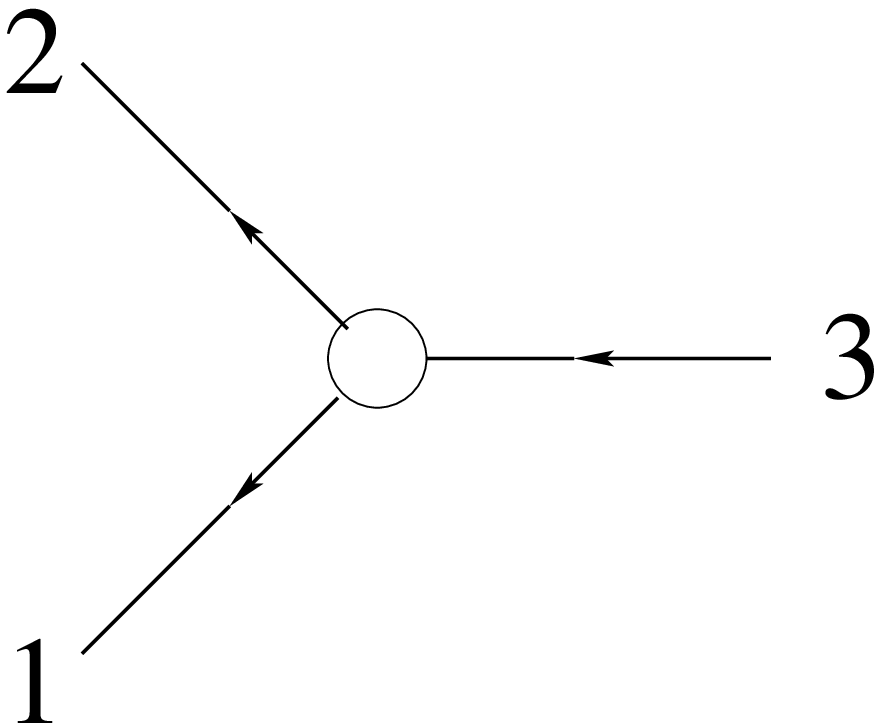}}}\:=
    \:\frac{\delta^{\mbox{\tiny $(2\times2)$}}\left(\lambda\cdot\tilde{\lambda}\right)
    \delta^{\mbox{\tiny $(1\times\mathcal{N})$}}\left(\alpha\cdot\tilde{\eta}\right)[1,2]^{4-\mathcal{N}}}{[1,2][2,3][3,1]}, \\
  &\mathcal{M}_3^{\mbox{\tiny $(\mathfrak{2})$}}\:=\:
    \raisebox{-1cm}{\scalebox{.30}{\includegraphics{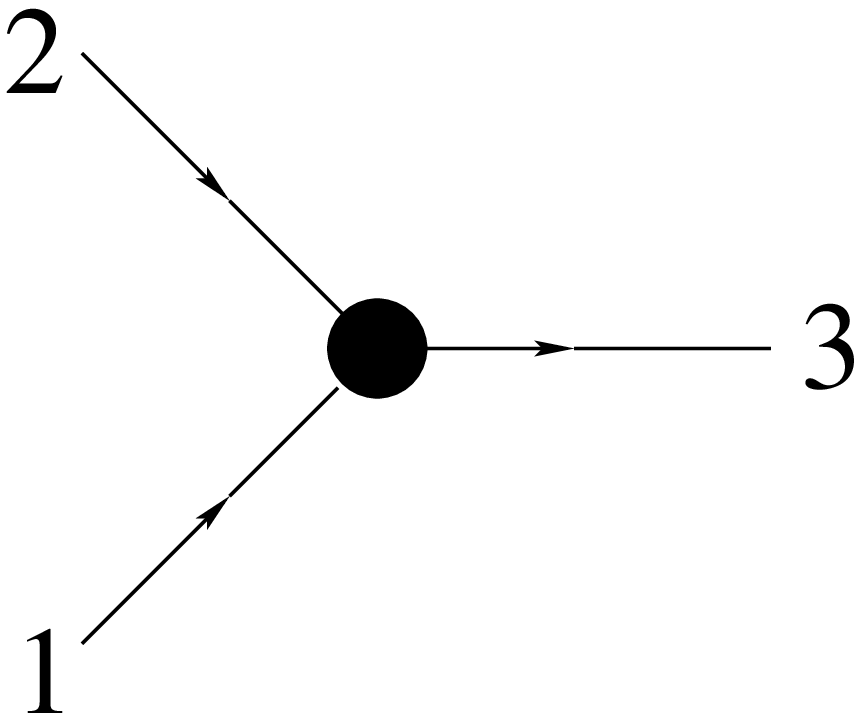}}}\:=
   \frac{\delta^{\mbox{\tiny $(2\times2)$}}\left(\lambda\cdot\tilde{\lambda}\right)
    \delta^{\mbox{\tiny $(2\times\mathcal{N})$}}\left(\lambda\cdot\tilde{\eta}\right)\langle1,2\rangle^{4-\mathcal{N}}}{\langle1,2\rangle\langle2,3\rangle\langle3,1\rangle},
  \end{split}
\end{equation}
where $\lambda\cdot\tilde{\lambda}\,\equiv\,\sum_i\lambda^{\mbox{\tiny $(i)$}}\tilde{\lambda}^{\mbox{\tiny $(i)$}}$, $\lambda\cdot\tilde{\eta}\,\equiv\,\sum_i\lambda^{\mbox{\tiny $(i)$}}\tilde{\eta}^{\mbox{\tiny $(i)$}}$,
      $\alpha\cdot\tilde{\eta}\,\equiv\,\sum_i[i-1,i]\tilde{\eta}^{\mbox{\tiny $(i+1)$}}$. Furthermore, 
      $\langle i,j\rangle\,\equiv\,\epsilon^{ab}\lambda_a^{\mbox{\tiny $(i)$}}\lambda_b^{\mbox{\tiny $(j)$}}$ and 
      $[i,j]\,\equiv\,\epsilon^{\dot{a}\dot{b}}\tilde{\lambda}_{\dot{a}}^{\mbox{\tiny $(i)$}}\tilde{\lambda}_{\dot{b}}^{\mbox{\tiny $(j)$}}$ are the Lorentz invariant combination of the spinors, while 
      $\tilde{\eta}^{\mbox{\tiny $(i)$}}_I$ are the Grassmann variables through which the supersymmetric coherent states are defined ($I\,=\,1,\,\ldots\,\mathcal{N}$ is the $SU(\mathcal{N})$ index).

The three-particle amplitude, whose rational part just depends on the holomorphic Lorentz invariants $\langle i,j\rangle$, are defined on a support where all the
$\tilde{\lambda}$'s are proportional to each other. The one whose rational part is just a function of the anti-holomorphic Lorentz invariants $[i,j]$, are instead defined on a support where all the $\lambda$'s are proportional
to each other.

The arrows in \eqref{eq:3ptAmpl} have been introduced to keep track of the helicity of each state. As a matter of convention, the incoming/outgoing arrows represent negative/positive helicity states, 
the trivalent black (white) nodes represent the three-particle amplitudes with all the $\tilde{\lambda}$'s ($\lambda$'s) proportional to each other.


Higher-point on-shell processes can be obtained by suitably gluing the three-particle amplitudes, {\it i.e.} integrating out the on-shell degrees of freedom of the edge along which the objects get glued. Depending on the number
of the degrees of freedom which the constraints are able to fix on the edges along which the on-shell processes get glued, the resulting process can be a {\it leading singularity}, if all the degrees of freedom are fixed; a 
singularity, if the constraints are more than the number of internal degrees of freedom (the result is a constraint on the external kinematic); or an on-shell form if there is some internal degrees of freedom which is left
unfixed (in this sense one can think of the leading singularities as on-shell $0$-forms). Higher-degree on-shell forms can be systematically generated from a lower-degree one via the {\it BCFW bridge}: given an $n$-point on-shell 
$p$-form $\mathcal{M}_n^{\mbox{\tiny $(p)$}}$, one can single out two adjacent\footnote{It is actually possible to consider two non-adjacent lines. In this case, a planar on-shell diagram -- which is embeddable into a disk --
can me mapped into a non-planar one, which is instead embeddable into an higher genus Riemann surface \cite{Franco:2015rma}.} external lines and connect them by gluing a {\it bridge} formed by two three-particle amplitudes of different
type -- the integration over the internal delta-functions leaves one degree of freedom unfixed, mapping the original $n$-point on-shell $p$-form $\mathcal{M}_n^{\mbox{\tiny $(p)$}}$ into an $n$-point on-shell $(p+1)$-form:
\begin{equation}\eqlabel{eq:BCFWbridge}
 \mathcal{M}_n^{\mbox{\tiny $(p)$}}\:=\:
  \begin{array}{r}
   \vspace{.8cm}\\
   \vspace{.8cm}\\
   {i+1}
  \end{array}
  \hspace{-1cm}\raisebox{-1.2cm}{\scalebox{.25}{\includegraphics{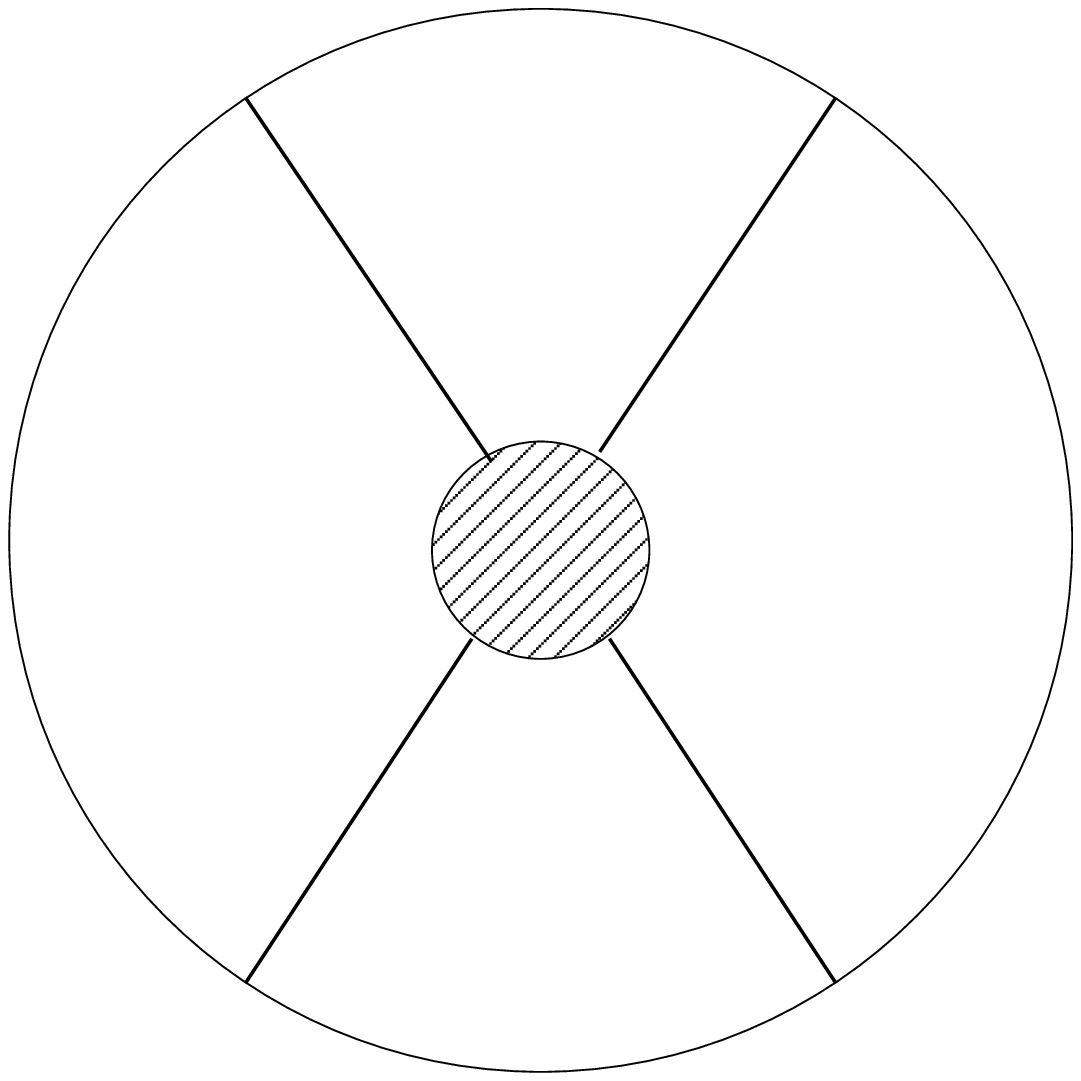}}}
  \begin{array}{l}
   \vspace{.8cm}\\
   \vspace{.8cm}\\
   {\hspace{-.7cm} i}
  \end{array}
  \quad\Longrightarrow\quad
  \begin{array}{r}
   \vspace{.8cm}\\
   \vspace{.8cm}\\
   {i+1}
  \end{array}
  \hspace{-.8cm}\raisebox{-1.2cm}{\scalebox{.25}{\includegraphics{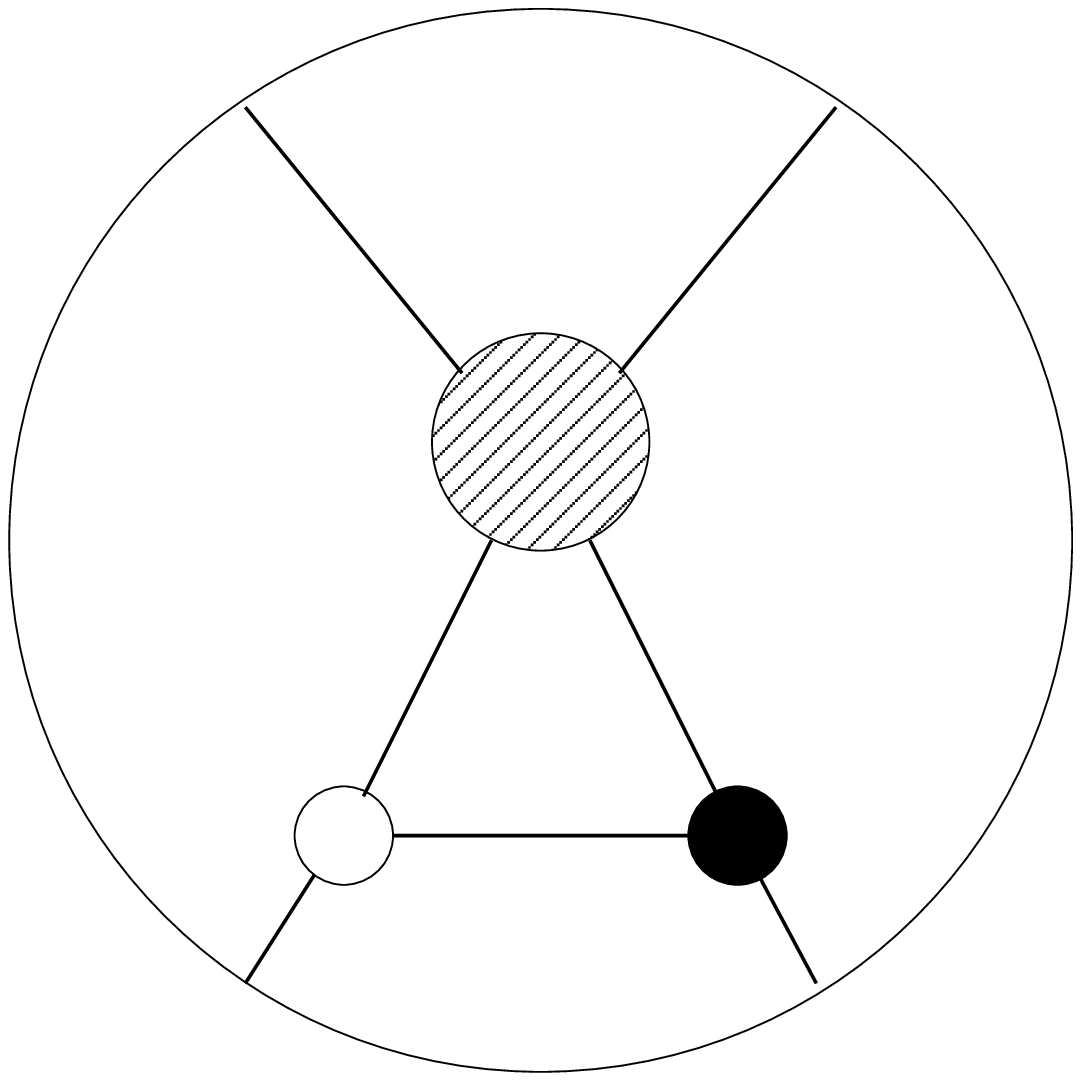}}}
  \begin{array}{l}
   \vspace{.8cm}\\
   \vspace{.8cm}\\
   {\hspace{-.7cm} i}
  \end{array}
  \hspace{-.2cm}=\:dz\,\mu(z)\,\mathcal{M}_n^{\mbox{\tiny $(p)$}}(z)\:\equiv\:\mathcal{M}_n^{\mbox{\tiny $(p+1)$}},
\end{equation}
with $z$ being the unfixed degree of freedom, $\mu(z)$ is the measure associated to the bridge and $\mathcal{M}_n^{\mbox{\tiny $(p)$}}(z)$ is the BCFW-deformed $p$-form whose lines which have been single out (and which have now
become internal) have momenta $p^{\mbox{\tiny $(i)$}}(z)\,=\,(\lambda^{\mbox{\tiny $(i)$}}+z\lambda^{\mbox{\tiny $(i+1)$}})\tilde{\lambda}^{\mbox{\tiny $(i)$}}$ and 
$p^{\mbox{\tiny $(i+1)$}}(z)\,=\,\lambda^{\mbox{\tiny $(i+1)$}}(\tilde{\lambda}^{\mbox{\tiny $(i+1)$}}-z\tilde{\lambda}^{\mbox{\tiny $(i)$}})$. 

The measure $\mu(z)$ depends on the helicity configuration of the BCFW bridge. Concretely,
given a certain helicity configuration for the on-shell $p$-form $\mathcal{M}_n^{\mbox{\tiny $(p)$}}$, the BCFW bridge is added in such a way that the external states $i$ and $i+1$ in the newly generated on-shell $(p+1)$-form
still have the original helicity. However, in the internal edges now we need to sum all the possible helicity states which can propagate. If the direction of the helicity arrows from the external states into the deformed internal ones
is preserved, than $\mu(z)\,=\,z^{-1}$, while if it is not $\mu(z)\,=\,z^{3-\mathcal{N}}$ \cite{Benincasa:2015zna}. This means that in the first case the original on-shell $p$-form is associated to the residue of the pole $z\,=\,0$, 
while in the second case the on-shell $p$-form is mapped into another on-shell $p$-form with a different helicity configuration which is no longer associated to a residue at $z\,=\,0$ given that such a pole is now absent (for 
$\mathcal{N}\,\le\,3$). 
Furthermore, in the latter case a multiple pole at infinity is introduced, which is a reflection of the change of the helicity arrow directions in the edges on which the BCFW bridge has been applied. 
Another way to think about this is that, if in the on-shell $(p+1)$-form \eqref{eq:BCFWbridge} the helicity states of the particles 
labelled by $(i,\,i+1)$ is $(+,\,-)$ ({\it i.e.} the on-shell diagram shows an incoming arrow in the white node and an outgoing one in the black node), the helicities of the internal states are fixed and there is a {\it helicity flow} 
from the external lines towards the on-shell $p$-form: in this case the BCFW measure is $\mathcal{\mu}(z)\,=\,z^{-1}$, and the bridge induces a BCFW-deformation on the internal $p$-form with no pole at infinity. If now, the external
states $(i,\,i+1)$ have helicity $(-,\,+)$, different coherent states can propagate in the internal edges, generating a counter-clockwise helicity flow in one case and a clockwise one in the other case. The counter-clockwise
helicity flow preserves the helicity states along the lines $i$ and $i+1$, the related BCFW measure is again $\mu(z)\,=\,z^{-1}$ and the BCFW bridge induces a deformation on the $p$-form with a multiple pole at infinity. 
The clockwise flow instead does not preserve the helicities along the lines $i$ and $i+1$, the related BCFW measure is now $\mu(z)\,=\,z^{3-\mathcal{N}}$ and the $p$-form which the BCFW bridge is attached has now a different
helicity configuration. 
The helicity flows thus keep track of the singularities in an on-shell process: given a sub-diagram of the form of the right-hand-side of \eqref{eq:BCFWbridge},
the presence of a helicity flow guarantees that the bridge can be removed and that the diagram left corresponds to the residue of the related simple pole; the presence of the helicity loops are instead
a manifestation of the existence of higher order poles.

The decoration associated to the helicities therefore introduces a perfect orientation on the on-shell processes and it is  a reflection of their singularity structure.


\subsection{Equivalence classes and equivalence operations}\label{subsec:EquivClass}

When we build complicated on-shell processes by gluing several three-particle amplitudes, not all of them turns out to be inequivalent. As an example, let us take an on-shell process which has two black (white) nodes connected to each 
other along one line as a sub-diagram. These black (white) nodes have all the $\tilde{\lambda}$'s ($\lambda$'s) proportional to each other, so that they can be equivalently merged together to form a four-valent black (white) node
and expanded again along a different line into two three-valent black (white) node:
\begin{equation*}
   \raisebox{-.8cm}{\scalebox{.35}{\includegraphics{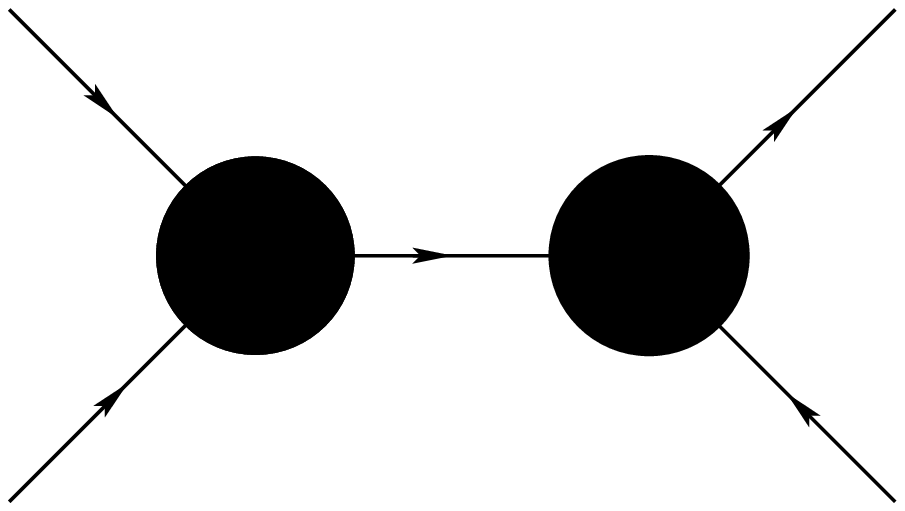}}}\quad\Longleftrightarrow\quad
   \raisebox{-.8cm}{\scalebox{.35}{\includegraphics{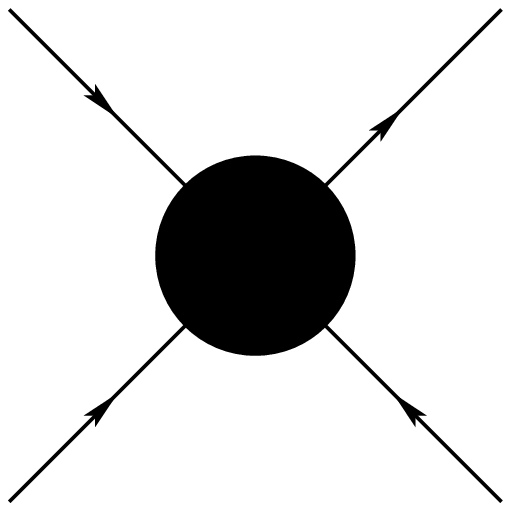}}}\quad\Longleftrightarrow\quad
   \raisebox{-1.4cm}{\scalebox{.35}{\includegraphics{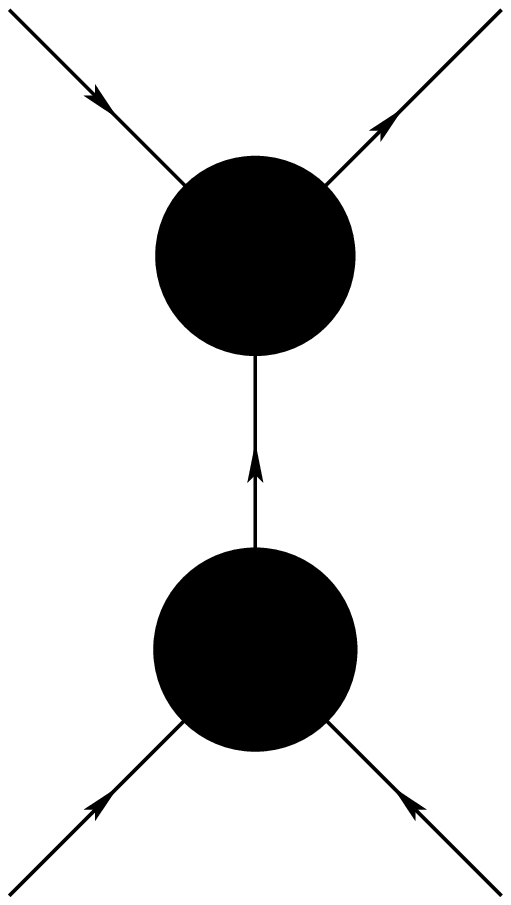}}}
\end{equation*}
This equivalence operation goes under the name of {\it merger}: Any two diagrams which can be mapped into each other by merging together two nodes of the same type and expanding them along a different channel are equivalent. 
Importantly, this equivalence operation does not depend on the particular helicity configuration because it is just related to the proportionality relations among spinors of the same type.

Let us now consider a $p$-valent black (white) node with just incoming (outgoing) helicity arrows. Notice that in general it cannot be opened up into a tree-like configuration as in the merger operation just described. However,
it can be open up into the sum of two $p$-gons with the two possible helicity loops:
\begin{equation*}
   \raisebox{-.8cm}{\scalebox{.35}{\includegraphics{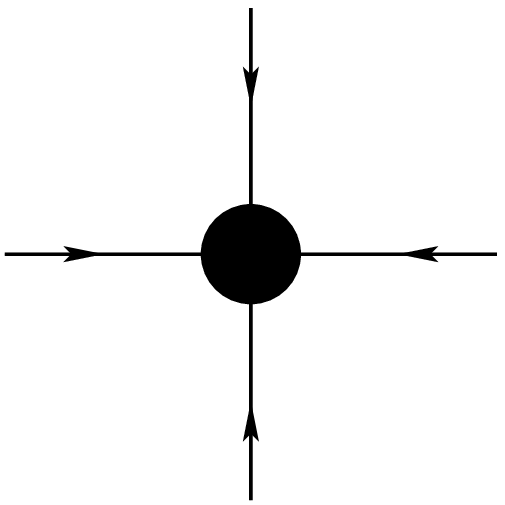}}}\quad\Longleftrightarrow\quad
   \raisebox{-1.5cm}{\scalebox{.35}{\includegraphics{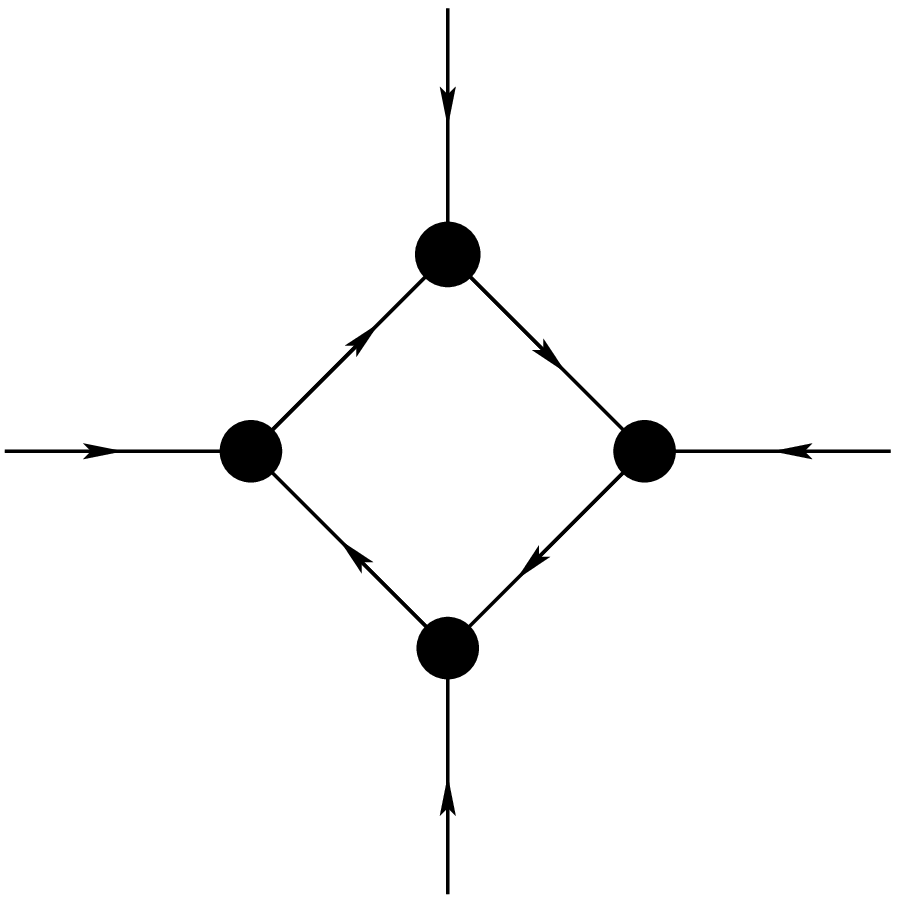}}}\quad+\quad
   \raisebox{-1.5cm}{\scalebox{.35}{\includegraphics{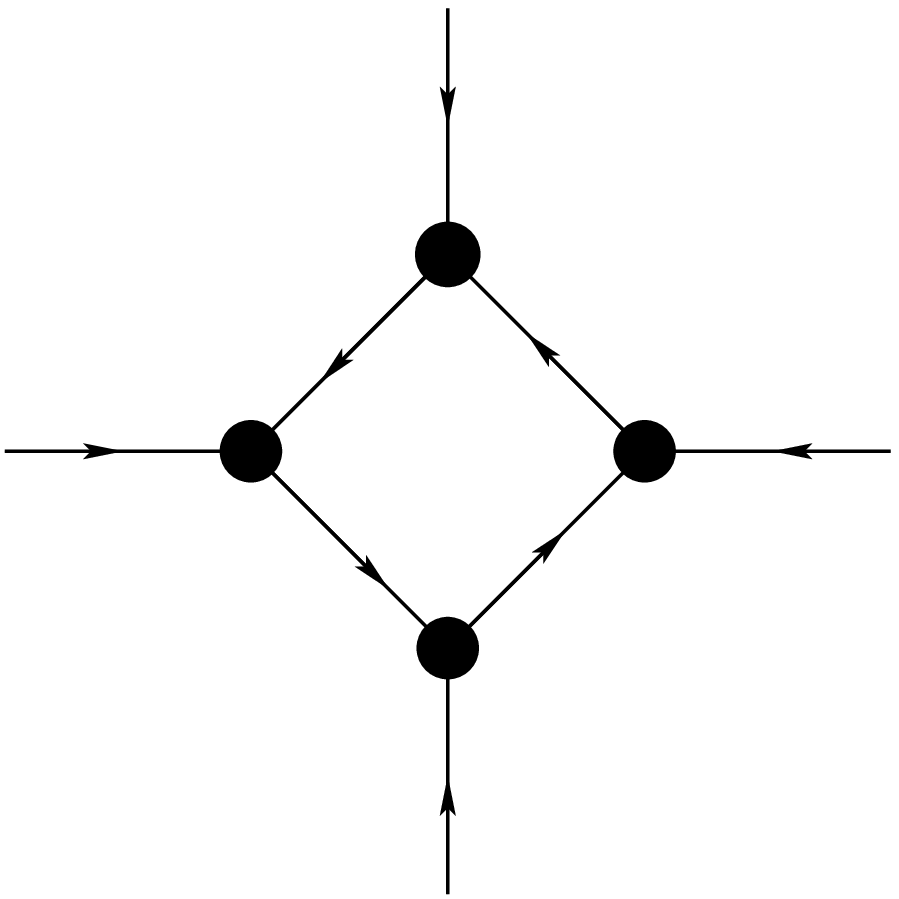}}}
\end{equation*}
This operation is named {\it blow-up}. It is important to stress that this operation is possible if and only if the $p$-valent node has all the helicity arrows with the same direction.
These are the only cases given the relation outlined before between helicity flows and singularity structure of an on-shell diagram. 

Let us now consider the on-shell diagram having the topology of a square with nodes (three-particle amplitudes) of alternating colour at the vertices. If the diagram is not decorated,
the two possible ordering for alternating black/white nodes are actually equivalent: they contain exactly the same sub-diagrams. This means that a diagram having a square with alternating
black and white nodes as a sub-diagram can be mapped into an equivalent diagram by exchanging the black and white nodes in the square. This equivalence operation is called {\it square move}.
For the decorated diagrams, the requirement that the square diagram contains the same sub-diagrams means that the helicity flows are preserved. This turns out to be true just if the external helicity
arrows with the same direction are adjacent
\begin{equation*}
 \raisebox{-1.2cm}{\scalebox{.25}{\includegraphics{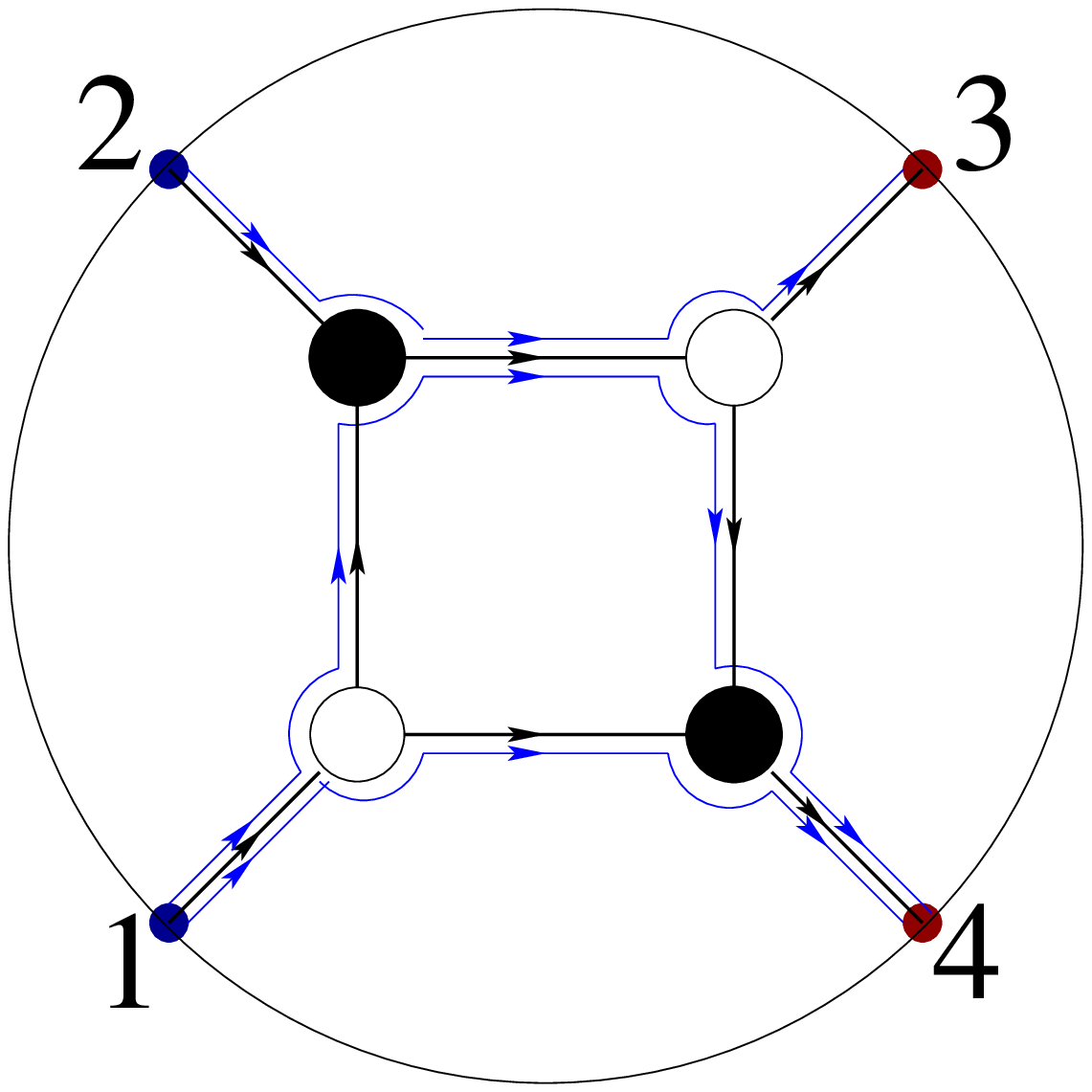}}}\quad\Longleftrightarrow\quad
 \raisebox{-1.2cm}{\scalebox{.25}{\includegraphics{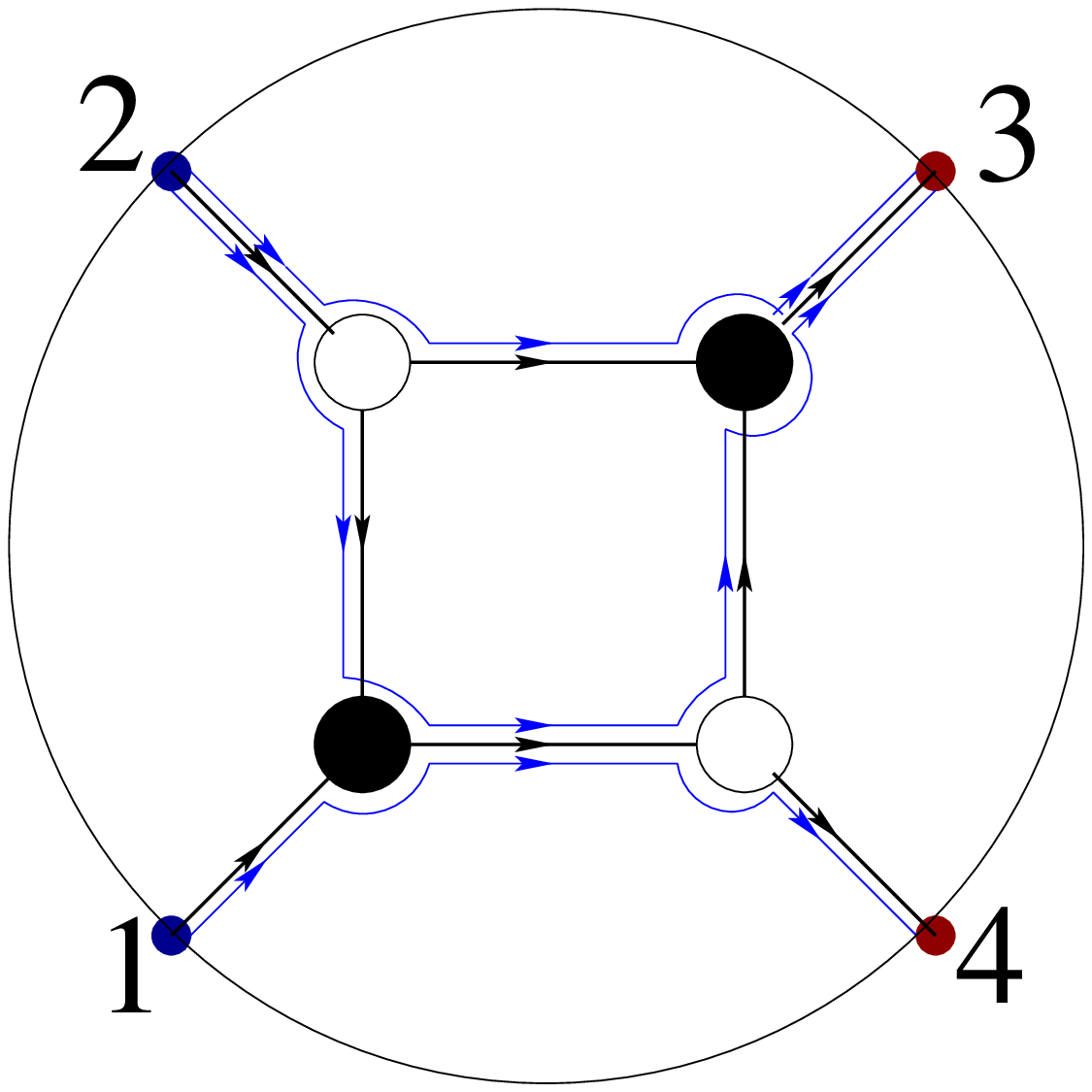}}}
\end{equation*}
If instead the states with the same helicity are not adjacent, the helicity flow structure is sensibly different, with one of the two configuration allowing for both the multiplet to propagate in the
internal lines generating helicity loops:
\begin{equation*}
 \raisebox{-1.2cm}{\scalebox{.25}{\includegraphics{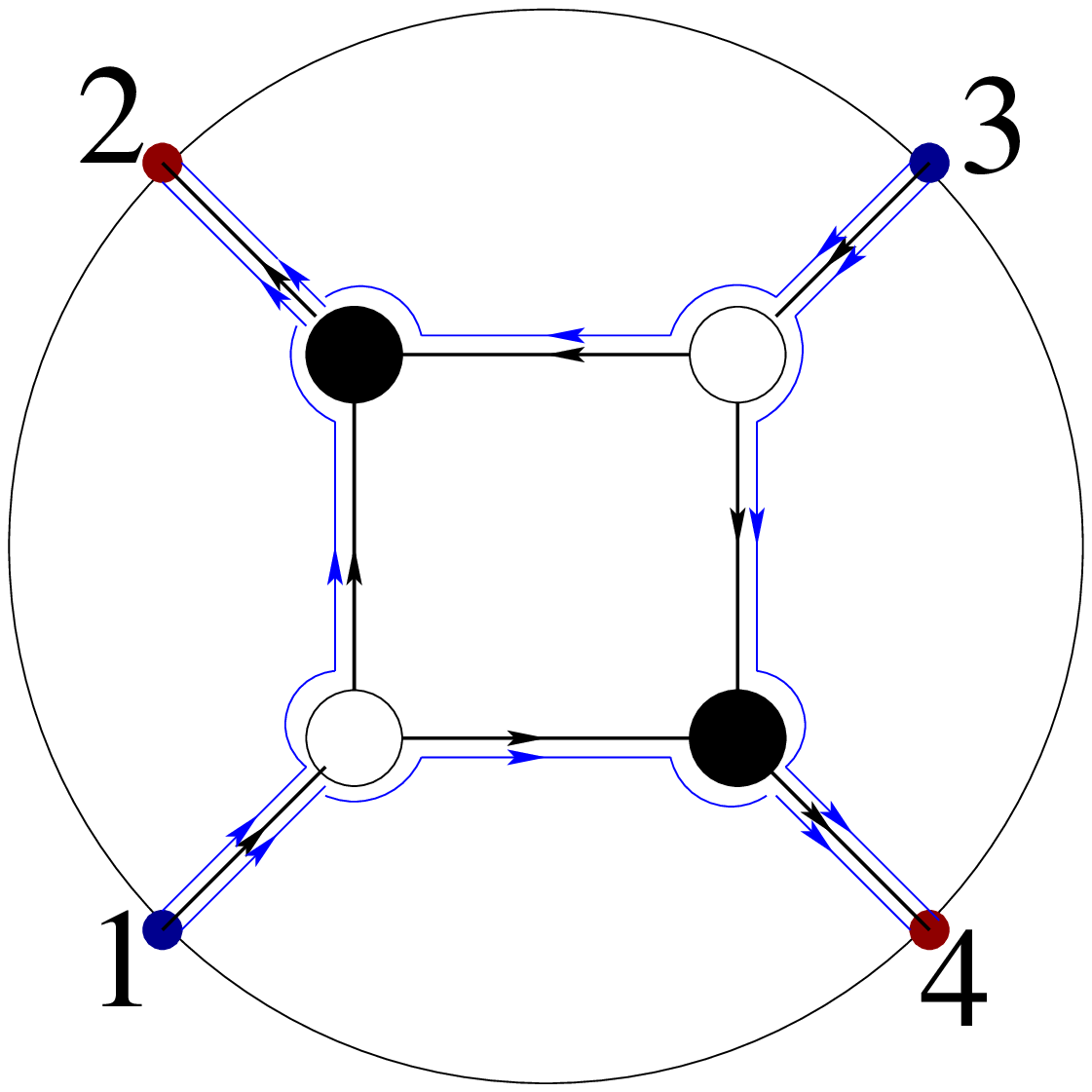}}}\quad\nLeftrightarrow\quad
 \raisebox{-1.2cm}{\scalebox{.25}{\includegraphics{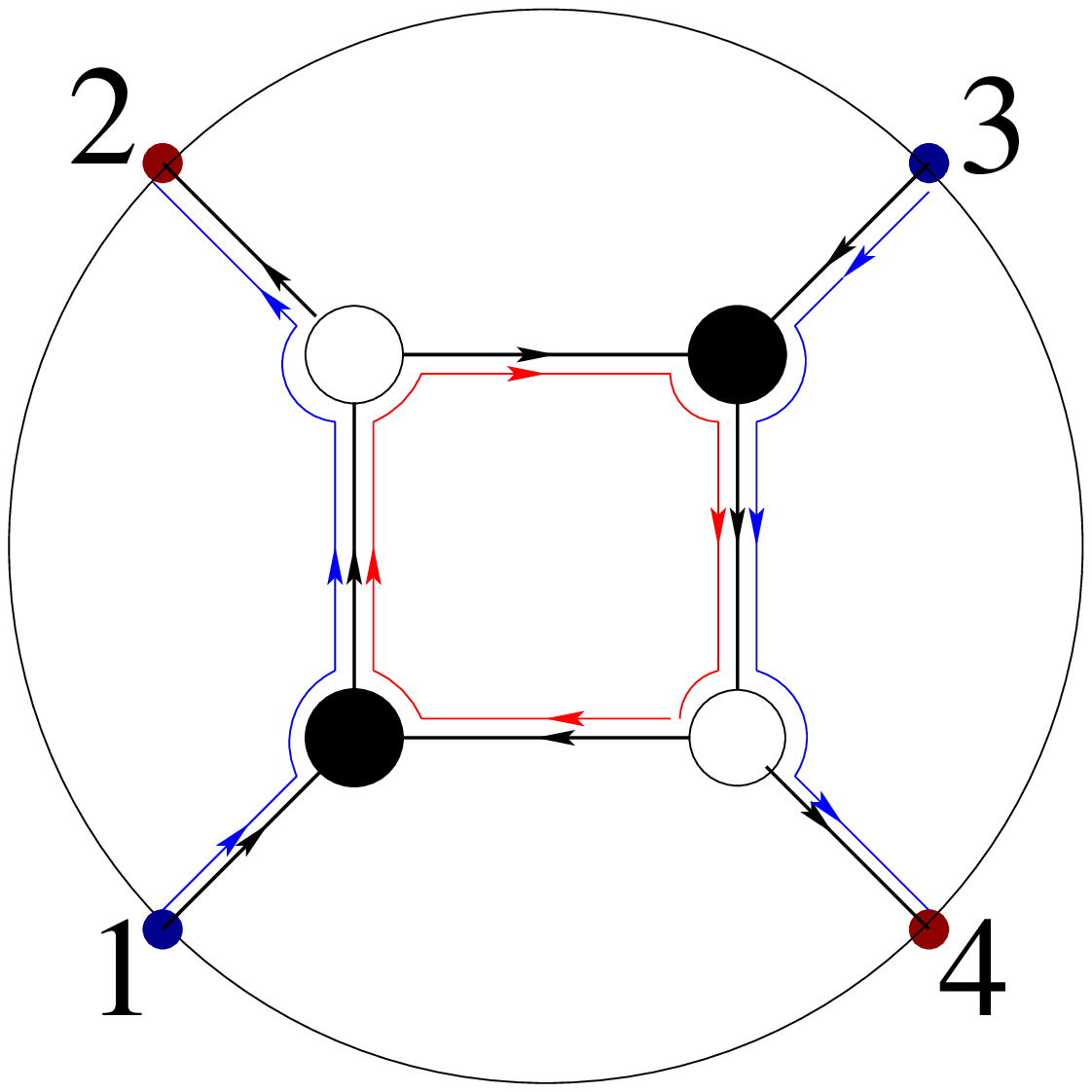}}}\,+\,
 \raisebox{-1.2cm}{\scalebox{.25}{\includegraphics{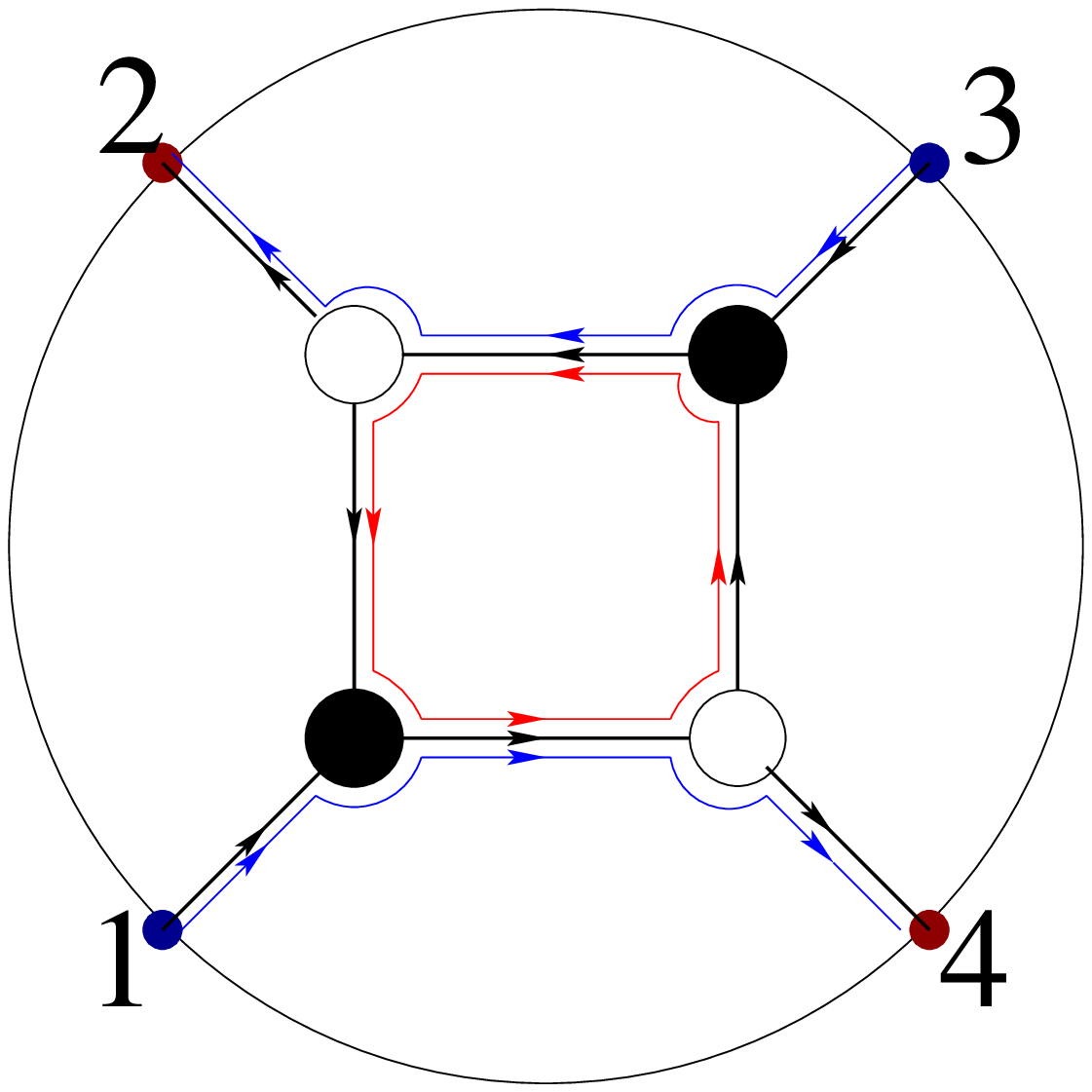}}}
\end{equation*}
Therefore, the square move is an equivalence relation if and only if the on-shell box shows two states with the same helicity direction as adjacent.

All the equivalence relation above can be viewed under the light of permutations. Disregarding for a moment the helicity flows, if one assigns the directed paths $i\,\longrightarrow\,i-1$ to the black nodes and
$i\,\longrightarrow\,i+1$ to the white nodes with the map $\sigma:\:\{n\}\,\longrightarrow\,\{2n\}$ such that $\sigma(i)\,\in\,[i,\,i+1]$ and the fixed points $\sigma(i)\,=\,i$ and $\sigma(i)\,=\,i+n$
corresponding to the black and white lollipop respectively, then a decorated permutation is assigned to each on-shell diagram and all the equivalence relations discussed above {\it do not} change the 
permutation. Thus, two equivalent on-shell diagrams belong to the same decorated permutation. Notice however that all the decorated on-shell box diagrams discussed above belong to the same decorated permutation
but, because of the different helicity flow structure, strictly speaking not all of them are equivalent. However, one can map them into each other by a helicity flow reversal operation \cite{Benincasa:2015zna}. Thus, the fact
that all those diagrams belong to the same permutation means that they are related by Ward identities. If we associate a point to each helicity configuration and an edge to the helicity flow reversal, the
resultant polytope represent the Ward identities relating different decorated on-shell diagrams
\begin{equation*}
  \raisebox{-1.8cm}{\scalebox{.65}{\includegraphics{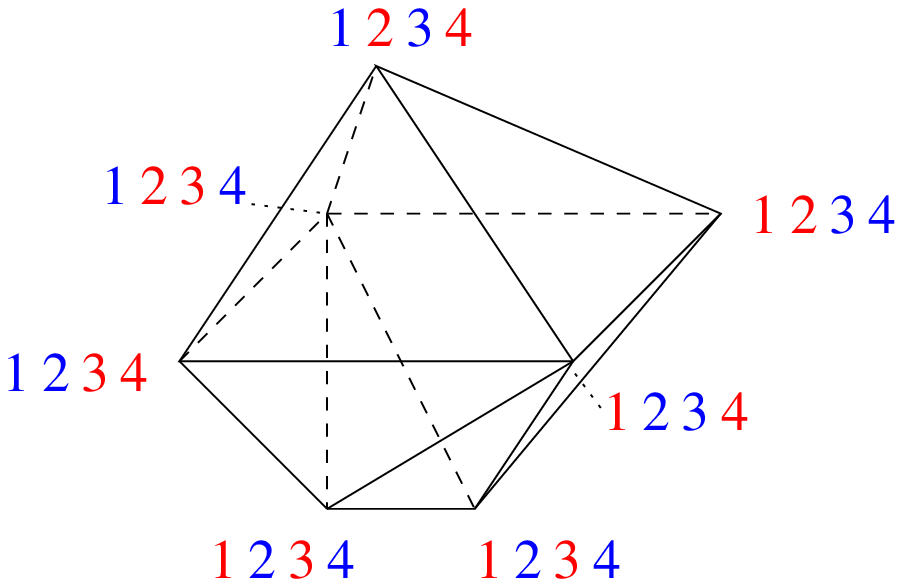}}}  \qquad\Longleftrightarrow\qquad
  \raisebox{-1.2cm}{\scalebox{.30}{\includegraphics{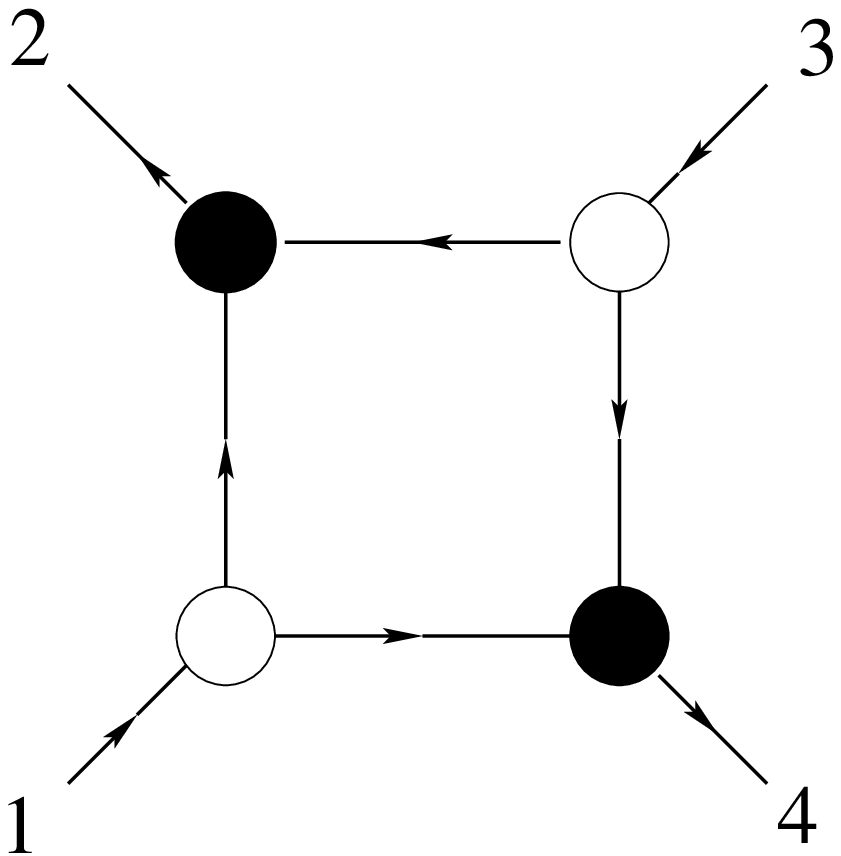}}},
\end{equation*}
where the {\color{blue} blue} ({\color{red} red}) labels indicate the incoming (outgoing) helicity arrows, while on the right it is represented the on-shell diagram which sits at the top vertex of the polytope. This diagram, as well as 
the ones sitting at the two 
lowest vertices, does not admit the square move equivalence relation, while the other four vertices of the polytope correspond to diagram which do admit such an equivalence relation. The polytope associated to on-shell boxes with 
exchanged white and black nodes can be obtained from the one above by contracting the two lowest vertices ${\color{red} 1}{\color{blue} 2}{\color{red} 3}{\color{blue} 4}$ and expanding the top vertex 
${\color{blue} 1}{\color{red} 2}{\color{blue} 3}{\color{red} 4}$ into two.

The polytope above, together with the operation of contraction and expansion of the vertices with alternating colors for its labels, represents all the possible functions with ordering $(1234)$ which can be defined on the top 
cell of $Gr(2,4)$ and can be completely constructed via BCFW bridges.

Finally, there is a further operation which maps a given diagram into another diagram with one face less singling out one degree of freedom. It can be performed whenever a diagram has, as a sub-diagram, a black
node and a white node connected through two edges forming a bubble. This diagrammatic operation goes under the name of {\it bubble reduction}. If the helicity arrows along these two lines have the same direction, 
than the bubble can be replaced by a single edge decorated with the same helicity arrow, factorising a $d\log$-form:
\begin{equation*}
 \raisebox{-.4cm}{\scalebox{.4}{\includegraphics{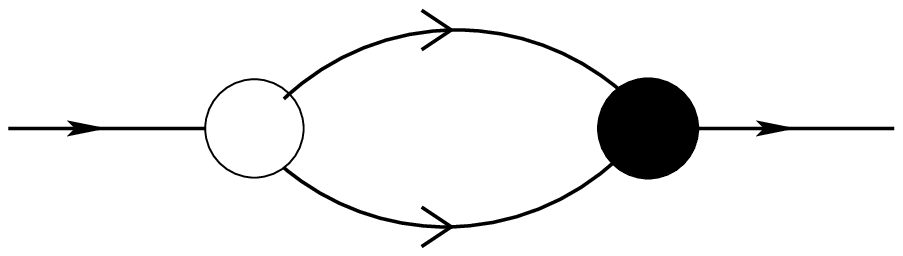}}}\quad =\quad
 \raisebox{-.0cm}{\scalebox{.4}{\includegraphics{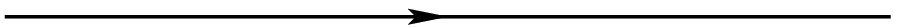}}}\;d\log{\zeta}
\end{equation*}
In this case just one coherent state is allowed to propagate. If instead both the two coherent states can run into the bubble, the diagram will show the two helicity loops signalising the appearance of a multiple pole.
The bubble reduction can be still performed, factorising a different differential form than the $d\log$:
\begin{equation*}
 \begin{split}
  &\raisebox{-.4cm}{\scalebox{.4}{\includegraphics{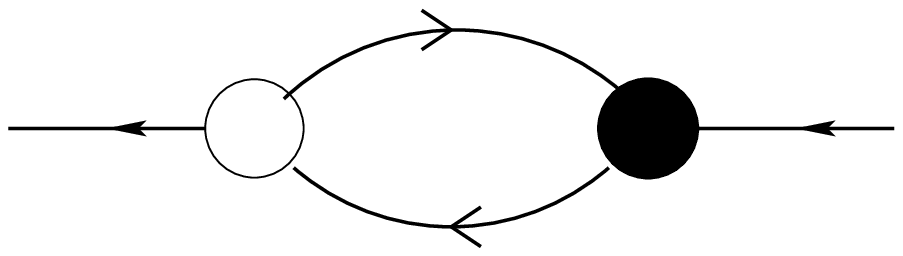}}}\quad +\quad
   \raisebox{-.4cm}{\scalebox{.4}{\includegraphics{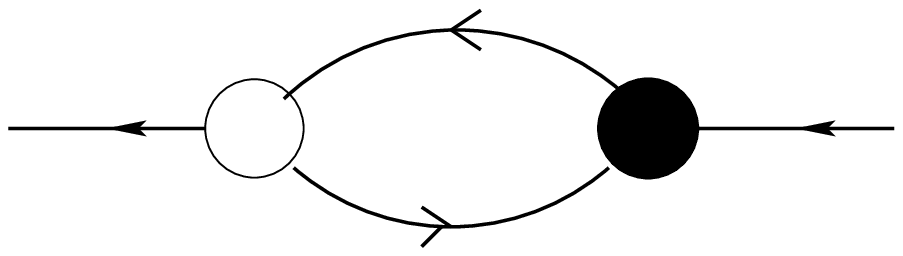}}}\quad =\\
  &\hspace{3cm}\raisebox{-.0cm}{\scalebox{.4}{\includegraphics{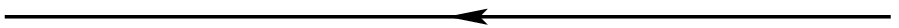}}}\;\frac{d\zeta}{\zeta}
   \left[
    \frac{1}{(1-\zeta)^{4-\mathcal{N}}}+\frac{(-\zeta)^{4-\mathcal{N}}}{(1-\zeta)^{4-\mathcal{N}}}
   \right]
 \end{split}
\end{equation*}
which, for $\mathcal{N}\,=\,3$, reduces to the $d\log{\zeta}$ form as in the maximally supersymmetric case.


\section{Decorated On-Shell Diagrams and the Grassmannian}\label{sec:GrassDOS}

The undecorated on-shell diagrams can be naturally associated to the total non-negative Grassmannian \cite{ArkaniHamed:2012nw}. In this section we will quickly review the basic definition of the Grassmannian and its 
properties\footnote{For a more exhaustive exposition see \cite{Postnikov:2006kva, ArkaniHamed:2012nw, Franco:2013nwa, Lam:2015uma}}
and then we will discuss the Grassmannian formulation for amplitudes in $\mathcal{N}\,<\,4$ SYM theories.


\subsection{Generalities on the Grassmannian}\label{subsec:GrassGen} 

The Grassmannian $Gr(k,n)$ is defined as the space of $k$-planes in $n$-dimensions intersecting at the origin. Any of its elements $ºC\,\in\,Gr(k,n)$ can be represented as a $k\times n$ matrix. Notice that the space spanned
by the $k$ rows is not changed by a $GL(k)$ transformation, so that the Grassmannian can be also defined as the space of $(k\times n)$ matrices modulo $GL(k)$ and, consequently, its dimension is $k(n-k)$. 

The degrees of freedom of $C\,\in\,Gr(k,n)$ can be parametrised via the so-called Pl{\"u}cker coordinates, which are nothing but the set of maximal minors $\left\{\Delta_I,\:I\,\in\,\begin{pmatrix} [n]\\ k \end{pmatrix}\right\}$.
Of all these minors, just a subset of them is really independent: The Grassmannian is defined on the subspace of the Pl{\"u}cker embedding defined by the so-called Pl{\"u}cker relations
\begin{equation}\eqlabel{eq:PluckRel}
 \begin{array}{c}
  Gr(k,n)\:\longrightarrow\:\hspace{.8cm}\mathbb{P}^{\mbox{\tiny $\begin{pmatrix} [n] \\ k\end{pmatrix}-1$}}\\
  \phantom{\ldots}
  \hspace{1cm}C\hspace{.4cm}\longrightarrow\hspace{.3cm}\left[\Delta_{\mbox{\tiny $I$}}\left(C\right)\right]_{I\,\in\,\mbox{\tiny $\begin{pmatrix} [n] \\ k \end{pmatrix}$}}
 \end{array}
 ,
 \hspace{1cm}
 \sum_{i=1}^{k+1}(-1)^{i-1}\Delta_{\mbox{\tiny $I_1\cup a_i$}}\Delta_{\mbox{\tiny $I_2\setminus a_i $}}\:=\:0,
\end{equation}
with $I_1$ and $I_2$ being respectively $(k-1)$- and $(k+1)$-element subsets of $[n]$ and $a_i\,\in\,I_2$. Importantly, the Pl{\"u}cker coordinates $\Delta_I$'s are $SL(k)$-invariant, so that an invariant way to parametrise
an element of the Grassmannian is through ratios of the Pl{\"u}cker coordinates themselves, which is instead $GL(k)$-invariant. They allow to define affine charts 
$\Omega_{I}\:\equiv\:\left\{C\,\in\,Gr(k,n)\,|\,\Delta_{I}(C)\,\neq\,0\right\}$,  whose representative $C^{\mbox{\tiny $\star$}}$ is given by a matrix with the $(k\times k)$ identity $\mathbb{I}_{\mbox{\tiny $(k\times k)$}}$ placed in 
the columns $I$, and the collection  of {\tiny{$\begin{pmatrix} n \\ k \end{pmatrix}$}} covers the whole $Gr(k,n)$. Choosing $\Omega_{I}$ such as $\Delta_{I}(C)\,\neq\,0$ is the first non-zero Pl{\"u}cker coordinates in lexicographic 
order, then the Grassmannian is decomposed in the so-called Schubert cells, for which all the Pl{\"u}cker coordinate lexicographically larger than $I$ are not constrained and thus can be zero or non-zero. The intersection of the Schubert
cells $\Omega_{I_{i}}^{\mbox{\tiny $(i)$}}$ ($i\,=\,1\,\ldots\,n$), with $i$ labelling where the counting for the lexicographic order starts, defines a positroid stratum of $Gr(k,n)$. The locus in $Gr(k,n)$ characterised by having
all the Pl{\"u}cker coordinates non-negative defines the so-called totally non-negative Grassmannian $Gr_{\mbox{\tiny $\ge\,0$}}(k,n)$ \cite{Postnikov:2006kva} and its intersection with the positroid stratification defines a positroid 
cell. The positroid cell having all the Pl{\"u}cker coordinates non-zero is named top-cell, which is the highest dimensional cell. Furthermore, to each positroid stratification it is naturally associated a rational top-form which is 
characterised by logarithmic poles at the boundary and the absence of zeros, and it is $GL(k)$-invariant
\begin{equation}\eqlabel{eq:GrTopForm} 
 \omega_{\mbox{\tiny $k,n$}}\:=\:\frac{d^{k\times n}C}{\mbox{Vol$\{GL(k)\}$}}\frac{1}{\Delta_{12\ldots k}\Delta_{23\ldots(k+1)}\ldots\Delta_{n1\ldots(k-1)}}.
\end{equation}
The Grassmannian degrees of freedom can be nicely parametrised via the (undecorated) on-shell diagrams by either assigning a weight $\alpha_e$ to each edge $e$ and fixing a perfect orientation  with
two (one) incoming arrows for the black (white) nodes, or by assigning a variable to each face into which the disc is divided by the diagram. In the first case, fixing a perfect orientation implies fixing the sources and
sinks at the boundary of the disc. Then the entry $c_{ij}$ of the Grassmannian representative matrix is given by the sum of the products of the edge variables $\alpha$'s along all the paths from the source $i$ to the sink $j$:
\begin{equation}\eqlabel{eq:EdgeVar}
 c_{ij}\:=\:-\sum_{\Gamma\in\{i\rightarrow j\}}\prod_{e\in\Gamma}\alpha_e
\end{equation}
This definition leaves unfixed a $GL(1)$ for each vertex, which can be used to further set some of the edge variables to one. The edge variables can be actually chosen in such a way that all the minors $\Delta_I(C)$ are
positive if they are themselves real and positive \cite{ArkaniHamed:2012nw}. 

The relation between the entries of the matrix representative of the Grassmannian and the face variables is instead given (minus) the sum of the products of the variables associated to the faces which will be inside the paths from
the source $i$ to the sink $j$ once they get closed clockwise:
\begin{equation}\eqlabel{eq:FaceVar}
 c_{ij}\:=\:-\sum_{\Gamma\in\{i\rightarrow j\}}\prod_{f\in\hat{\Gamma}}(-f).
\end{equation}   
Again, not all the parameters turn out to be independent, rather they are linked by the relation $\prod (-f)\,=\,1$, with the index of the product running over all the faces.

The choice of a given perfect orientation amounts to the choice of a coordinate patch which (partially) covers the Grassmannian. Any equivalence relation amounts to a change of coordinates in the same patch. Notice also that
the parametrisation of the Grassmannian via the on-shell diagrams does not rely of any amplitude interpretation.


\subsection{On-shell processes for $\mathcal{N}\,<\,4$ SYM on the Grassmannian}\label{subsec:GrasSA}

On-shell processes can be seen as an integral over the Grassmannian (partially) localised on some kinematic support:
\begin{equation}\eqlabel{eq:GrassAmpl}
 \mathcal{M}_{k,n}^{\mbox{\tiny (OD)}}\:=\:\int_{\Pi}\omega_{k,n}(C)\,\mathcal{F}^{\mbox{\tiny $(4k|k\mathcal{N})$}}(\delta)\,f(\Delta(C)),
\end{equation}
where $\omega_{k,n}$ is the canonical top form defined in \eqref{eq:GrTopForm}, $\mathcal{F}^{\mbox{\tiny $(4k|k\mathcal{N})$}}(\delta)$ is the kinematic support, whose explicit expression depends on the space where the kinematics
is defined. Finally, the function $f(\Delta)$ of the Pl{\"u}cker coordinates $\Delta$ guarantees that the integrand transform properly under $C\,\longrightarrow\,t\,C$. For concreteness, we will consider for the time being the 
kinematics either defined in momentum space or in twistor space:
\begin{equation}\eqlabel{eq:KinSpace}
 \begin{split}
  &\left.\mathcal{F}^{\mbox{\tiny $(4k|k\mathcal{N})$}}(\delta)\right|_{\mbox{\tiny $(\lambda,\tilde{\lambda})$}}\:=\:\delta^{\mbox{\tiny $(2\times(n-k))$}}\left(\lambda\cdot C^{\mbox{\tiny $\perp$}}\right)
    \delta^{\mbox{\tiny $(2\times k)$}}\left(C\cdot\tilde{\lambda}\right)\delta^{\mbox{\tiny $(k\times\mathcal{N})$}}\left(C\cdot\tilde{\eta}\right),\\
  &\left.\mathcal{F}^{\mbox{\tiny $(4k|k\mathcal{N})$}}(\delta)\right|_{\mbox{\tiny $\mathcal{W}$}}\:=\:\delta^{\mbox{\tiny $(4k|k\mathcal{N})$}}\left(C\cdot\mathcal{W}\right),
 \end{split}
\end{equation}
where $C^{\mbox{\tiny $\perp$}}$ in the first line is the orthogonal complement of $C$ defined by $C\cdot C^{\mbox{\tiny $\perp$}}\,=\,0$, while $\mathcal{W}\,\equiv\,(\tilde{\mu},\,\tilde{\lambda},\,\tilde{\eta})^{\mbox{\tiny T}}$ 
encodes the kinematics in twistor space, and $\tilde{\mu}$ is defined through the twistor transform $\int d^{2\times n}\,e^{i\lambda\cdot\tilde{\mu}}$. The $k$-plane $C$ is thus orthogonal to the $2$-plane $\tilde{\lambda}$ 
(and contains the $2$-plane $\lambda$) in momentum space, while it is orthogonal to $\mathcal{W}$ in twistor space. Notice that the $\delta$-functions localise $2n-4$ degrees of freedom of $C$, so that the cells of the
Grassmannian with exactly $2n-4$ degrees of freedom correspond to rational functions of the kinematic data (up to the momentum conserving delta-function support), while lower cells will have some $\delta$-function support and higher
cells will be some differential form. Notice that the top cell, which has $k(n-k)$ dimensions is fully localised just in the $k\,=\,2$ and $k\,=\,n-2$ case, {\it i.e.} in the MHV and $\bar{\mbox{MHV}}$ sectors. 
In twistor space one actually has distribution constraining the twistor data.

Let us now consider the external kinematic data in twistor space, so that the generic form for an on-shell diagram can be written as
\begin{equation}\eqlabel{eq:ODtwist}
 \mathcal{M}_{\mbox{\tiny $k,n$}}^{\mbox{\tiny (OD)}}\:=\:\int_{\Pi}\omega_{k,n}\,\delta^{\mbox{\tiny $(4k|k\mathcal{N})$}}\left(C\cdot\mathcal{W}\right)f(\Delta).
\end{equation}
A way to constrain the function $f(\Delta)$ is the requirement that the integrand is invariant under the transformation $C\:\longrightarrow\:t\,C$. While the top form $\omega_{k,n}$ is indeed invariant under such a 
transformation, the maximal minors $\Delta_{\mbox{\tiny $I$}}$ are mapped into $t^{k}\Delta_{I}$, while the twistor space $\delta$-function provide a factor $t^{-k(4-\mathcal{N})}$. Thus, in order for the integrand to
be invariant, the function $f(\Delta)$ needs to transform as: $f(\Delta)\,\longrightarrow\,f(t^{k}\Delta)\,=\,t^{k(4-\mathcal{N})}f(\Delta)$, {\it i.e.} it needs to transform as a maximal minor with some power. 
Its general structure therefore becomes 
\begin{equation}\eqlabel{eq:fgenstruct}
 f(\Delta)\:=\:\Delta_{s_1\ldots s_k}^{4-\mathcal{N}}\mathfrak{f}\left(\frac{\Delta_I}{\Delta_J}\right),
\end{equation}
where the indices $s_1,\ldots,\,s_k$ indicates the sources of the helicity arrows in the decorated on-shell diagrammatics, while $\mathfrak{f}(\Delta_I/\Delta_J)$ is just a function of ratios of the Pl{\"u}cker coordinates.
The little group covariance also implies that $\mathfrak{f}$ needs to be invariant under a little group transformation. On the Grassmannian such a transformation can be seen as just the rescaling of a given column of $C$:
$c^{\mbox{\tiny $(i)$}}\,\longrightarrow\,t_i\,c^{\mbox{\tiny $(i)$}}$. Thus, in order for $\mathfrak{f}$ to be invariant under the little group, it needs to be a (sum of) ratio(s) of the Pl{\"u}cker coordinates such that numerators
and denominators have the same indices but shuffled. Notice that, being a function of ratios of Pl{\"u}cker coordinates, $\mathfrak{f}$ introduces new (higher order) singularities, breaking the general logarithmic structure 
at the boundary.

Finally, coming to the parametrisation of the Grassmannian, even if in principle one could keep using whichever parametrisation coming from assigning a perfect orientation to the undecorated counterpart, the helicity arrows
provide a physical perfect orientation which becomes the preferred way to parametrise $C\,\in\,Gr(k,n)$ given that makes the physical structure related to the helicity flows manifest. With this choice, irrespectively of the type
(edge or face) of the variables that one can use, the Pl{\"u}cker coordinates related to the sources only is set to one.

For future reference, let us write here the explicit Grassmannian form for the three-particle building blocks:
\begin{equation}\eqlabel{eq:3ptGrass}
 \begin{split}
  &\mathcal{M}_{\mbox{\tiny $1,3$}}\:=\:\raisebox{-1cm}{\scalebox{.30}{\includegraphics{M3-2b.eps}}}\:=\:
   \int\frac{d^{\mbox{\tiny $(1\times3)$}}C}{\mbox{Vol}\{GL(1)\}}\frac{\delta^{\mbox{\tiny $(4|\mathcal{N})$}}(C\cdot\mathcal{W})\Delta_{3}^{4-\mathcal{N}}}{\Delta_{1}\Delta_{2}\Delta_{3}},\\
   &\mathcal{M}_{\mbox{\tiny $2,3$}}\:=\:\raisebox{-1cm}{\scalebox{.30}{\includegraphics{M3-1b.eps}}}\:=\:
   \int\frac{d^{\mbox{\tiny $(2\times3)$}}C}{\mbox{Vol}\{GL(2)\}}\frac{\delta^{\mbox{\tiny $(8|2\mathcal{N})$}}(C\cdot\mathcal{W})\Delta_{12}^{4-\mathcal{N}}}{\Delta_{12}\Delta_{23}\Delta_{31}},
 \end{split}
\end{equation}
where the Grassmannian measure (the top form) $\omega_{\mbox{\tiny $k,3$}}$ has been explicitly written. Notice that the two functions defining these three-point objects, 
$\delta^{\mbox{\tiny $(4k|k\mathcal{N})$}}(C\cdot\mathcal{W})\Delta_{I}^{4-\mathcal{N}}$, are defined on the top cell of $ G(k,3)$. Furthermore, $\mathfrak{f}\,=\,1$ is the only possible function which is $GL(k)$-invariant and
has the correct little group behaviour at the same time. As a last comment, notice that any higher-point on-shell diagram, being built gluing three-particle amplitude is characterised by 
$\mathfrak{f}\,=\,\left[\mathfrak{h}(\Delta_{\mbox{\tiny $I$}}/\Delta_{\mbox{\tiny $J$}})\right]^{4-\mathcal{N}}$.



\section{Amplitudes, Singularities and the Grassmannian}\label{sec:SingGrass}

In the previous section we discussed the general structure of decorated on-shell diagrams as defined by the usual delta-functions times a rational function of the Pl{\"u}cker coordinates. Their general form is constrained 
by the little group covariance and the invariance under $GL(k)$ transformations, and it can be computed by the usual amalgamation and projection operations on the three-particle amplitudes \eqref{eq:3ptGrass}. The appearance
of such a rational function introduces new singularities, which will be the main subject of this section.


As a first step, let us build the Grassmannian representation for the decorated on-shell boxes with non-adjacent sources. The starting point is the following singularity diagram
\begin{equation}\eqlabel{eq:4ptGrass}
 \raisebox{-0.7cm}{\scalebox{.25}{\includegraphics{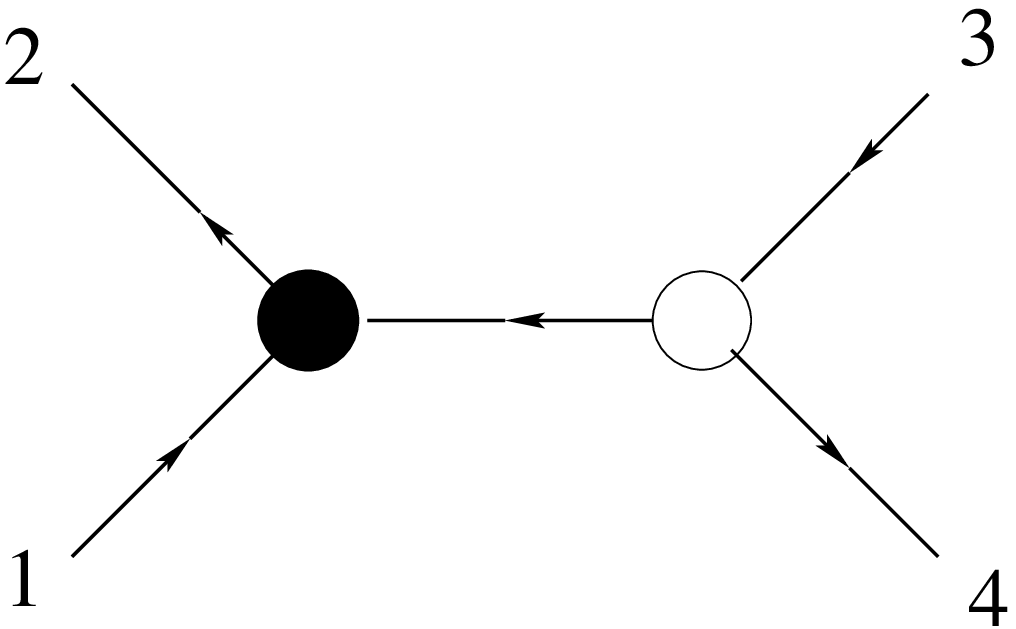}}}\:=\:
  \int\frac{d^{\mbox{\tiny $(2\times4)$}}C}{\mbox{Vol}\{GL(2)\}}\,\frac{\delta^{\mbox{\tiny $(8|2\mathcal{N})$}}(C\cdot\mathcal{W})\delta(\Delta_{34})\Delta_{13}^{4-\mathcal{N}}}{\Delta_{12}\Delta_{23}\Delta_{41}},
\end{equation}
from which the on-shell boxes can be constructed via a BCFW bridge, whose degree of freedom gets localised by the $\delta$-function support in \eqref{eq:4ptGrass}. Concretely, the on-shell box with no internal helicity loops is obtained
by applying a BCFW to \eqref{eq:4ptGrass} in the $(4,1)$-channel:
\begin{equation}\eqlabel{eq:4ptGrass2}
 \begin{split}
  \raisebox{-0.9cm}{\scalebox{.25}{\includegraphics{4ptDiag2b.eps}}}\:&=\:
   \int\frac{dz}{z}\,\int\frac{d^{\mbox{\tiny $(2\times4)$}}C}{\mbox{Vol}\{GL(2)\}}\,\frac{\delta^{\mbox{\tiny $(8|2\mathcal{N})$}}(C\cdot\mathcal{W})
    \delta(\Delta_{34}-z\Delta_{13})\Delta_{13}^{4-\mathcal{N}}}{\Delta_{12}\Delta_{23}\Delta_{41}}\:=\\
  &=\:\int \frac{d^{\mbox{\tiny $(2\times4)$}}C}{\mbox{Vol}\{GL(2)\}}\frac{\delta^{\mbox{\tiny $(8|2\mathcal{N})$}}(C\cdot\mathcal{W})\Delta_{13}^{4-\mathcal{N}}}{\Delta_{12}\Delta_{23}\Delta_{34}\Delta_{41}},
 \end{split}
\end{equation}
while adding a BCFW bridge in the $(2,3)$-channel returns the on-shell boxes with the internal helicity loops:
\begin{equation}\eqlabel{eq:4ptGrass3}
 \begin{split}
  \raisebox{-0.9cm}{\scalebox{.25}{\includegraphics{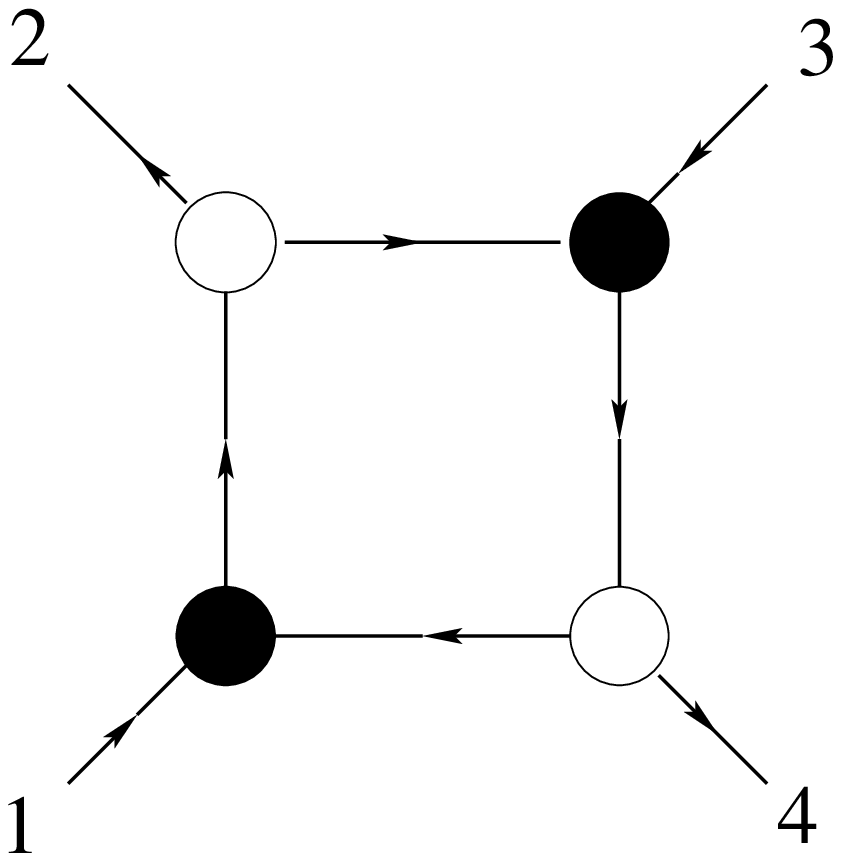}}}\:&=\:
   \int\frac{dz}{z}\,\int\frac{d^{\mbox{\tiny $(2\times4)$}}C}{\mbox{Vol}\{GL(2)\}}\,\frac{\delta^{\mbox{\tiny $(8|2\mathcal{N})$}}(C\cdot\mathcal{W})
    \delta(\Delta_{34}+z\Delta_{24})(\Delta_{13}+z\Delta_{12})^{4-\mathcal{N}}}{\Delta_{12}\Delta_{23}\Delta_{41}}\:=\\
  &=\:\int \frac{d^{\mbox{\tiny $(2\times4)$}}C}{\mbox{Vol}\{GL(2)\}}\frac{\delta^{\mbox{\tiny $(8|2\mathcal{N})$}}(C\cdot\mathcal{W})\Delta_{13}^{4-\mathcal{N}}}{\Delta_{12}\Delta_{23}\Delta_{34}\Delta_{41}}
    \left(\frac{\Delta_{23}\Delta_{14}}{\Delta_{13}\Delta_{24}}\right)^{4-\mathcal{N}},\\
  \raisebox{-0.9cm}{\scalebox{.25}{\includegraphics{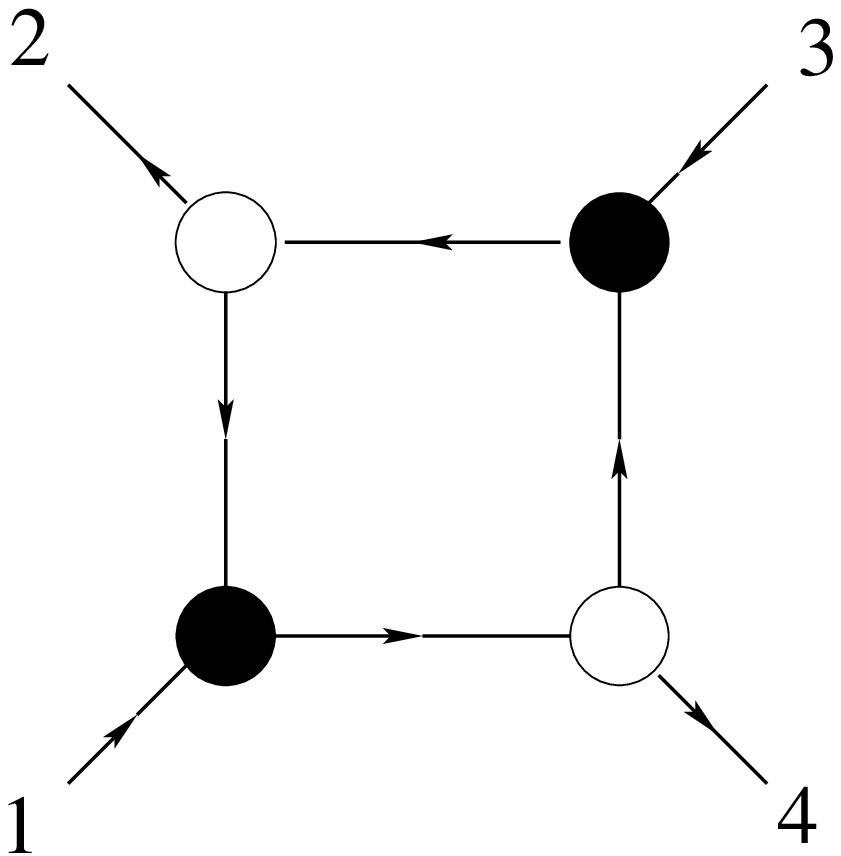}}}\:&=\:
   \int dz\,z^{3-\mathcal{N}}\int\frac{d^{\mbox{\tiny $(2\times4)$}}C}{\mbox{Vol}\{GL(2)\}}\,\frac{\delta^{\mbox{\tiny $(8|2\mathcal{N})$}}(C\cdot\mathcal{W})
    \delta(\Delta_{34}+z\Delta_{24})\Delta_{12}^{4-\mathcal{N}}}{\Delta_{12}\Delta_{23}\Delta_{41}}\:=\\
  &=\:\int \frac{d^{\mbox{\tiny $(2\times4)$}}C}{\mbox{Vol}\{GL(2)\}}\frac{\delta^{\mbox{\tiny $(8|2\mathcal{N})$}}(C\cdot\mathcal{W})\Delta_{13}^{4-\mathcal{N}}}{\Delta_{12}\Delta_{23}\Delta_{34}\Delta_{41}}
    \left(\frac{\Delta_{12}\Delta_{34}}{\Delta_{13}\Delta_{24}}\right)^{4-\mathcal{N}}.
 \end{split}
\end{equation}
Notice that in the second diagram one needs to take into account a Jacobian related to the change in the helicities of the states which the BCFW bridge is applied on: This is related to the fact that the BCFW bridge
introduces a new multiple pole at infinity and, consequently, the singularity \eqref{eq:4ptGrass} is not part of the boundary of the second diagram in \eqref{eq:4ptGrass3}. The Grassmannian expressions \eqref{eq:4ptGrass2} and
\eqref{eq:4ptGrass3} have the exact structure as predicted in \eqref{eq:ODtwist} and \eqref{eq:fgenstruct} from $GL(2)$-invariance and little group covariance. Actually in the present cases, even the form of 
$\mathfrak{h}(\Delta_{I}/\Delta_{J})$ can be exactly predicted directly from the on-shell diagrams. More precisely, the helicity flows encode the information about the singularity structure of the on-shell diagrams. For the 
on-shell box \eqref{eq:4ptGrass2}, the helicity flow structure guarantees that all the four sub-diagrams have exactly the same helicity configuration as the full-diagram: all of them belongs to the boundary of the on-shell box
and thus represent its singularities.  In other words, all the strata defined by $\Delta_{\mbox{\tiny $i,i+1$}}\,=\,0$ encode the singularity information of the amplitude. Consequently, $\mathfrak{h}(\Delta_{I}/\Delta_{J})$ cannot
have in the numerator any of the minors formed by two consecutive columns of $C$ and the only function which can fulfil the requirement of invariance under both $GL(2)$ and little group transformations is $1$.
 
Considering the two diagrams with helicity loops, the helicity flow structure shows that just two sub-diagrams really belongs to the boundary of each of the on-shell diagrams. These two diagrams correspond to the two complex
factorisations in a given channel (the $s$-channel for the first on-shell box in \eqref{eq:4ptGrass3}, and the $t$-channel for the second one). In other words, on the Grassmannian the decorated on-shell boxes have just
two poles of the form $\Delta_{\mbox{\tiny $i,i+1$}}\,=\,0$. Consequently, the function $\mathfrak{h}(\Delta_I/\Delta_J)$ needs to suppress the other two possible singularities, so that its numerator can correspond only
to $\Delta_{23}\Delta_{14}$ in the on-shell box with clockwise helicity loop and $\Delta_{12}\Delta_{34}$ in the on-shell diagram with counter-clockwise helicity loop. Then, the little group invariance allows the denominator
to be only $\Delta_{13}\Delta_{24}$ in both cases. Having this extra singularity corresponds to the presence of the helicity loops in the diagrammatics.

Notice also that summing the two on-shell boxes in \eqref{eq:4ptGrass3} for $\mathcal{N}\,=\,3$ and using the Pl{\"u}cker identities, one obtains \eqref{eq:4ptGrass2}, which represents the four-particle amplitude at tree-level.

Finally, let us keep focusing on the diagrams with the helicity loops. As mentioned earlier, these on-shell diagrams have higher order singularities which are graphically identified via the helicity loops themselves. The
residue related to this higher-order pole can be also diagrammatically identified by following the helicity flows. Concretely,
\begin{equation}\eqlabel{eq:HOres}
 \begin{split}
   \raisebox{-0.9cm}{\scalebox{.25}{\includegraphics{4ptDiag3b.eps}}}\qquad&\Longrightarrow\qquad\raisebox{-0.9cm}{\scalebox{.25}{\includegraphics{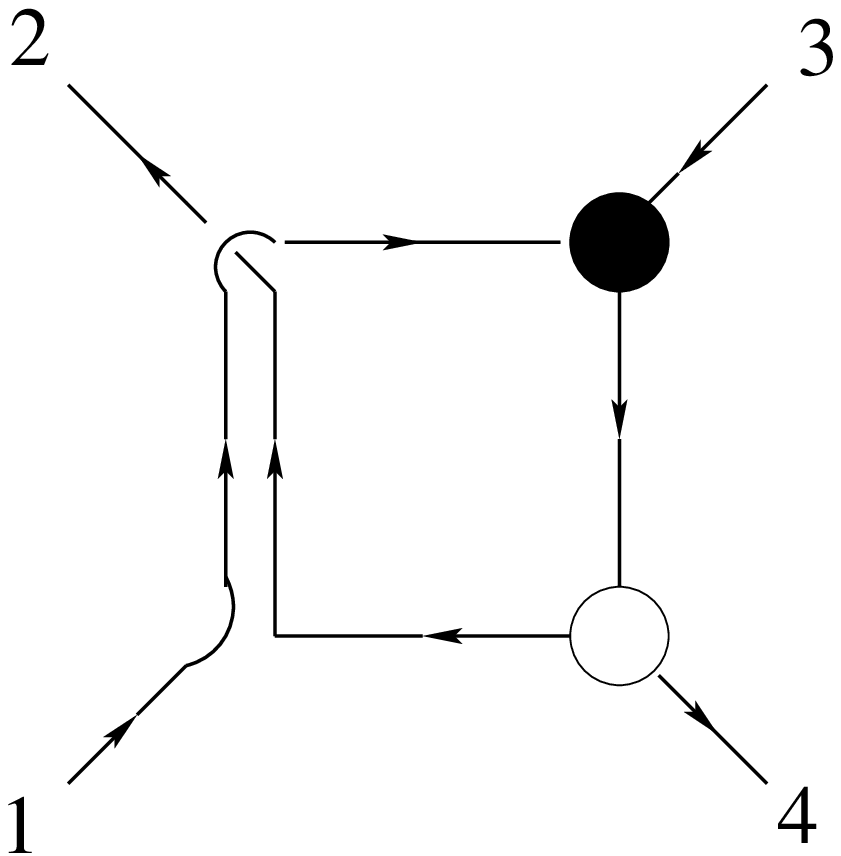}}}\:\equiv\:
    \raisebox{-0.9cm}{\scalebox{.25}{\includegraphics{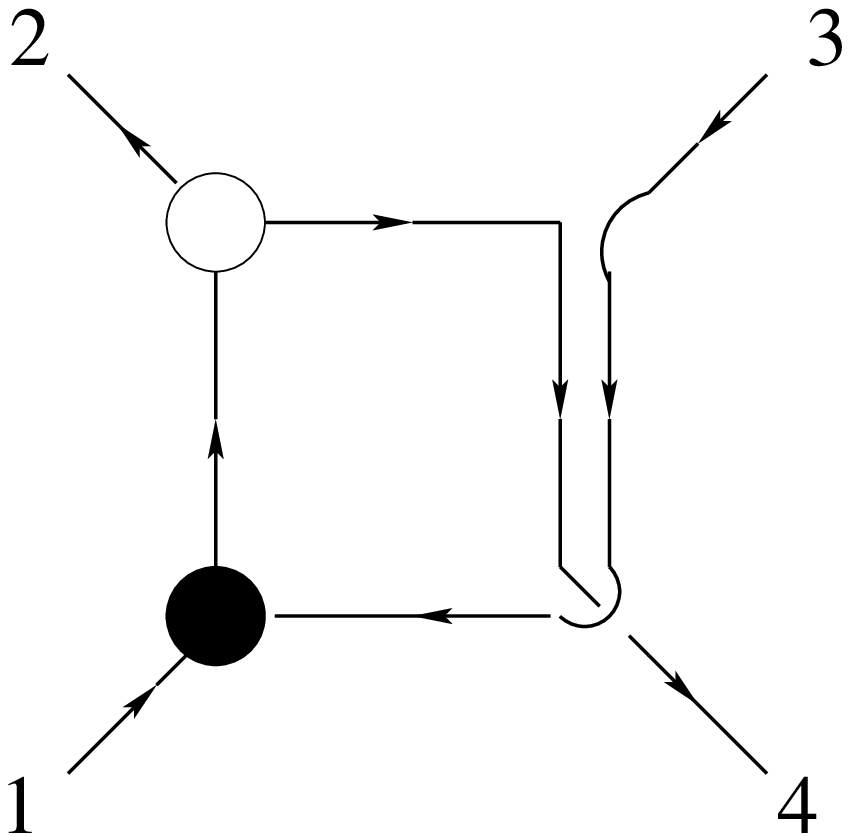}}}\\
   \raisebox{-0.9cm}{\scalebox{.25}{\includegraphics{4ptDiag4b.eps}}}\qquad&\Longrightarrow\qquad\raisebox{-0.9cm}{\scalebox{.25}{\includegraphics{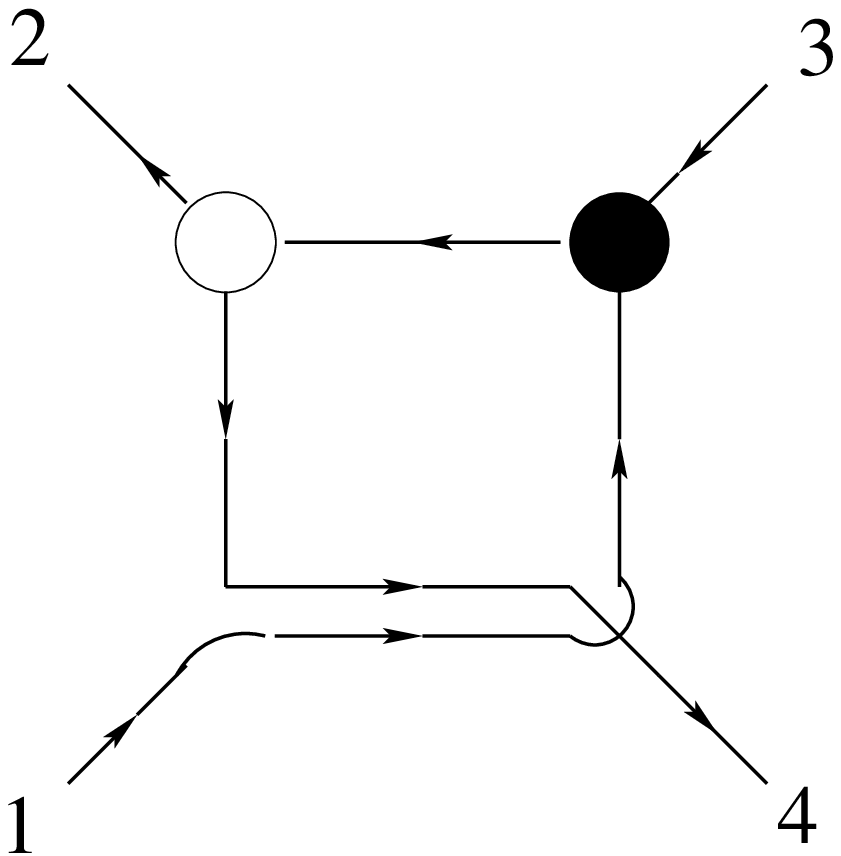}}}\:\equiv\:
    \raisebox{-0.9cm}{\scalebox{.25}{\includegraphics{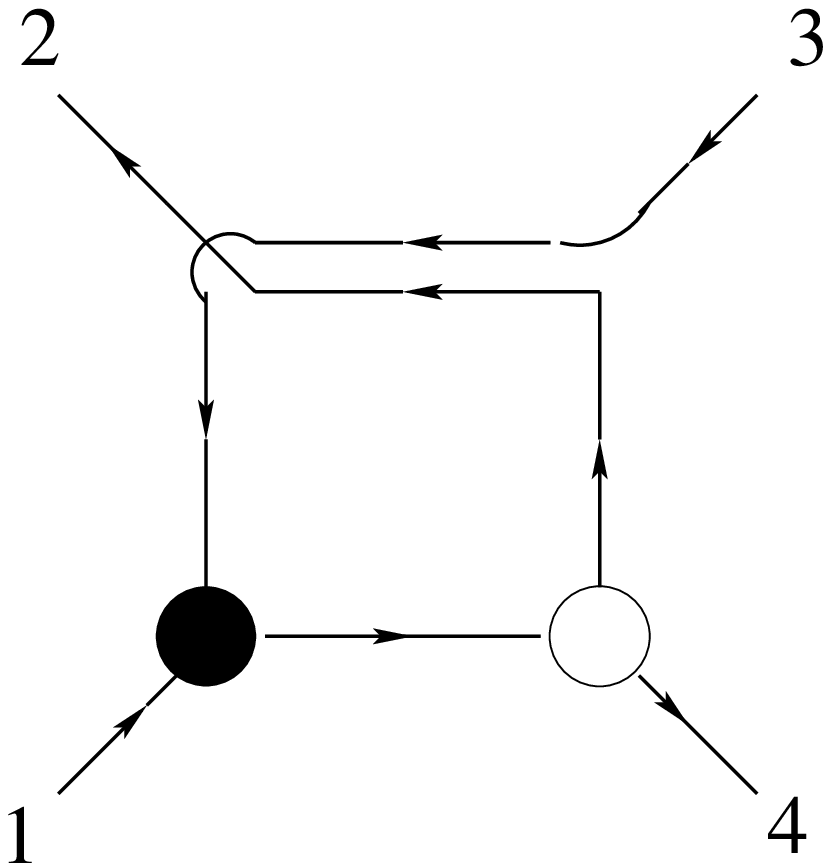}}}
 \end{split}
\end{equation}
which are $u$-channel singularities with support on a $(3-\mathcal{N})$-derivative delta-function:
\begin{equation}\eqlabel{eq:HOresGrass}
 \begin{split}
  &\raisebox{-0.9cm}{\scalebox{.25}{\includegraphics{4ptNPsingS1b.eps}}}\:=\:\int\frac{d^{\mbox{\tiny $(2\times4)$}}C}{\mbox{Vol}\{GL(2)\}}\,
    \frac{\delta^{\mbox{\tiny $(2\times4|2\times\mathcal{N})$}}\left(C\cdot\mathcal{W}\right)
    \left(\Delta_{23}\Delta_{41}\right)^{\mbox{\tiny $3-\mathcal{N}$}}}{\Delta_{34}\Delta_{21}}\delta^{\mbox{\tiny $(3-\mathcal{N})$}}\left(\Delta_{24}\right),\\
  &\raisebox{-0.9cm}{\scalebox{.25}{\includegraphics{4ptNPsingT1b.eps}}}\:=\:\int\frac{d^{\mbox{\tiny $(2\times4)$}}C}{\mbox{Vol}\{GL(2)\}}\,
    \frac{\delta^{\mbox{\tiny $(2\times4|2\times\mathcal{N})$}}\left(C\cdot\mathcal{W}\right)
    \left(\Delta_{12}\Delta_{34}\right)^{\mbox{\tiny $3-\mathcal{N}$}}}{\Delta_{23}\Delta_{14}}\delta^{\mbox{\tiny $(3-\mathcal{N})$}}\left(\Delta_{24}\right),
 \end{split}
\end{equation}
where, contrarily to the notation used so far, the apex in the last delta function indicates the number of derivatives on the $\delta$-function\footnote{More precisely, the apex of the $\delta$-function in \eqref{eq:HOresGrass} includes
a factor $(-1)^{(3-\mathcal{N})}/(3-\mathcal{N})!$, so that:
\begin{equation*}
 \delta^{\mbox{\tiny $(3-\mathcal{N})$}}(x)\:\overset{\mbox{\tiny def}}{=}\:\frac{(-1)^{3-\mathcal{N}}}{(3-\mathcal{N})!}\left(\frac{d}{dx}\right)^{3-\mathcal{N}}\delta(x).
\end{equation*}}. 
Notice that 
the two lines in \eqref{eq:HOresGrass} differ just in their lexicographic order ($2\,<\,1\,<\,3\,<\,4$ in the first line and $4\,<\,2\,<\,3\,<\,1$)  if it is seen still as a embedded in a disk, while it can be equivalently seen as a 
non-planar diagram if we embed it into an annulus. As shown in \cite{Franco:2015rma}, a given on-shell diagram can be equivalently embedded into different surfaces, the embedding not being a property of the diagrams themselves. 

A comment is now in order. The natural direct construction of the singularity diagrams in \eqref{eq:HOres} would be via the gluing of a black and a white nodes, which would return an object with support on $\delta(\Delta_{24})$. 
However, for the ones we are dealing with now, this is strictly true in the $\mathcal{N}\,=\,3$ case only. Furthermore, all the on-shell diagrams constructed by gluing black and white nodes live on $\delta$-function supports or on a 
constant support. Objects with support on distributions which are derivative of $\delta$-functions can be naturally defined from higher dimensional {\it planar} diagrams with helicity loops, as we just saw, and they correspond to
residues of (higher order) poles involving Pl{\"u}cker coordinates made out of non-adjacent columns, without the need of setting any other Pl{\"u}cker coordinate to zero. In a sense, they are {\it non-planar} objects with support on a 
derivative $\delta$-function. In the present context,
the presence of non-planar looking structures can be only due to the presence of some higher order poles in the higher dimensional diagram. Furthermore, they can be seen as {\it unwinding} a helicity loop. The presence
of factors such as the ones in the round brackets with power $3-\mathcal{N}$ are needed for both preserving the $GL(k)$-invariance of the integrand and for having the correct little group behaviour, and their specific form is
related to the direction of the helicity loop (which determines how it can be unwinded) -- indeed, all this comes naturally from the computation of the higher order residue, but it can be understood on these more general grounds.
Thus, along all the paper, any non-planar diagram will be understood as either having support on a derivative delta-function or having a higher-order pole and a constant support (it depends on its dimensionality), and it will 
be represented in a way that makes manifest which and how a helicity loop has been unwinded. Indeed, this is not completely satisfactory given that the proposed association between diagrams and functions with support on
derivative $\delta$-function makes sense just when we go from a higher dimensional object to a lower one, while it is still not clear how to construct them from lower dimensional quantities. However, starting from the higher-order
singularities \eqref{eq:HOresGrass}, it is possible to construct the $0$-forms \eqref{eq:HOres} via a BCFW bridge: applying a BCFW bridge in the $s$-channel to the higher order singularity in the first line of \eqref{eq:HOresGrass},
one obtains the on-shell box with clockwise helicity loop, a BCFW bridge in the $t$-channel to the one in the second line of \eqref{eq:HOresGrass} returns the on-shell box with counter-clockwise helicity loop. More precisely,
these BCFW-bridges on the higher-order singularities return diagrams which are {\it equivalent} to the on-shell boxes with helicity loops. Such an equivalence can be demonstrated via a residue theorem, as we will show in the
subsection \ref{subsec:IdGr24}.

Indeed these higher order non-planar structures are present in the Grassmannian $Gr(k,n)$ for any $k\,\ge\,2$ and any $n\,\ge\,4$. In the $k\,=\,2$ sector, as the case we just discussed, these structures are related to
higher order poles of type $\Delta_{i,i+2}\,=\,0$, {\it i.e.} they are identified by a single Pl{\"u}cker coordinate vanishing. As soon as we move away from this sector, new poles appear. More precisely, depending on the
helicity arrow configurations, these higher order poles can either be located at $\Delta_{i_1\ldots i_{k}}\,=\,0$ or when some special relation within the Pl{\"u}cker coordinates is satisfied, as we will show in Section
\ref{sec:Gr36}. This is exactly what happens for non-planar on-shell diagrams in $\mathcal{N}\,=\,4$ SYM \cite{Franco:2015rma, Bourjaily:2016mnp}, which is interpreted as a signature of the impossibility of expressing all the leading 
singularities as linear combination of the planar ones \cite{ArkaniHamed:2012nw, Franco:2015rma}. In the present case, the interpretation is quite different: they are a signature of the inequivalence between the two leading 
singularities in a given channel (which also reflects in the general non-validity of the square move) and thus of the existence of sub-leading singularities which cannot be expressed in terms of the leading ones. This is what lies 
behind the failure of representing an amplitude in terms of standard on-shell diagrams under any BCFW recursion. These on-shell diagrams with higher-order $\delta$-function support allow to complete such recursions and realise, as 
discussed in Appendix \ref{app:1Lint}, a direct link to the new singularities which arise in the on-shell $4L$-forms representing the $L$-loop integrand \cite{Benincasa:2015zna}.

\subsection{Identities on $Gr(2,4)$}\label{subsec:IdGr24}

Let us now consider the on-shell $1$-form which is obtained by applying a BCFW bridge to the on-shell box \eqref{eq:4ptGrass2} in the following fashion
\begin{equation}\eqlabel{eq:4ptGrass1form}
  \raisebox{-1.3cm}{\scalebox{.20}{\includegraphics{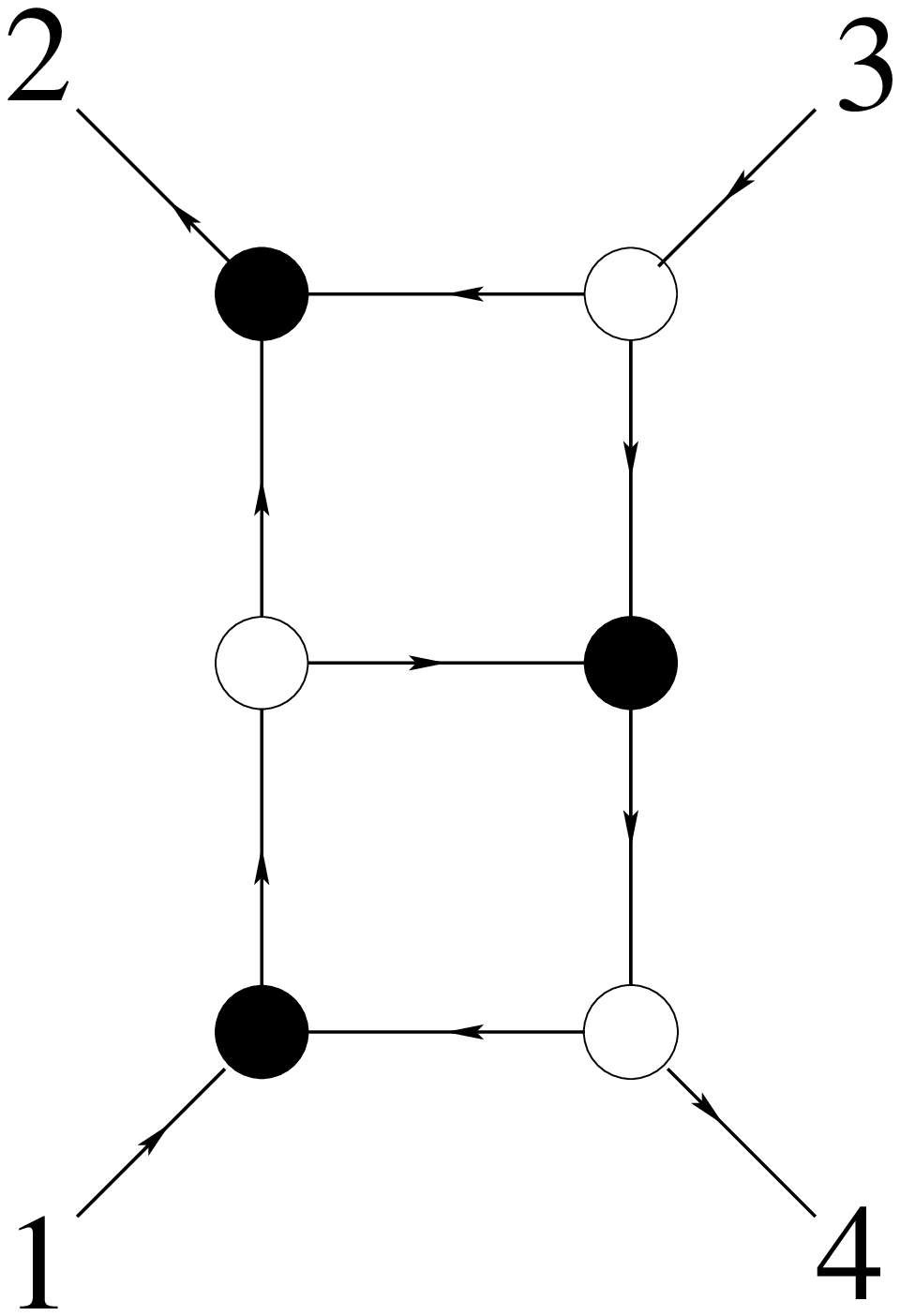}}}\:=\:
  \frac{dz'}{z'}\,\int\frac{d^{\mbox{\tiny $(2\times4)$}}C}{\mbox{Vol}\{GL(2)\}}\,\frac{\delta^{\mbox{\tiny $(8|2\mathcal{N})$}}(C\cdot\mathcal{W})(\Delta_{13}-z'\Delta_{34})^{4-\mathcal{N}}}{(\Delta_{12}-z'\Delta_{24})
   \Delta_{23}\Delta_{34}\Delta_{41}},
\end{equation}
where the lower box shows a clockwise helicity loop. In principle we would need to sum over the two possible helicity loops. However, for the present discussion we will just focus for the time being on the diagram in 
\eqref{eq:4ptGrass1form}. If we integrate such a $1$-form over the Riemann sphere, the residue theorem returns a relation between the original diagram (the residue of the pole at $z\,=\,0$, which corresponds to diagrammatically
remove the lower horizontal line in the left-hand-side. of \eqref{eq:4ptGrass1form}), the residue of the pole at $z\,=\,\Delta_{12}/\Delta_{23}$, which corresponds to remove the upper horizontal line leaving an on-shell box with a 
clockwise helicity loop and finally the residue of the multiple pole at infinity. Let us focus on the latter and compute its residue explicitly. First, notice that the on-shell one-form \eqref{eq:4ptGrass1form} can be also parametrised 
as
\begin{equation}\eqlabel{eq:4ptGrass1form2}
  \raisebox{-1.3cm}{\scalebox{.20}{\includegraphics{TripleCut1bb.eps}}}\:=\:
  \frac{dz}{z}\,\int\frac{d^{\mbox{\tiny $(2\times4)$}}C}{\mbox{Vol}\{GL(2)\}}\,\frac{\delta^{\mbox{\tiny $(8|2\mathcal{N})$}}(C\cdot\mathcal{W})\Delta_{13}^{4-\mathcal{N}}}{(\Delta_{12}+z\Delta_{13})
   \Delta_{23}\Delta_{34}\Delta_{41}}
   \left[
    \frac{\Delta_{23}\Delta_{41}}{\Delta_{13}(\Delta_{24}+z\Delta_{34})}
   \right]^{4-\mathcal{N}}\hspace{-.5cm},
\end{equation}
so that the multiple pole at infinity is mapped into a multiple pole at $z\,=\,-\Delta_{24}/\Delta_{34}$. The parametrisation \eqref{eq:4ptGrass1form2} is obtained by applying a standard BCFW bridge in the $(2,3)$-channel
($c_2\,\longrightarrow\,c_2+z c_3$) to the on-shell box with a clockwise helicity loop and it is related to the previous one in \eqref{eq:4ptGrass1form} via a M{\"o}bius transformation \cite{Benincasa:2015zna}. As discussed in the 
previous subsection, such a pole corresponds to taking collinear two non-adjacent particles but has support on a $(3-\mathcal{N})$-derivative delta-function.

Thus, the integration of the on-shell one-form \eqref{eq:4ptGrass1form2} over the full Riemann sphere returns the following identity:
\begin{equation}\eqlabel{eq:4ptId}
 0\:=\:\raisebox{-0.9cm}{\scalebox{.25}{\includegraphics{4ptDiag3b.eps}}}\:-\:\raisebox{-0.9cm}{\scalebox{.25}{\includegraphics{4ptDiag2b.eps}}}\:+\:
       \raisebox{-1.3cm}{\scalebox{.20}{\includegraphics{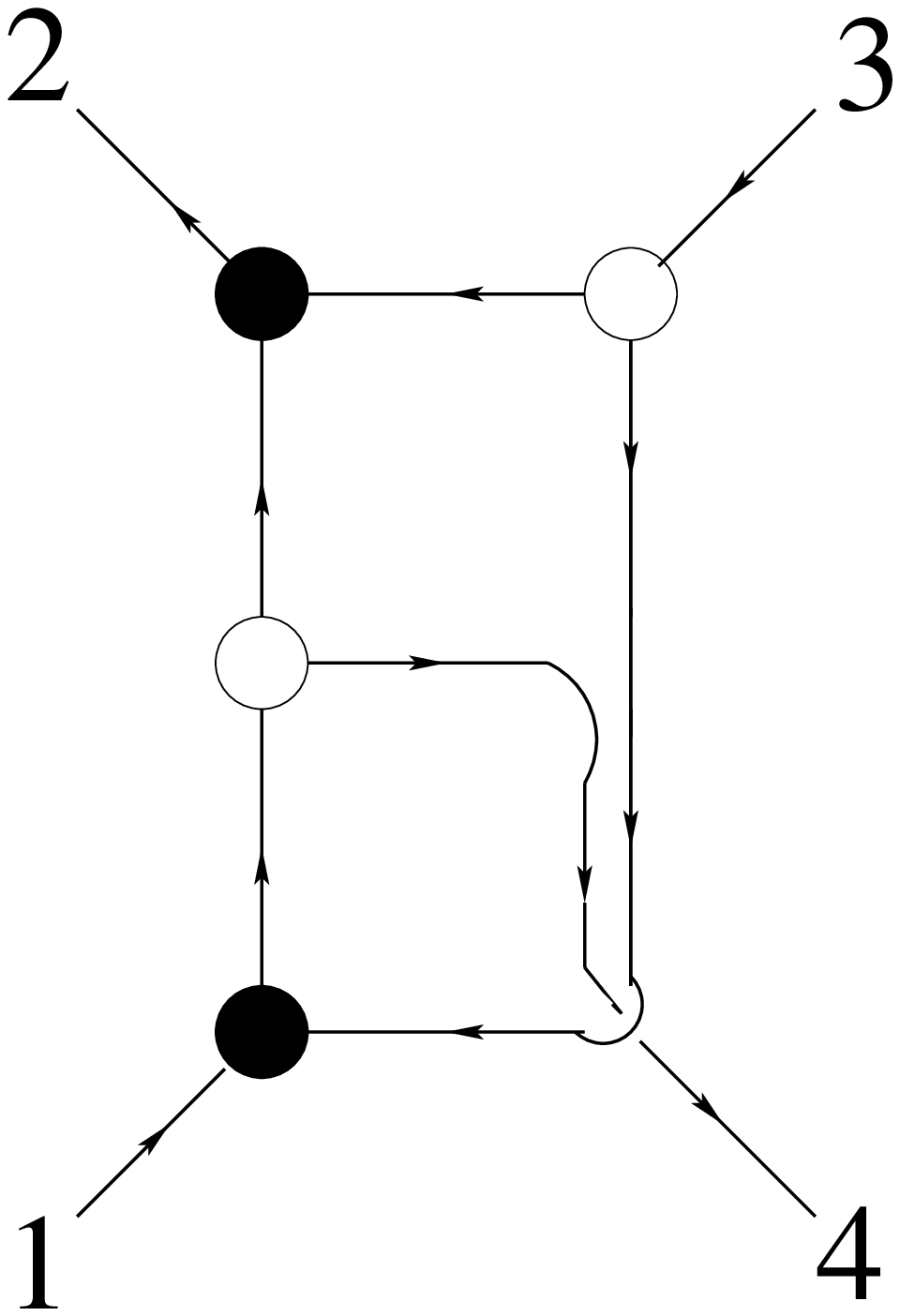}}},
\end{equation}
where
\begin{equation}\eqlabel{eq:4ptHP}
 \raisebox{-1.3cm}{\scalebox{.20}{\includegraphics{4ptHPsing.eps}}}
  \:=\:\int\frac{d^{\mbox{\tiny $2\times4$}}C}{\mbox{Vol}\{GL(2)\}}\frac{\delta^{\mbox{\tiny $(8|2\mathcal{N})$}}\left(C\cdot\mathcal{W}\right)\Delta_{13}^{4-\mathcal{N}}}{\Delta_{14}\Delta_{42}\Delta_{23}\Delta_{31}}
 \sum_{r=0}^{3-\mathcal{N}}\left(\frac{\Delta_{14}\Delta_{23}}{\Delta_{13}\Delta_{24}}\right)^r
\end{equation}
which correspond exactly to the four-particle amplitude with lexicographic order $1\,<\,4\,<\,2\,<\,3$ just for the case $\mathcal{N}\,=\,3$.

The one-form \eqref{eq:4ptGrass1form2} with counter-clockwise helicity loops has two poles in $z$, one at $0$ (whose residue is the on-shell box with counter-clockwise helicity loop) and the multiple pole at 
$z\,=\,-\Delta_{24}/\Delta_{34}$. Consequently, the identity obtained by integrating this one-form over the Riemann sphere establishes the equivalence between the on-shell box with counter-clockwise helicity loop
and (higher order) on-shell box with ordering $(1423)$:
\begin{equation}\eqlabel{eq:4ptGrass1form3}
 \begin{split}
  &\raisebox{-1.3cm}{\scalebox{.20}{\includegraphics{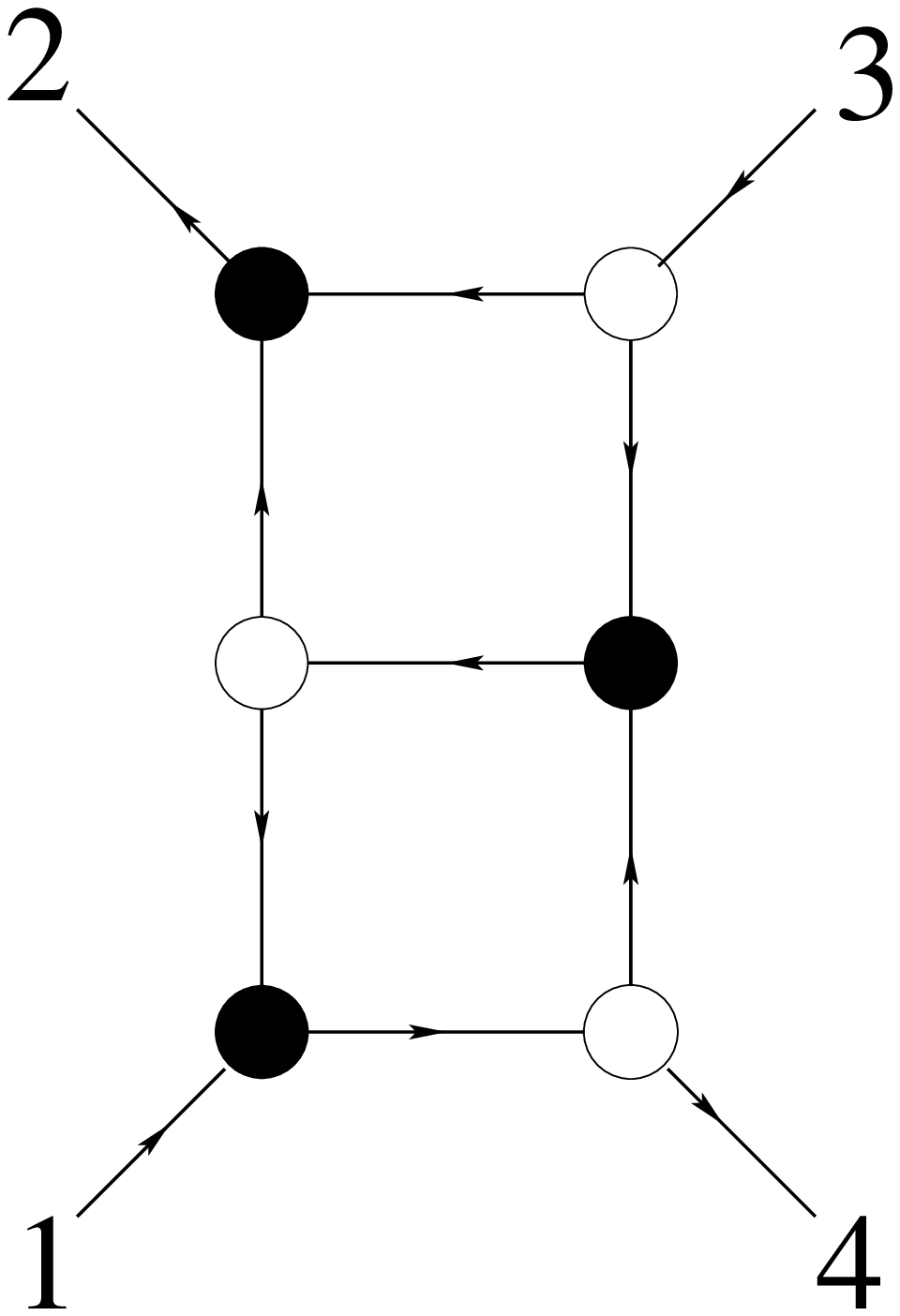}}}\:=\:
   \frac{dz}{z}\,\int\frac{d^{\mbox{\tiny $(2\times4)$}}C}{\mbox{Vol}\{GL(2)\}}\,\frac{\delta^{\mbox{\tiny $(8|2\mathcal{N})$}}(C\cdot\mathcal{W})\Delta_{13}^{4-\mathcal{N}}}{(\Delta_{12}+z\Delta_{13})
   \Delta_{23}\Delta_{34}\Delta_{41}}
   \left[
    \frac{(\Delta_{12}+z\Delta_{13})\Delta_{34}}{\Delta_{13}(\Delta_{24}+z\Delta_{34})}
   \right]^{4-\mathcal{N}}\hspace{-.5cm},\\
  &\hspace{8cm}\Downarrow\\
  &\hspace{5cm}0\:=\:\raisebox{-0.9cm}{\scalebox{.25}{\includegraphics{4ptDiag4b.eps}}}\:+\: \raisebox{-1.3cm}{\scalebox{.20}{\includegraphics{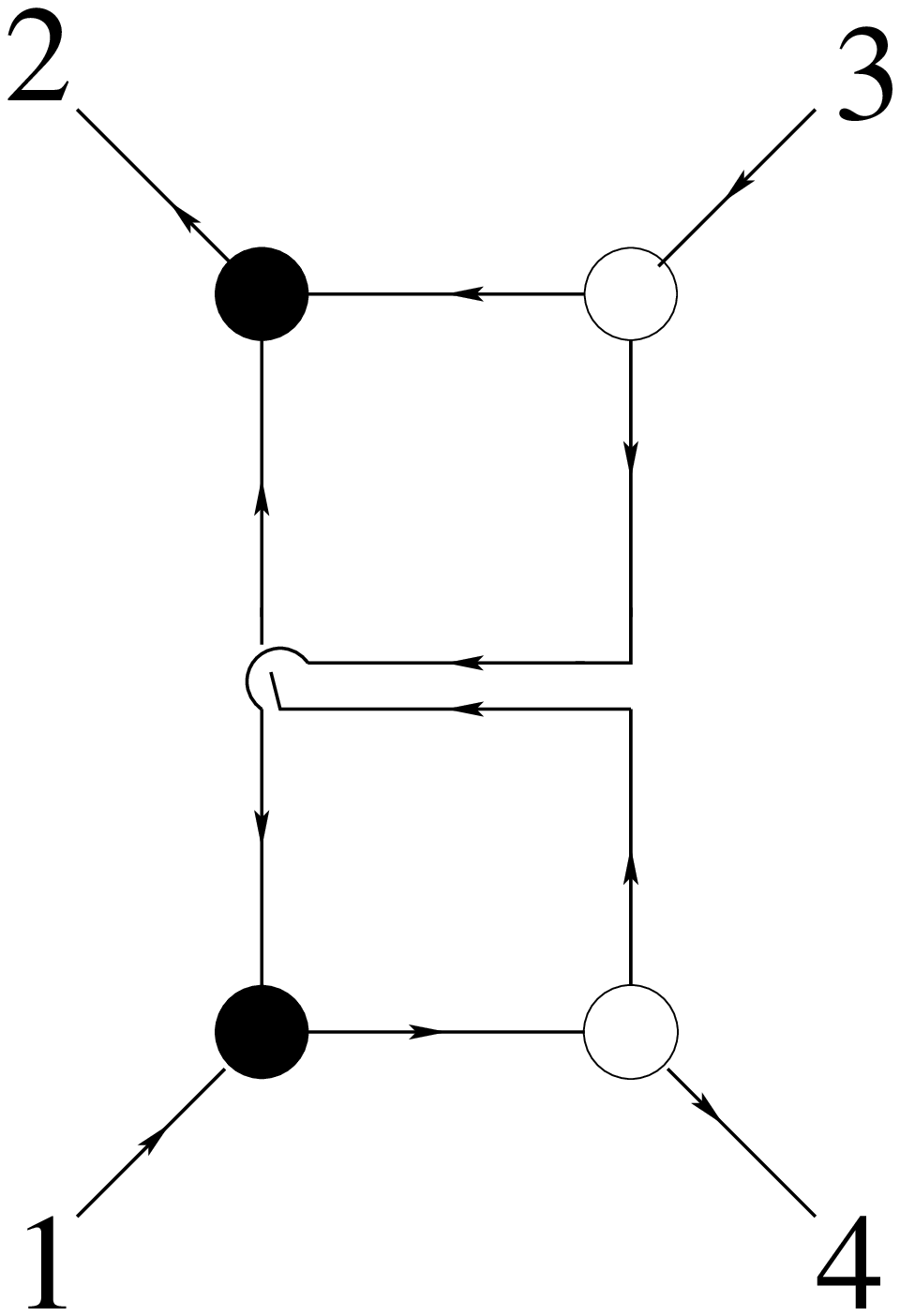}}}
 \end{split}
\end{equation}

If one applies a BCFW bridge in the $(3,4)$-channel to the on-shell boxes with internal helicity loops, one obtains the same identities but with the label exchanges $2\,\longleftrightarrow\,4$ and $1\,\longleftrightarrow\,3$.
Let us stress here that, in principle, the two non-planar-like diagrams in \eqref{eq:4ptGrass1form2} and \eqref{eq:4ptGrass1form3} are {\it in principle} topologically equivalent. What marks the difference between the two diagrams
are the way the legs are winded, which here keeps track of the origin of the diagram, {\it i.e.} one comes from opening up a clockwise helicity loop while the other one from a counter-clockwise helicity loop.

Finally, notice that, just in the $\mathcal{N}\,=\,3$, the pole associated to the helicity loop is a simple pole and its residue is exactly (up to a sign) the four-particle amplitude with a different ordering, which depends
on the orientation on the loop. Such an identity is given by a residue theorem of the type of \eqref{eq:4ptGrass1form3}.


\section{On-shell functions on $Gr(3,6)$}\label{sec:Gr36}

Let us focus on the simplest non-trivial example which is given by $Gr(3,6)$. In the planar sector, there is a unique top cell which the following undecorated bipartite diagram is associated to
\begin{equation}\eqlabel{eq:Gr36tc}
 \raisebox{-1.5cm}{\scalebox{.25}{\includegraphics{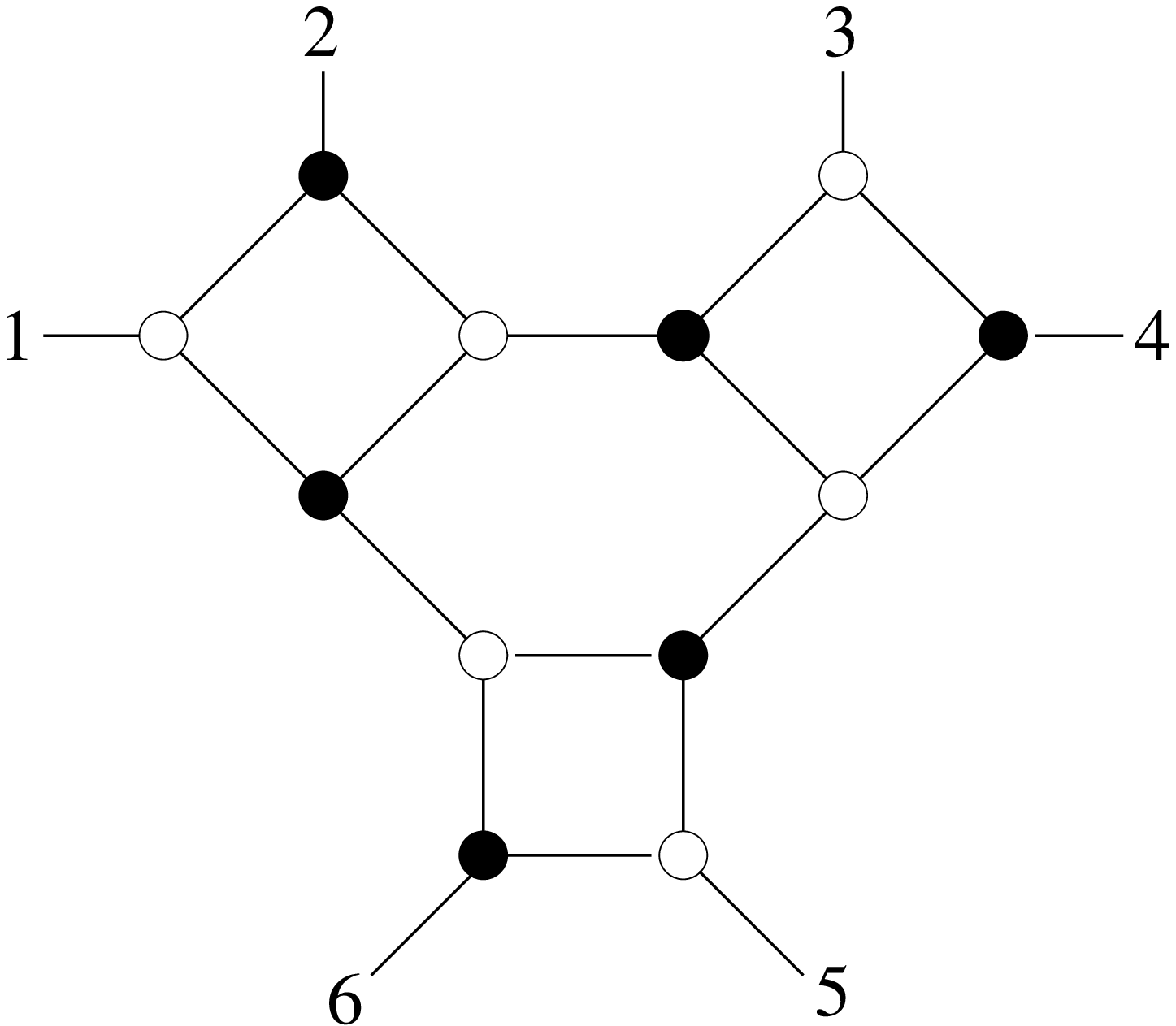}}}\quad\longleftrightarrow\:
 \omega_{\mbox{\tiny $3,6$}}\:\equiv\:\frac{d^{\mbox{\tiny $(3\times6)$}}C}{\mbox{Vol}\{GL(3)\}}\,\frac{1}{\Delta_{123}\Delta_{234}\Delta_{345}\Delta_{456}\Delta_{561}\Delta_{612}}.
\end{equation}
Through equivalence relations such as square-moves and mergers, the diagram drawn above can be mapped into other equivalent diagrams. However, once these diagrams get decorated with helicity arrows (which are fixed for
the external lines) they generate inequivalent on-shell functions, which can share at most a subset of simple poles and in general differ for the location of the multiple pole. For definiteness, let us choose the external
helicity arrows to have alternating directions -- all the other configurations in the NMHV sector can be obtained via the helicity flow reversal operation. In general, the (super)-momentum conserving $\delta$-functions fix
all the degrees of freedom of the Grassmannian but one which can be used to obtain identities among on-shell diagrams of codimension-$1$ by integrating it over the Riemann sphere.


\subsection{Poles and non-Pl{\"u}cker relations}\label{subsec:PnPr}

The on-shell function returned by the very same diagram in \eqref{eq:Gr36tc}
with such a choice shows helicity loops in the central hexagon only and thus just one higher order pole:
\begin{equation}\eqlabel{eq:Gr36tc2}
 \begin{split}
  &\raisebox{-1.2cm}{\scalebox{.25}{\includegraphics{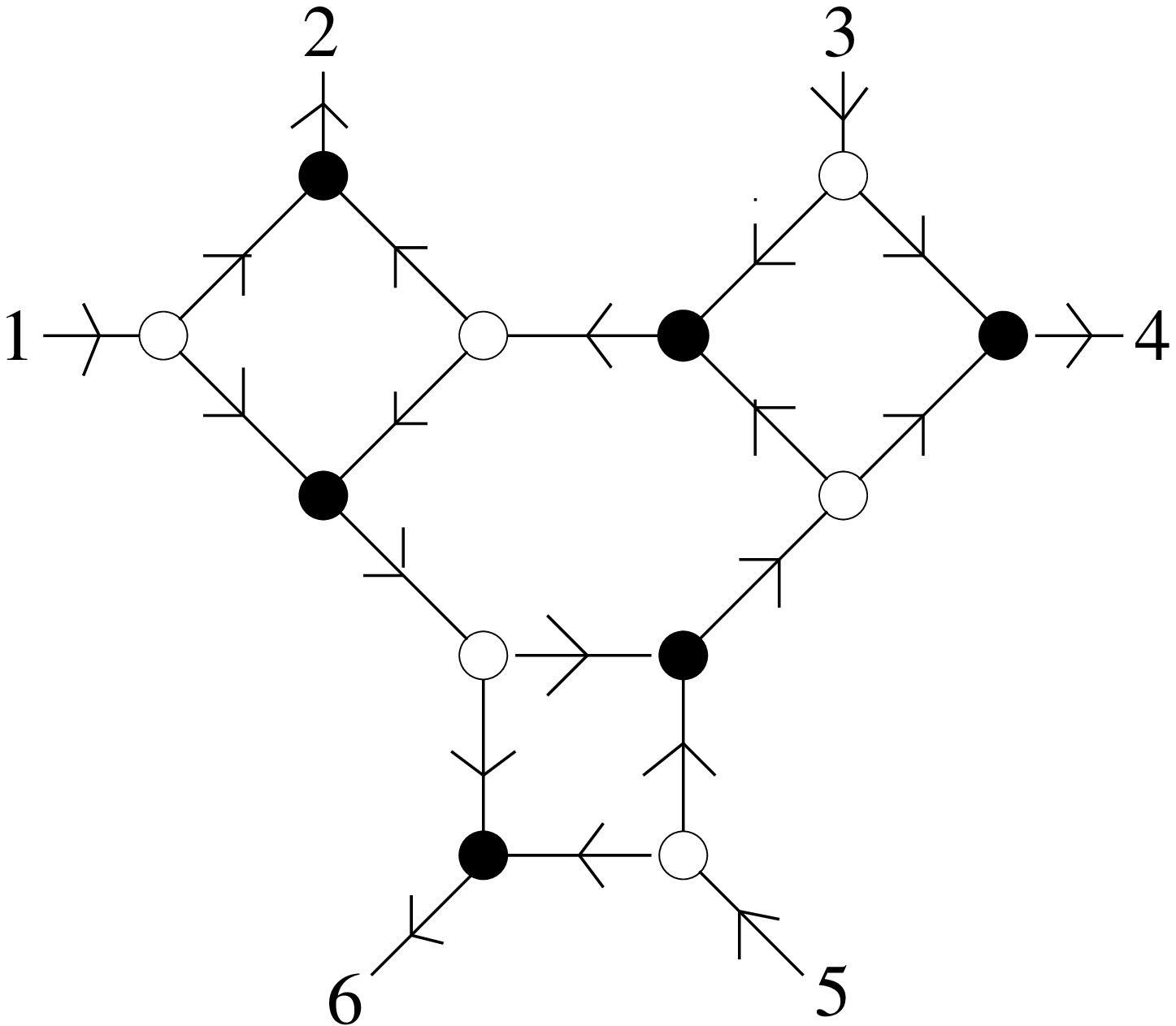}}}\hspace{-.5cm}=\:
   \int\frac{d^{\mbox{\tiny $(3\times6)$}}C}{\mbox{Vol}\{GL(3)\}}\,\frac{\delta^{\mbox{\tiny $(3\times4|3\mathcal{N})$}}\left(C\cdot\mathcal{W}\right)
    }{
    \Delta_{123}\Delta_{234}\Delta_{345}\Delta_{456}\Delta_{561}\Delta_{612}}
    \left[
     \frac{ 
            \Delta_{134}\Delta_{356}\Delta_{512}
          }{\Delta_{346}\Delta_{512}-\Delta_{345}\Delta_{612}}
    \right]^{4-\mathcal{N}}\\
   &\raisebox{-1.2cm}{\scalebox{.25}{\includegraphics{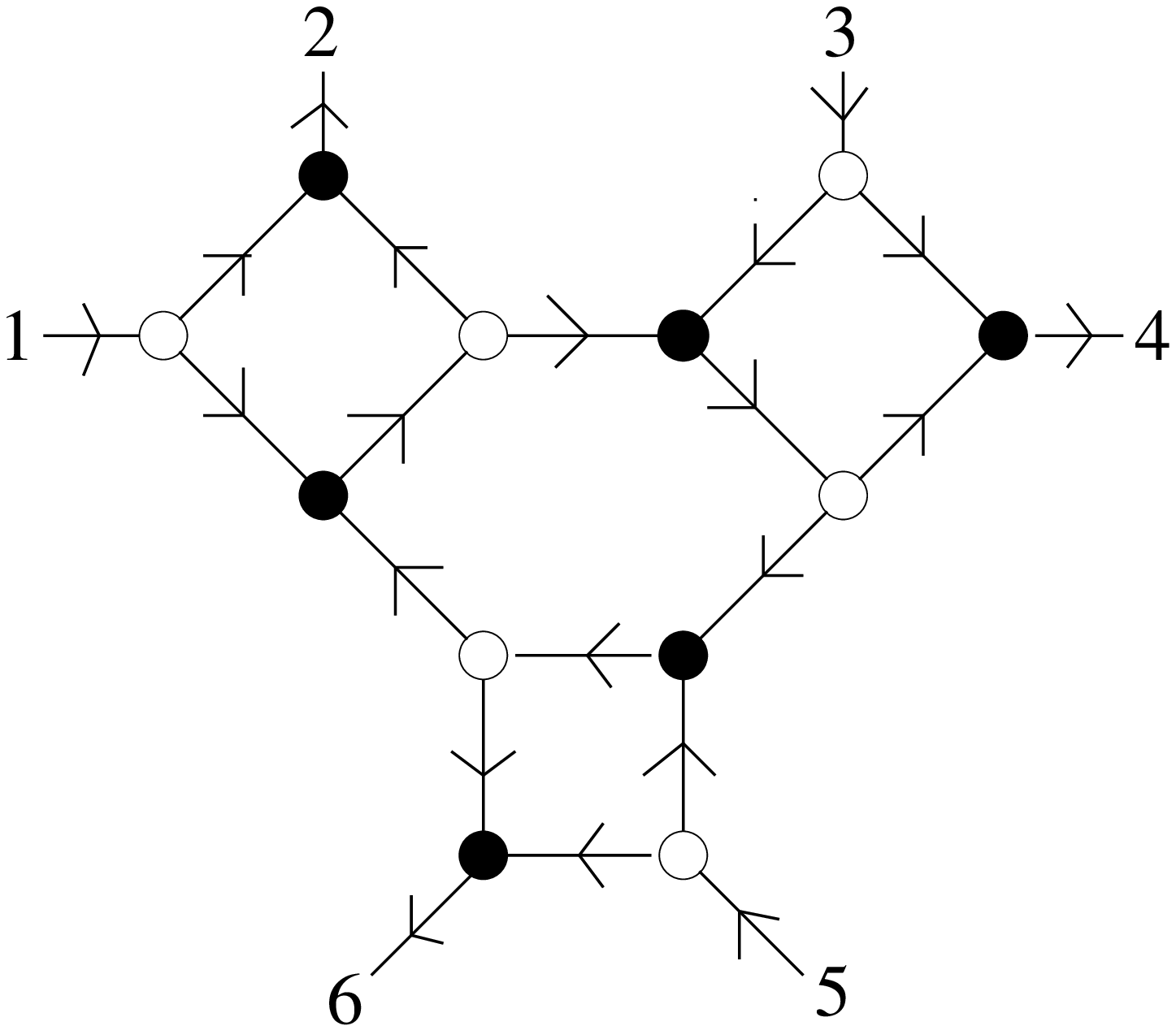}}}\hspace{-.5cm}=\:
   \int\frac{d^{\mbox{\tiny $(3\times6)$}}C}{\mbox{Vol}\{GL(3)\}}\,\frac{\delta^{\mbox{\tiny $(3\times4|3\mathcal{N})$}}\left(C\cdot\mathcal{W}\right)
    }{
    \Delta_{123}\Delta_{234}\Delta_{345}\Delta_{456}\Delta_{561}\Delta_{612}}
    \left[
     \frac{
           \Delta_{123}\Delta_{345}\Delta_{561}
          }{\Delta_{346}\Delta_{512}-\Delta_{345}\Delta_{612}}
    \right]^{4-\mathcal{N}}
 \end{split}
\end{equation}

Notice that in the diagram with counter-clockwise helicity loop in the first line, no equivalence move holds: the on-shell function on the right-hand-side in the first line is uniquely associated to this perfect orientation.
The helicity flows define all the possible {\it removable edges}, which coincide with taking residues in the on-shell functions: and edge is said to be removable if no helicity flow is broken by such an operation. 
All the simple poles are related to the boundary measurements from the top form. Taking the residues
at $\Delta_{123}\,=\,0$, $\Delta_{345}\,=\,0$ and $\Delta_{561}\,=\,0$ corresponds to remove one of the edges shared between the internal hexagon and one of the boxes. However, notice that the helicity flows allow to remove
such edges just in the first diagram in \eqref{eq:Gr36tc2}. Indeed, this is reflected into the Grassmannian representation which shows, for the second line of \eqref{eq:Gr36tc2}, a numerator proportional to 
$\Delta_{123}\Delta_{345}\Delta_{561}$.

In the first diagrams, instead, when one takes these type of residues the higher order
pole disappears. Diagrammatically, this is a reflection of the disappearance of the internal helicity loop when any of these edges gets removed:
\begin{equation}\eqlabel{eq:Gr36subcell}
 \raisebox{-1.2cm}{\scalebox{.25}{\includegraphics{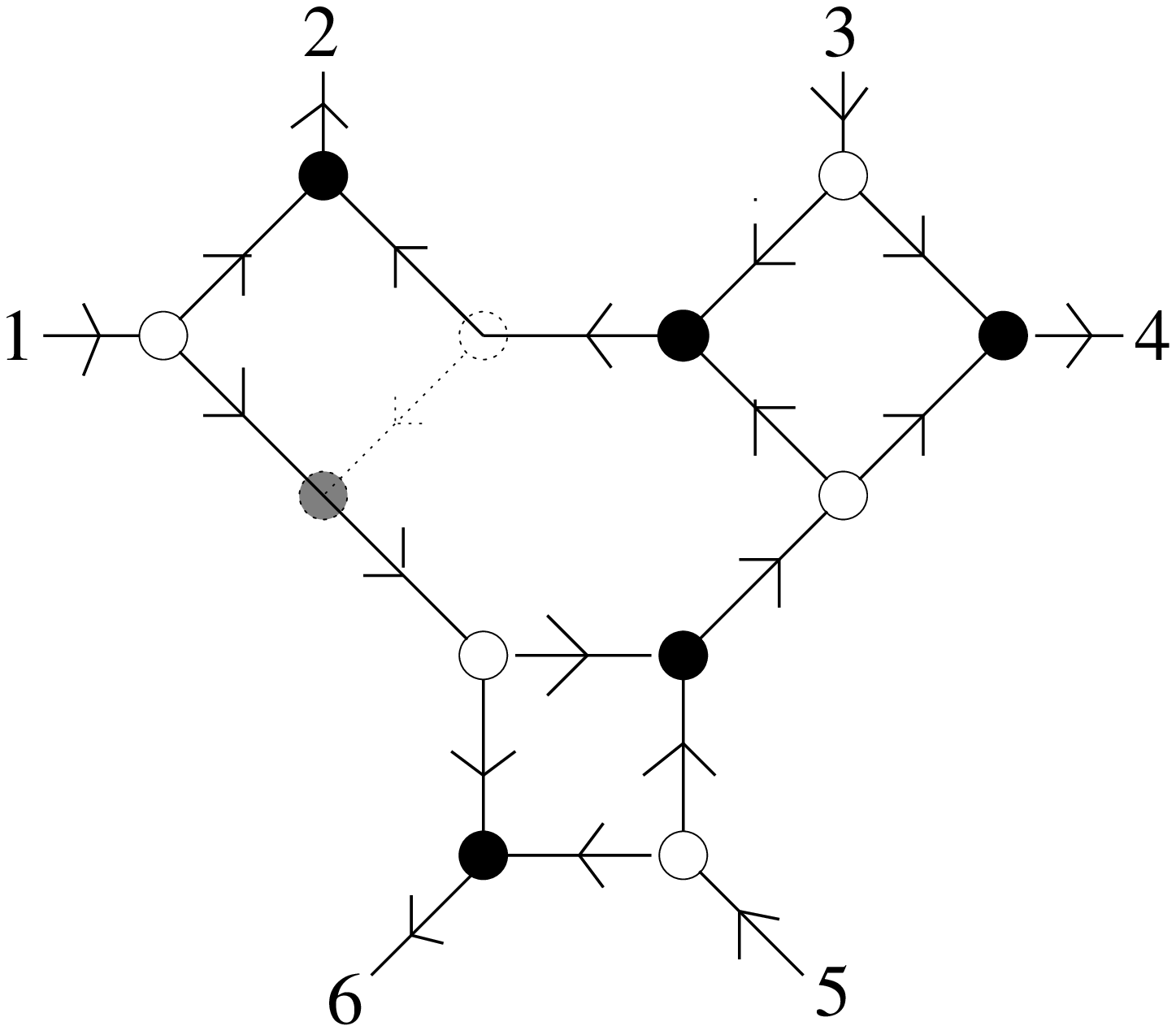}}}\:\equiv\hspace{-.2cm}  
 \raisebox{-2.0cm}{\scalebox{.20}{\includegraphics{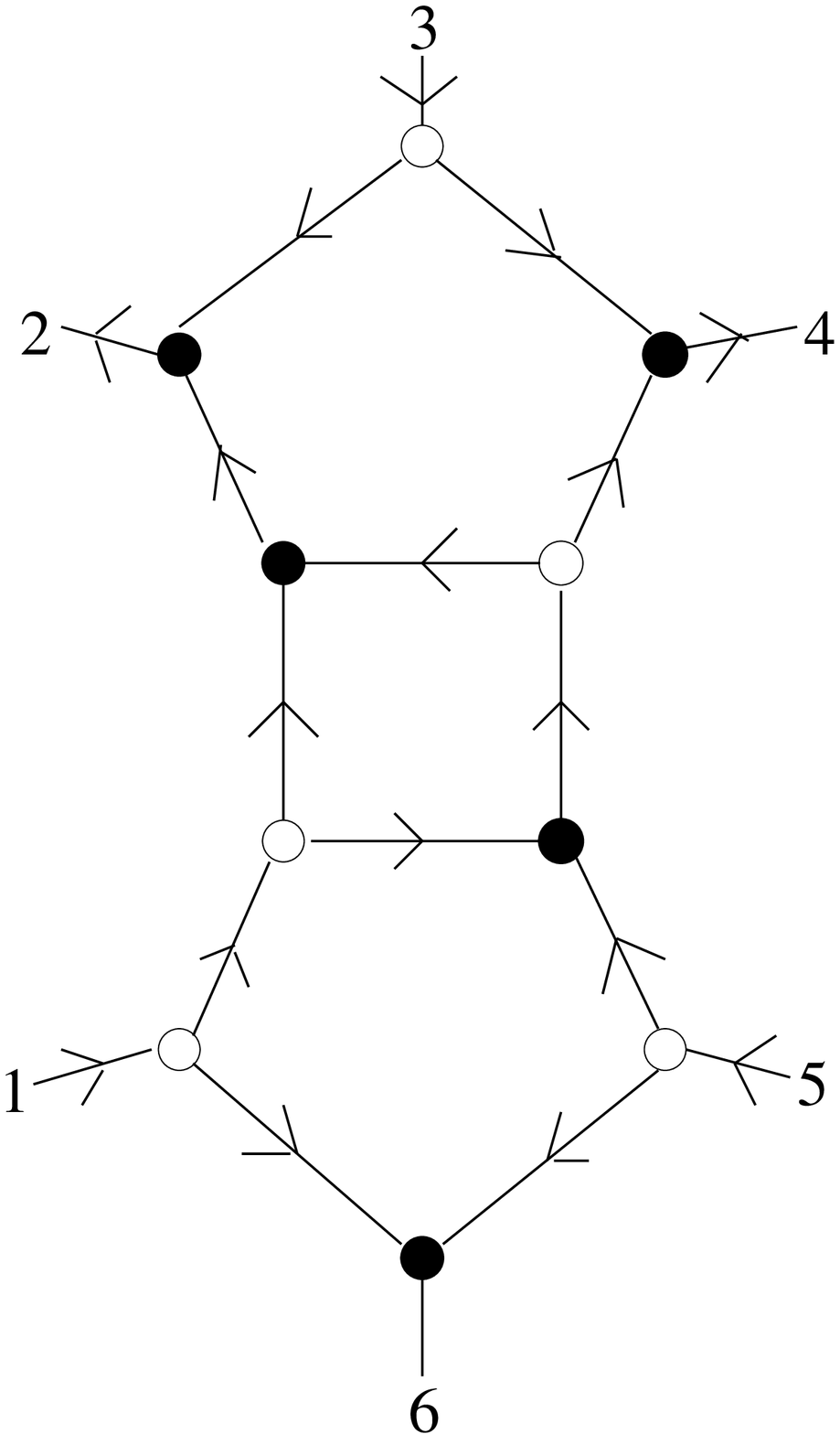}}}\hspace{-.5cm}=\:   
  \int\frac{d^{\mbox{\tiny $(3\times6)$}}C}{\mbox{Vol}\{GL(3)\}}\,\frac{\delta^{\mbox{\tiny $(3\times4|3\mathcal{N})$}}\left(C\cdot\mathcal{W}\right)\Delta_{135}^{4-\mathcal{N}}}{
   \Delta_{123}\Delta_{234}\Delta_{345}\Delta_{456}\Delta_{612}}\delta(\Delta_{561})
\end{equation}
The other two residues can be obtained from \eqref{eq:Gr36subcell} by shifting the labels: $i\:\longrightarrow\:i\,\mp\,2$.

The other three poles $\{\Delta_{234}\,=\,0,\,\Delta_{456}\,=\,0,\,\Delta_{612}\,=\,0\}$ of the top form $\omega_{\mbox{\tiny $3,6$}}$ are actually poles of the full integrand and their residues diagrammatically correspond 
to remove the outer edge of one of the boxes. The resulting diagrams still show an internal helicity loop and thus their Grassmannian representation still show a higher order pole:
\begin{equation}\eqlabel{eq:Gr36subcell2}
 \begin{split}
  &\raisebox{-1.2cm}{\scalebox{.25}{\includegraphics{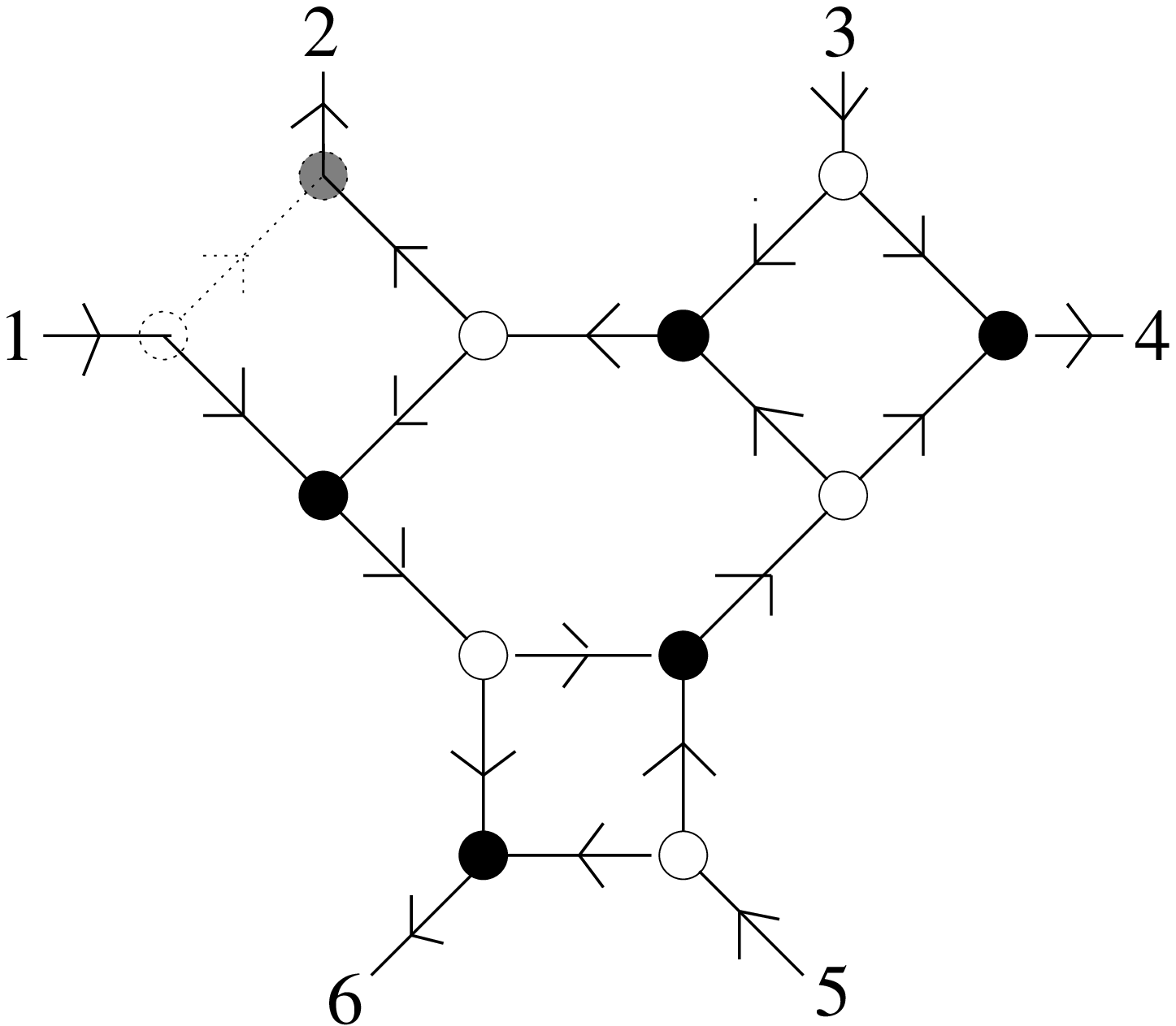}}}\hspace{-1cm}=\:
    \int\frac{d^{\mbox{\tiny $(3\times6)$}}C}{\mbox{Vol}\{GL(3)\}}\,\frac{\delta^{\mbox{\tiny $(3\times4|3\mathcal{N})$}}\left(C\cdot\mathcal{W}\right)
             }{
     \Delta_{123}\Delta_{345}\Delta_{456}\Delta_{561}\Delta_{612}}
     \left[
      \frac{
            \Delta_{356}\Delta_{512}
           }{\Delta_{256}}
    \right]^{4-\mathcal{N}}\hspace{-.5cm}\delta(\Delta_{234}),\\
  &\raisebox{-1.2cm}{\scalebox{.25}{\includegraphics{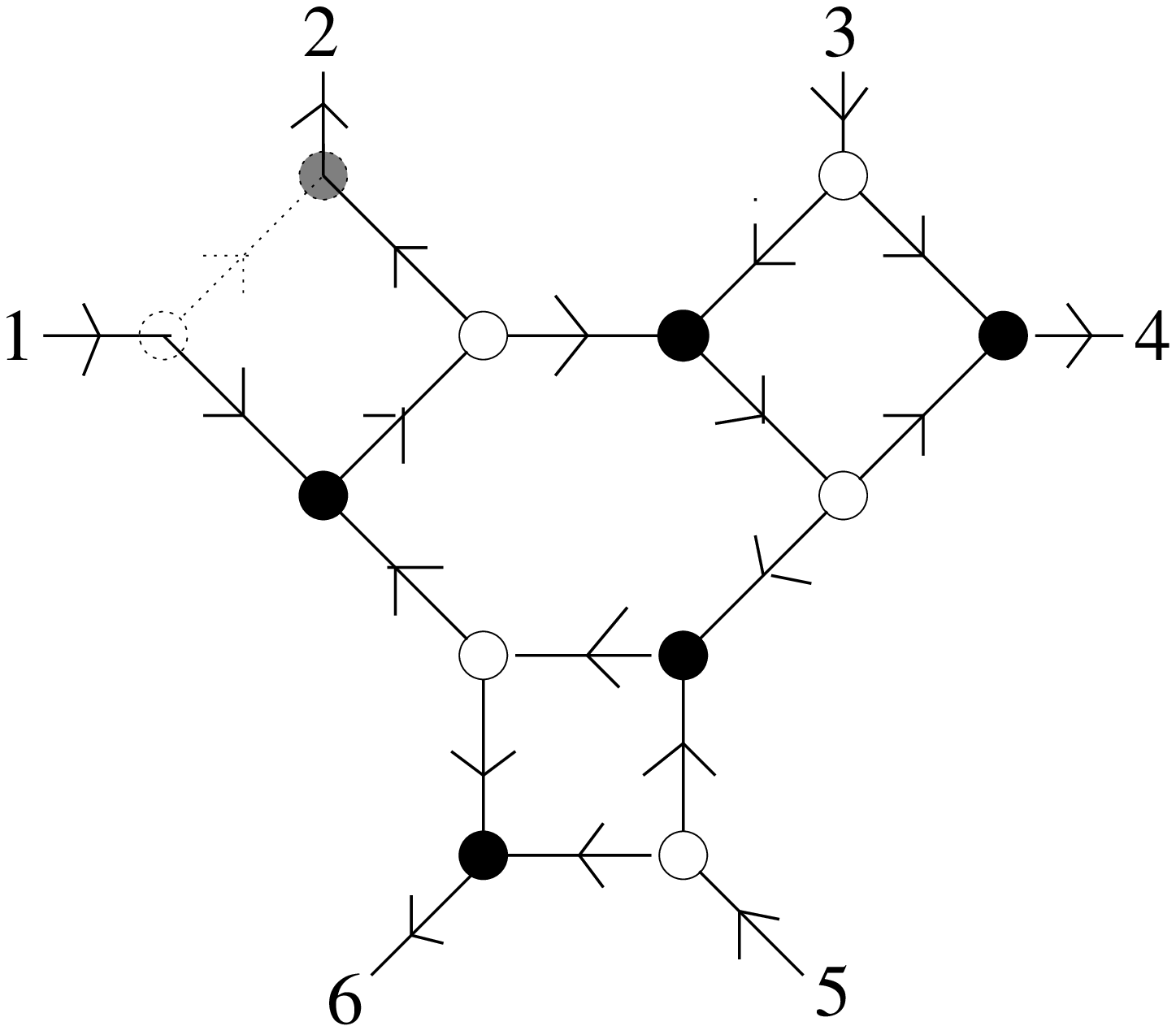}}}\hspace{-1cm}=\:
    \int\frac{d^{\mbox{\tiny $(3\times6)$}}C}{\mbox{Vol}\{GL(3)\}}\,\frac{\delta^{\mbox{\tiny $(3\times4|3\mathcal{N})$}}\left(C\cdot\mathcal{W}\right)}{
     \Delta_{123}\Delta_{345}\Delta_{456}\Delta_{561}\Delta_{612}}
     \left[
      \frac{
            \Delta_{123}\Delta_{345}\Delta_{561}
           }{\Delta_{134}\Delta_{256}}
    \right]^{4-\mathcal{N}}\hspace{-.5cm}\delta(\Delta_{234}),
 \end{split}
\end{equation}
where now the higher order pole has the more standard form of a single Pl{\"u}cker coordinate vanishing. From the expression above, it seems that there are actually two of such type of poles. However, it is easy to check that
$\Delta_{134}\,=\,0$ is not actually a singular point of the integrand\footnote{Notice that in the subcell $\Delta_{234}\,=\,0$, the identity $\Delta_{123}\Delta_{345}\,=\,\Delta_{134}\Delta_{235}$ holds, from which the statement
that $\Delta_{134}\,=\,0$ is not a pole follows. This identity can be obtained by rewriting the column $c_4$ as a linear combination of $c_2$ and $c_3$, as implied by $\Delta_{234}\,=\,0$.} and, thus, the higher order pole is localised 
at $\Delta_{256}\,=\,0$.

Finally, let us discuss the new structures which characterise the decorated diagrammatics and reflects into the appearance of new poles. In particular, for the on-shell functions in \eqref{eq:Gr36tc2}, the location of such
poles is no longer given by a single Pl{\"u}cker coordinate vanishing, rather by a relation among some of them. For the case at hand \eqref{eq:Gr36tc2}, it consists in
\begin{equation}\eqlabel{eq:NPpoleRel}
 0\:=\:\Delta_{346}\Delta_{512}-\Delta_{345}\Delta_{612}\:\equiv\:\Delta_{134}\Delta_{562}-\Delta_{234}\Delta_{561}\:\equiv\:\Delta_{124}\Delta_{563}-\Delta_{123}\Delta_{456},
\end{equation}
where the last two expressions have been obtained via Pl{\"u}cker relations. 
Geometrically, it means that, looking at the columns of $C\,\in\,Gr(3,6)$ as points in $\mathbb{C}^3$, the points $3$ and $4$ become collinear with the point identified by the intersection of the straight line determined by the points 
$(1,\,2)$ with the one passing through $(5,6)$:
\begin{equation*}\eqlabel{eq:Gr36NPR}
 \raisebox{-1.9cm}{\scalebox{.20}{\includegraphics{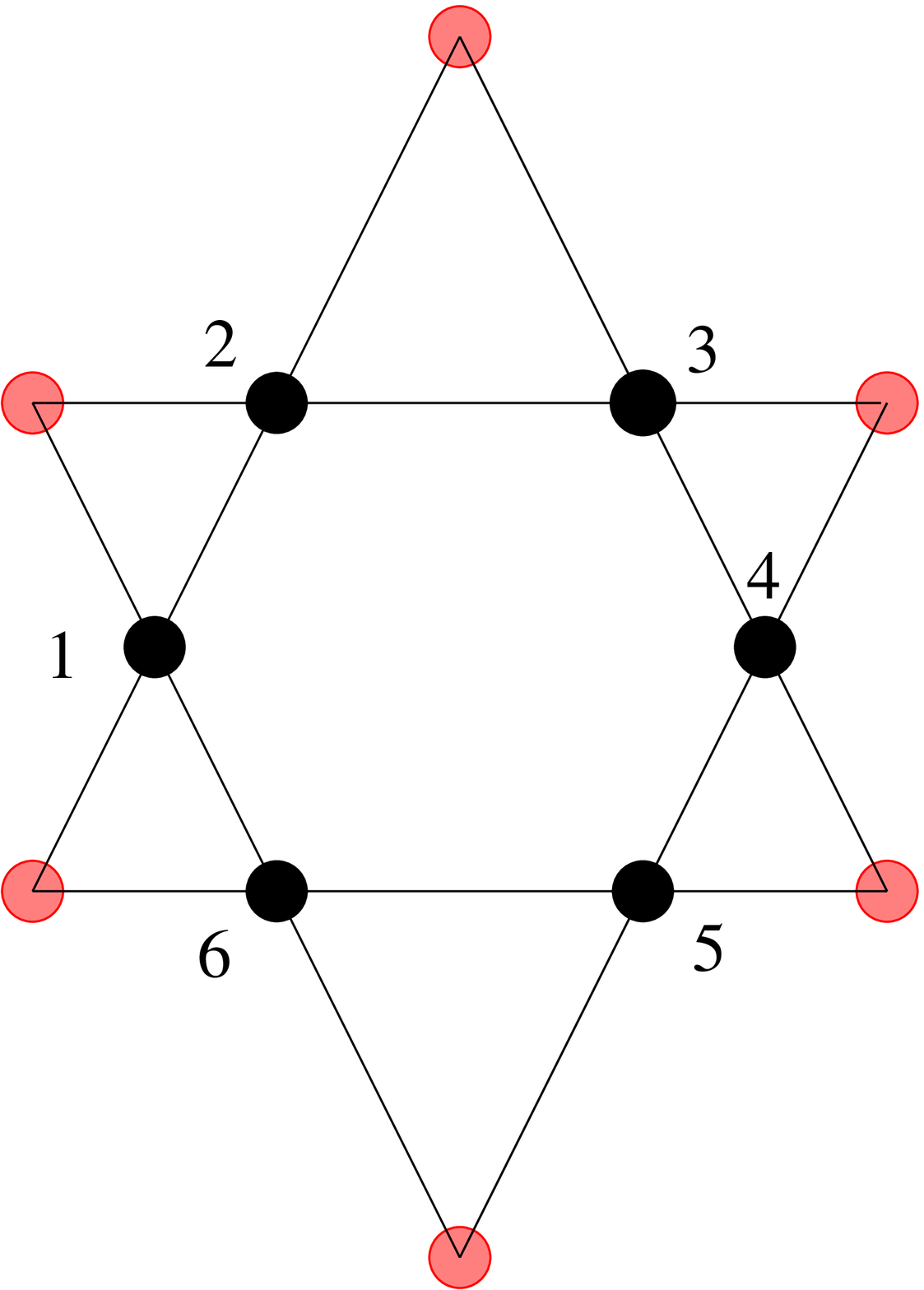}}}\qquad\Longrightarrow\qquad
 \raisebox{-1.0cm}{\scalebox{.20}{\includegraphics{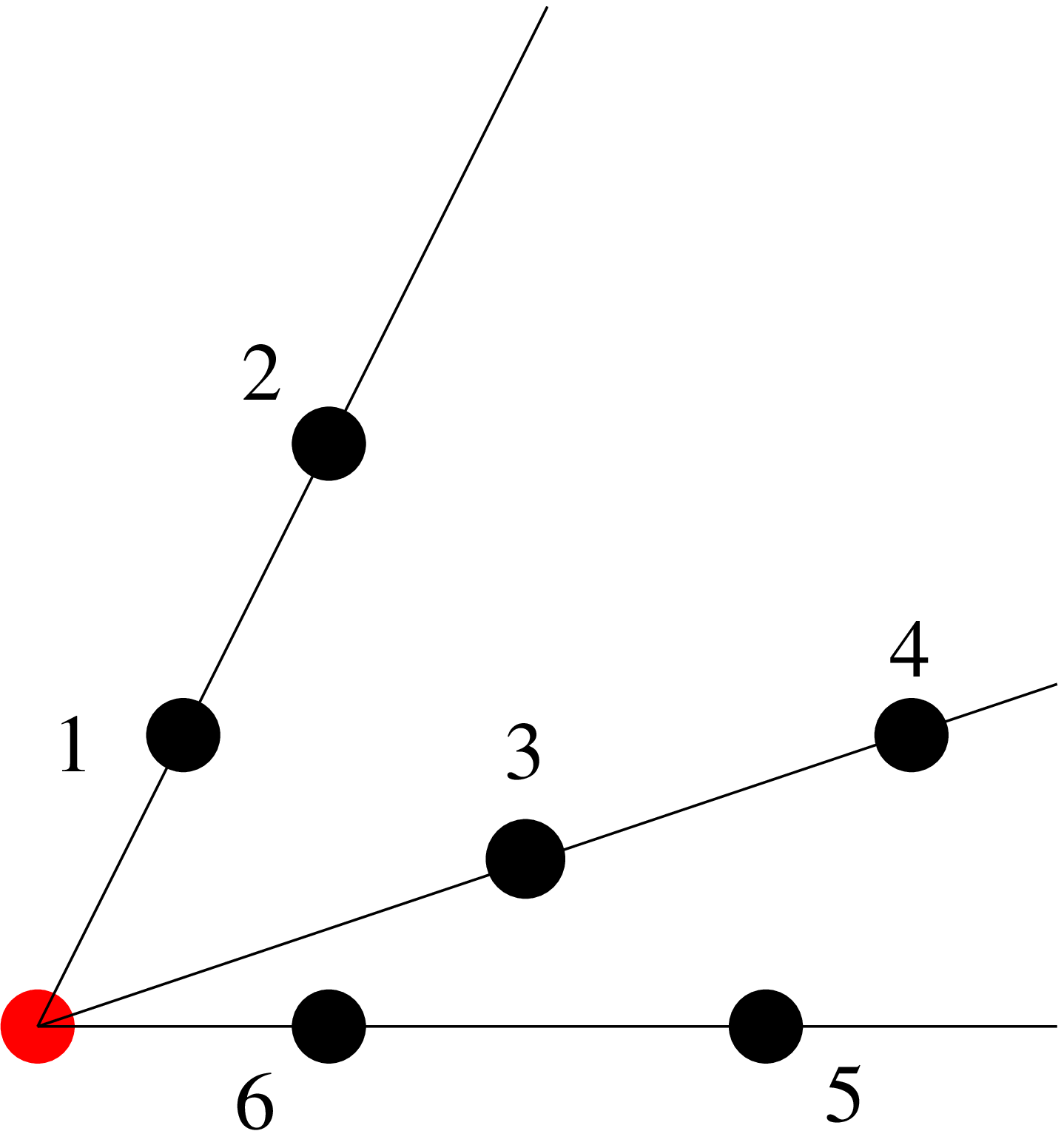}}} 
\end{equation*}
where in shaded {\color{red} red} nodes on the left hand side are the points in $\mathbb{C}^3$ identified by the intersection of (at least) two straight lines passing through two consecutive points, while the right-hand side
represents the geometry which realises the non-Pl{\"u}cker constraint \eqref{eq:NPpoleRel}, with three of the red points in the left-hand-side collapsing into one which lies simultaneously on the three straight lines identified
by the pairs of points $(1,2)$, $(3,4)$ and $(5,6)$. Identifying the intersection between the lines $\{(12),\,(34)\}$, $\{(3,4),\,(5,6)\}$ and $\{(56),\,(61)\}$ as $\mathfrak{a}$, $\mathfrak{b}$ and $\mathfrak{c}$ respectively,
this singularity can be indicated as $0\,=\,\Delta_{12\mathfrak{b}}\,=\,\Delta_{34\mathfrak{c}}\,=\,\Delta_{56\mathfrak{a}}$.

As we emphasised at the beginning, this type of singularity is a feature of a non-planar diagrams, and thus it is not a surprise that the on-shell diagram with support on 
$\delta^{\mbox{\tiny $(3-\mathcal{N})$}}(\Delta_{34\mathfrak{c}})$ is genuinely non-planar:
\begin{equation}\eqlabel{eq:Gr36resNP}
 \raisebox{-1.2cm}{\scalebox{.25}{\includegraphics{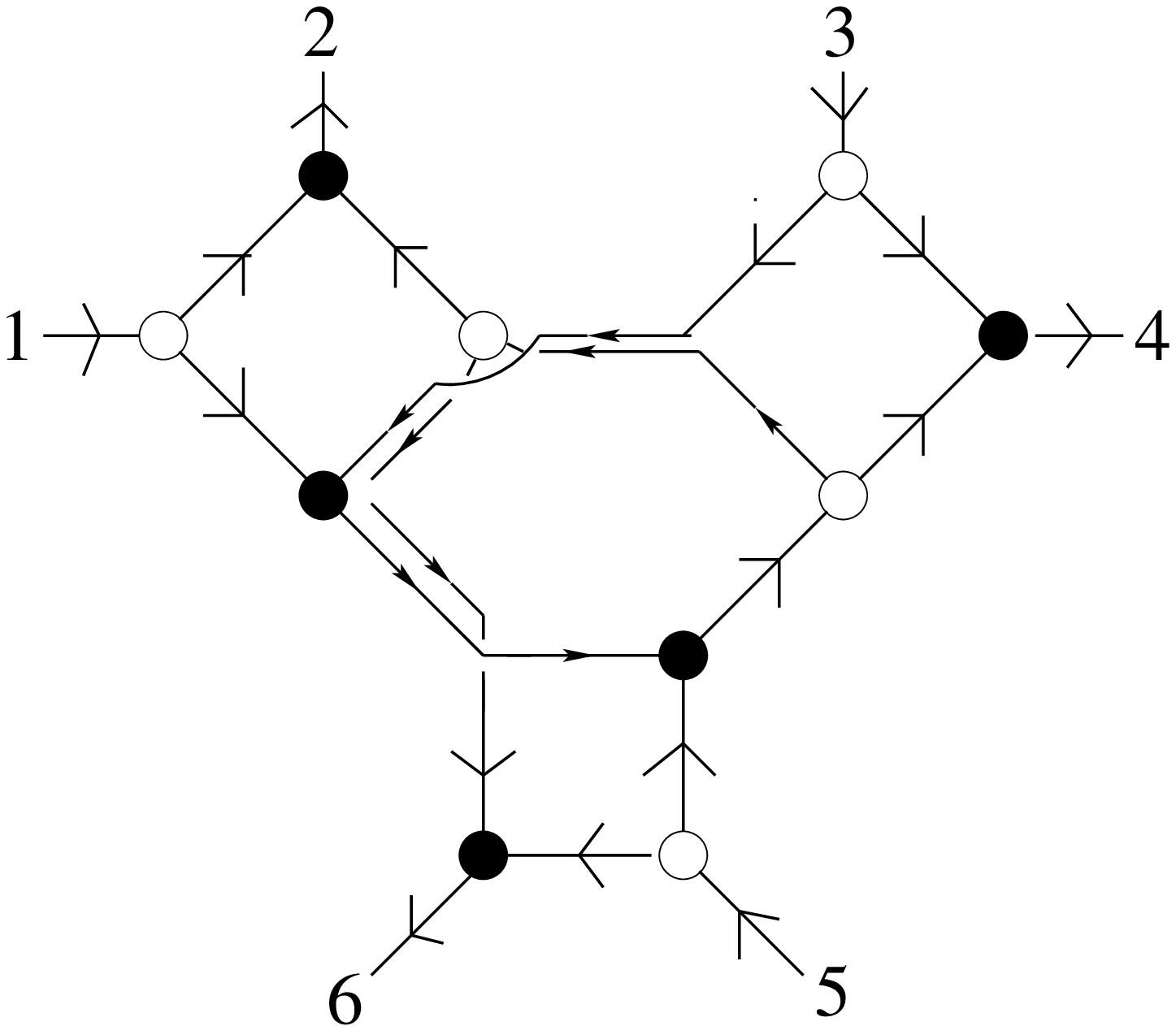}}}\hspace{-1cm}=\:
  \int\omega_{\mbox{\tiny $3,6$}}(C)\delta^{\mbox{\tiny $(3\times4|3\times\mathcal{N})$}}(C\cdot\mathcal{W})\left(\Delta_{134}\Delta_{356}\Delta_{512}\right)^{\mbox{\tiny $(4-\mathcal{N})$}}
   \delta^{\mbox{\tiny $(3-\mathcal{N})$}}\left(\Delta_{34\mathfrak{c}}\right).
\end{equation}

\subsection{Standard non-planar-like pole}\label{subsec:Pstd}

Let us now consider the undecorated diagram \eqref{eq:Gr36tc} but with the lower box with white and black node exchanged and let us decorate the external edges with exactly the same choice as above. There are three possible
helicity arrow assignment for the internal edges:
\begin{equation}\eqlabel{eq:Gr36tc3}
 \raisebox{-1.2cm}{\scalebox{.25}{\includegraphics{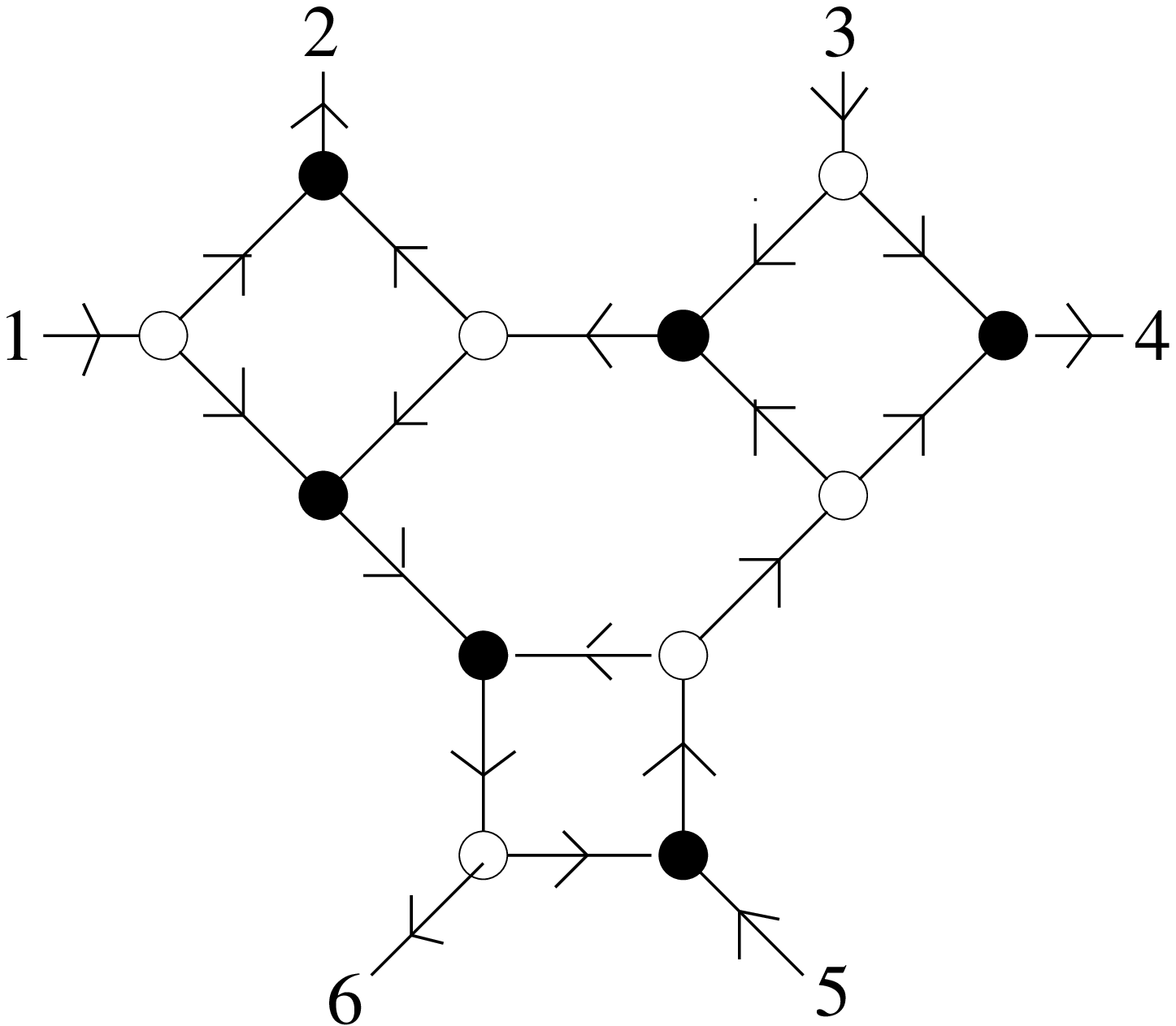}}}\:+\:
 \raisebox{-1.2cm}{\scalebox{.25}{\includegraphics{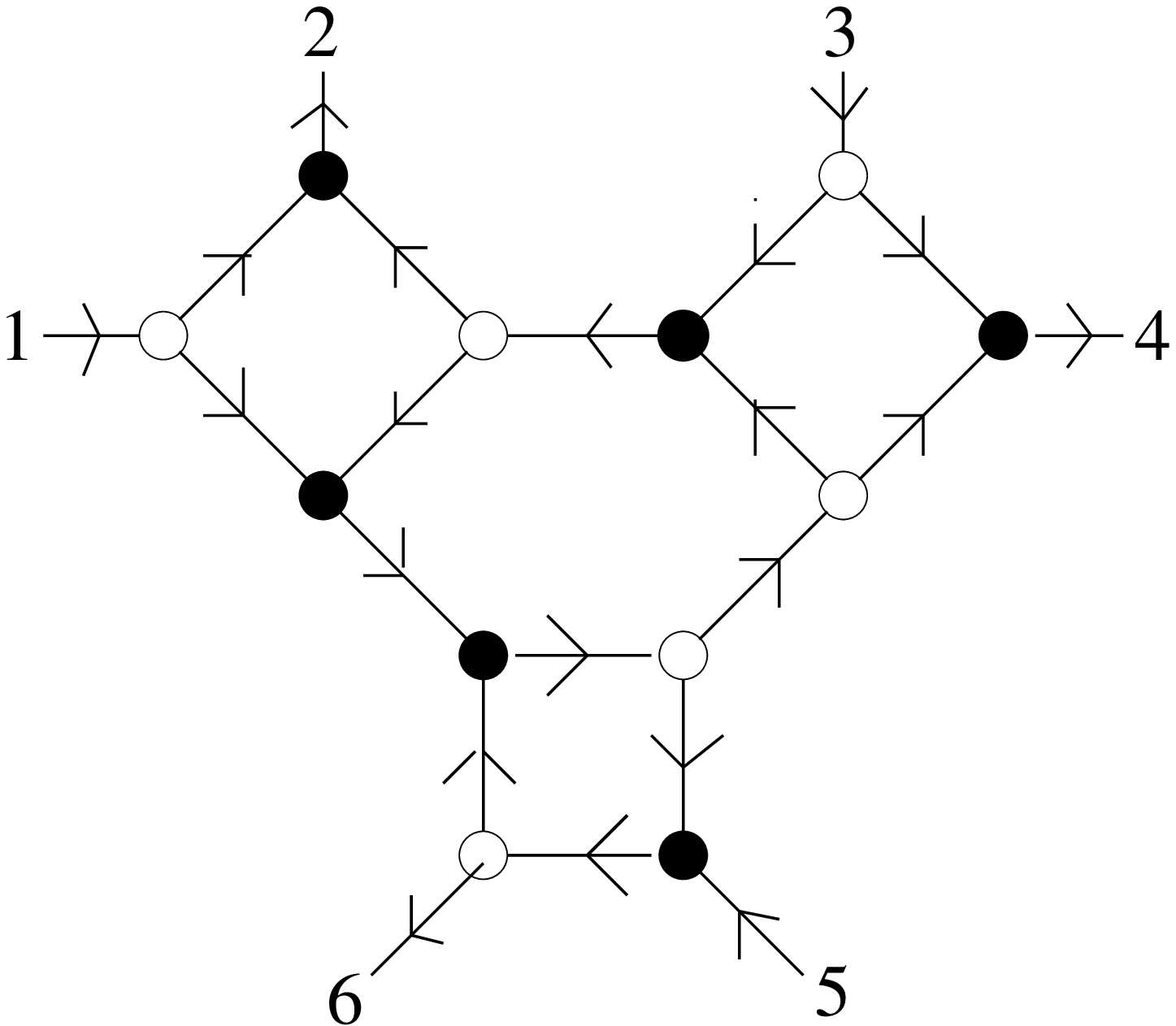}}}\:+\:
 \raisebox{-1.2cm}{\scalebox{.25}{\includegraphics{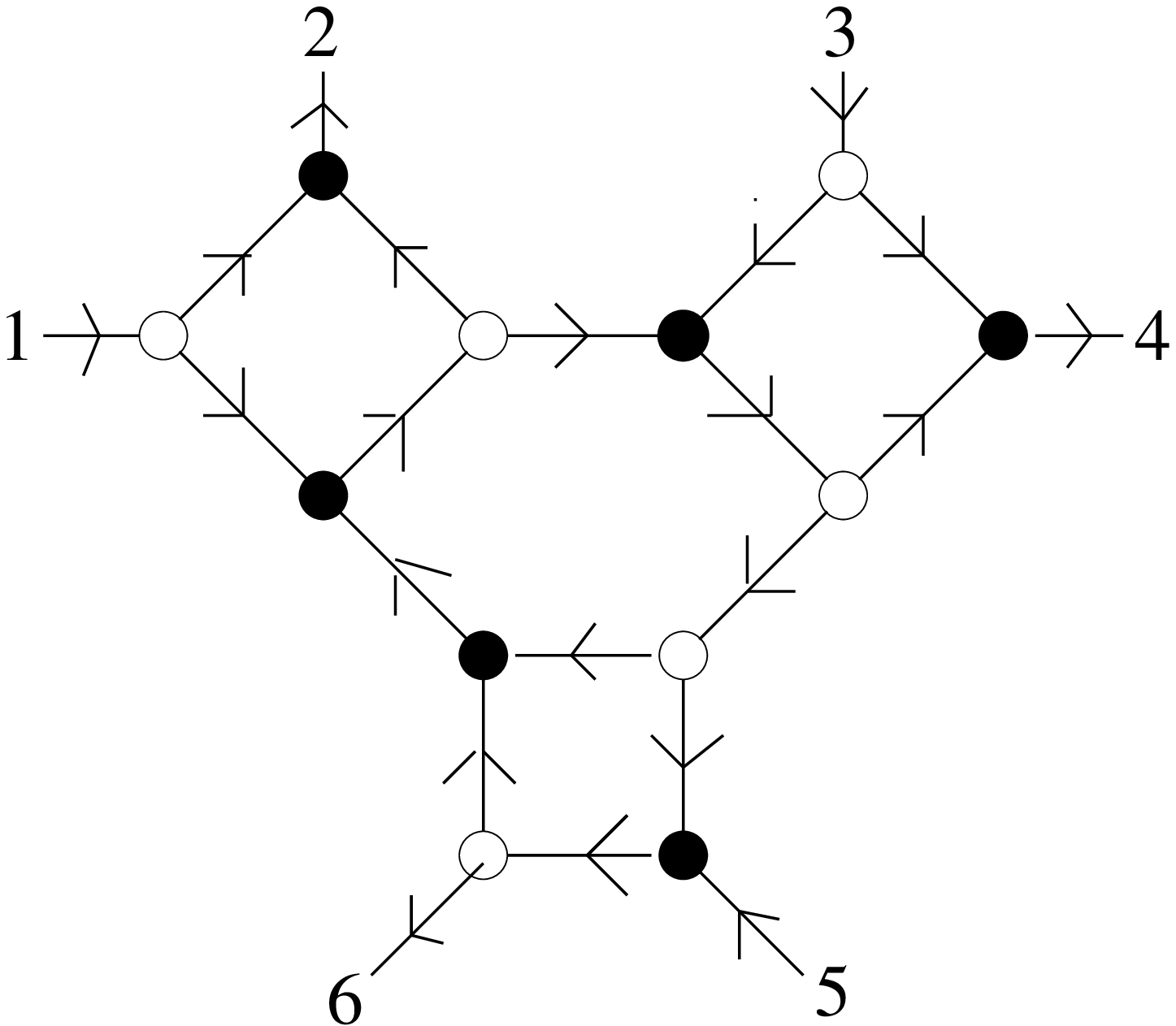}}}
\end{equation}
The first and the third diagrams have a single helicity loop (in the lower box and in the inner hexagon respectively), while the second one shows a helicity flow in both the lower box and the inner hexagon whose orientation
is opposite with respect to the one in the other diagrams. Notice that the third diagram is equivalent to the second one discussed in the previous section upon square move, which holds for the lower box only. For the first two 
diagrams, the only equivalence move possible is the merger.

The presence of the helicity loop in the outer sub-diagram, as in the first and second diagrams \eqref{eq:Gr36tc3}, indicates that the related higher order pole is located at $\Delta_{ijk}\,=\,0$, while its presence in the inner
sub-diagrams is an indication that the location of the related singularity is given by a special relation among Pl{\"u}cker coordinates, as in the example discussed in the previous section. Notice that the second diagram has
both type of helicity loop and, hence, it shows both type of singularities.

The explicit expressions for the on-shell functions on $Gr(3,6)$ represented by the first two diagrams in \eqref{eq:Gr36tc3} (the one for the third diagram is given in the second line of \eqref{eq:Gr36tc2}) are given by
\begin{equation}\eqlabel{eq:Gr36tc3b}
 \begin{split}
  &\raisebox{-1.2cm}{\scalebox{.25}{\includegraphics{Gr36topcellDec4.eps}}}\hspace{-.7cm}=
    \int\hspace{-.2cm}\frac{d^{\mbox{\tiny $(3\times6)$}}C}{\mbox{Vol}\{GL(3)\}}\frac{\delta^{\mbox{\tiny $(3\times4|3\times\mathcal{N}$)}}\left(C\cdot\mathcal{W}\right)}{\Delta_{123}\Delta_{234}\Delta_{345}\Delta_{456}\Delta_{561}\Delta_{612}}\hspace{-.1cm}
    \left[
     \frac{\Delta_{134}\Delta_{356}}{\Delta_{346}}
    \right]^{4-\mathcal{N}}\\
  &\raisebox{-1.2cm}{\scalebox{.25}{\includegraphics{Gr36topcellDec4b.eps}}}\hspace{-.5cm}=\:
    \int\frac{d^{\mbox{\tiny $(3\times6)$}}C}{\mbox{Vol}\{GL(3)\}}\frac{\delta^{\mbox{\tiny $(3\times4|3\times\mathcal{N}$)}}\left(C\cdot\mathcal{W}\right)\Delta_{612}^{4-\mathcal{N}}}{\Delta_{123}\Delta_{234}\Delta_{345}
     \Delta_{456}\Delta_{561}\Delta_{612}}\times\\
  &\hspace{7cm}\times
    \left[
     \frac{\Delta_{345}}{\Delta_{346}}\frac{\Delta_{134}\Delta_{356}}{\Delta_{346}\Delta_{512}-\Delta_{345}\Delta_{612}}
    \right]^{4-\mathcal{N}}.
 \end{split}
\end{equation}
As predicted, they share a higher order pole, located at $\Delta_{346}\,=\,0$, whose residue can be diagrammatically depicted by unwinding the helicity loop according to its direction, as seen in Section \ref{sec:SingGrass}:
\begin{equation}\eqlabel{eq:Gr36cell346}
 \begin{split}
  &\raisebox{-1.2cm}{\scalebox{.25}{\includegraphics{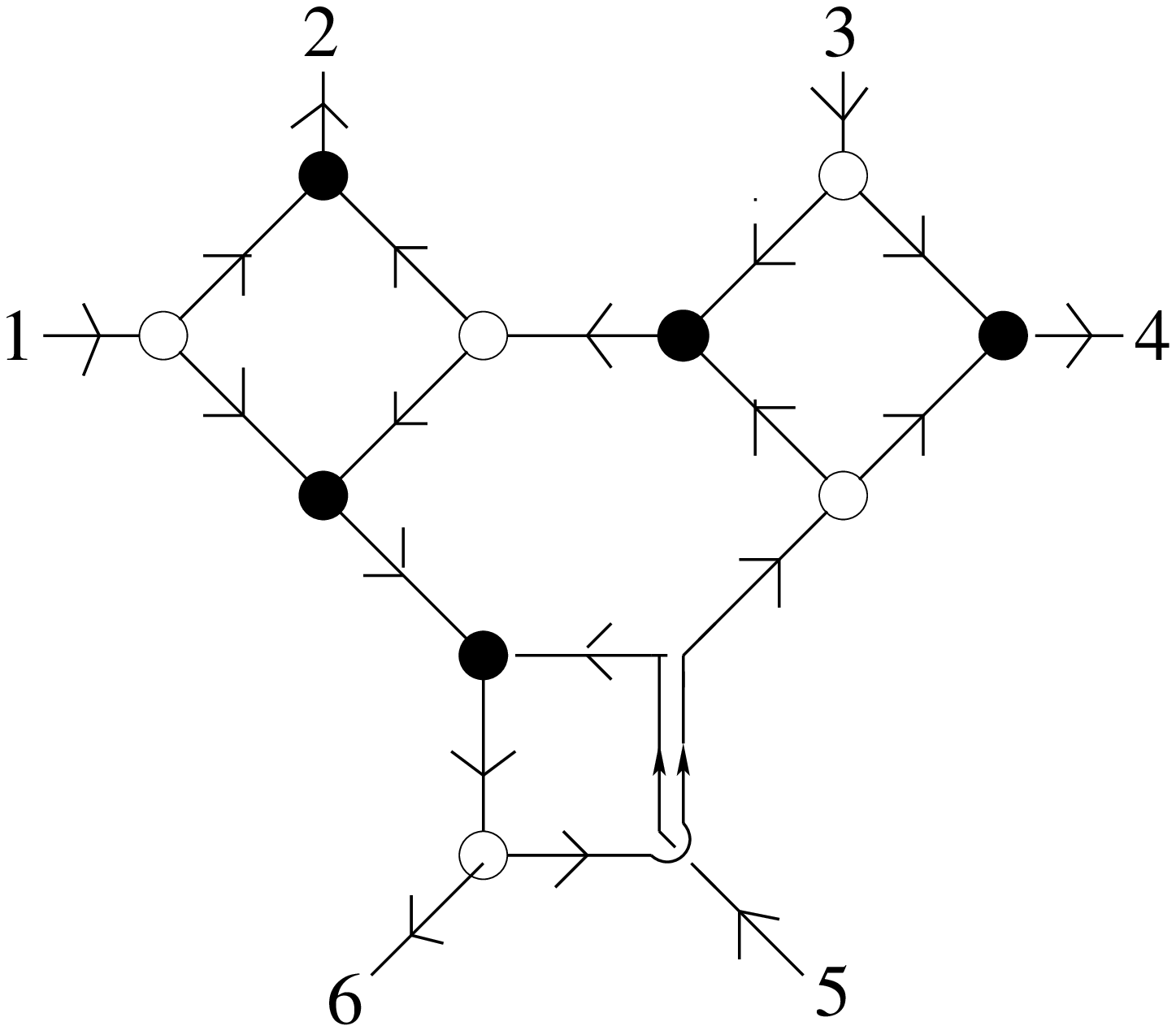}}}\hspace{-.5cm}=\: \int\frac{d^{\mbox{\tiny $(3\times6)$}}C}{\mbox{Vol}\{GL(3)\}}\frac{\delta^{\mbox{\tiny $(3\times4|3\times\mathcal{N}$)}}\left(C\cdot\mathcal{W}\right)\,\left(\Delta_{134}\Delta_{356}\right)^{4-\mathcal{N}}}{\Delta_{123}\Delta_{234}\Delta_{345}\Delta_{456}\Delta_{561}\Delta_{612}}\delta^{\mbox{\tiny $(3-\mathcal{N})$}}\left(\Delta_{346}\right)\\
  &\raisebox{-1.2cm}{\scalebox{.25}{\includegraphics{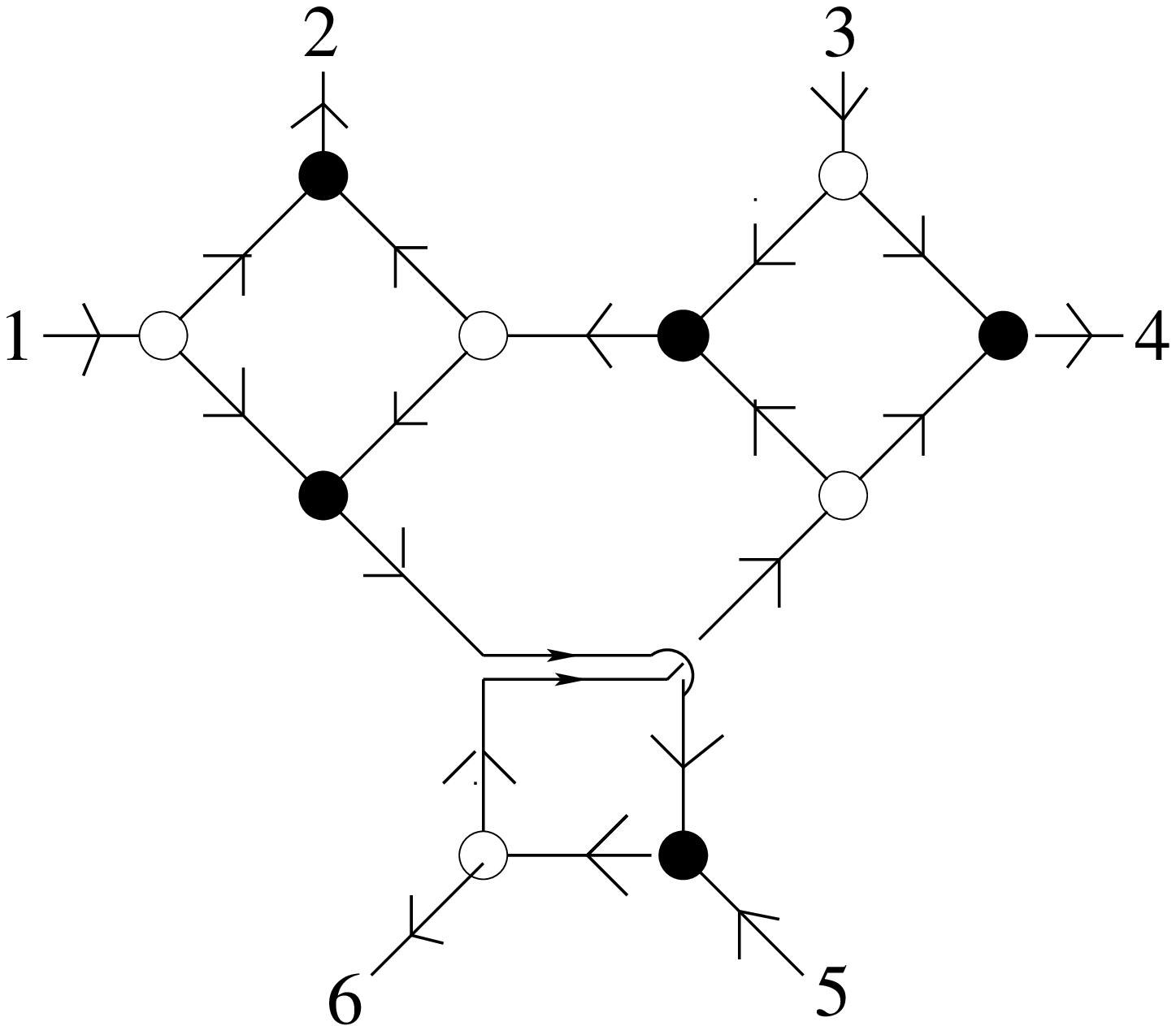}}}\hspace{-.5cm}=\: \int\frac{d^{\mbox{\tiny $(3\times6)$}}C}{\mbox{Vol}\{GL(3)\}}\frac{\delta^{\mbox{\tiny $(12|3\mathcal{N}$)}}\left(C\cdot\mathcal{W}\right)\left(\Delta_{612}\Delta_{345}\right)^{4-\mathcal{N}}}{\Delta_{123}\Delta_{234}\Delta_{345}\Delta_{456}\Delta_{561}\Delta_{612}}\times\\
  &\hspace{5cm}\times\left[
     \frac{\Delta_{134}\Delta_{356}}{\Delta_{346}\Delta_{512}-\Delta_{345}\Delta_{612}}
    \right]^{4-\mathcal{N}}\delta^{(3-\mathcal{N})}\left(\Delta_{346}\right)
 \end{split}
\end{equation}


\subsection{Identities among on-shell diagrams}\label{subsec:OSids}

As seen for the $Gr(2,4)$ case, also on $Gr(3,6)$ the residue theorem returns a set of identities among on-shell functions. In this case, the (super)-momentum conserving delta-functions are able to fix eight out of the nine
degrees of freedom of the top-cell and, therefore, the free parameter can be used as integration variable. The integration over the Riemann sphere, because of Cauchy's theorem, tells us that the sum over all the residues
needs to vanish. As a first example, let us consider the on-shell function in the first line of \eqref{eq:Gr36tc3b}:
\begin{align}\eqlabel{eq:Gr36Id1}
  0\:&=\raisebox{-1.2cm}{\scalebox{.25}{\includegraphics{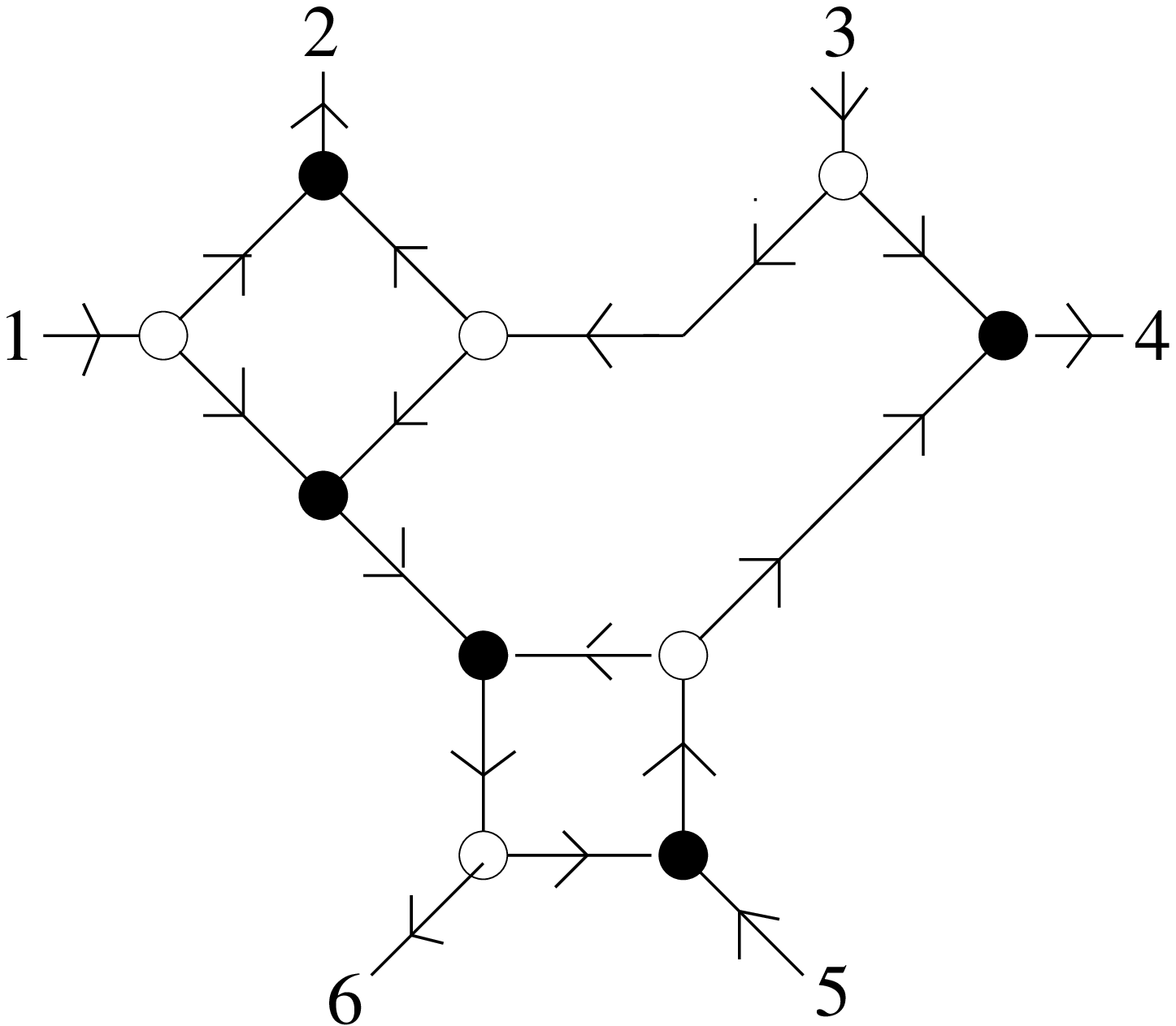}}}-\:\raisebox{-1.2cm}{\scalebox{.25}{\includegraphics{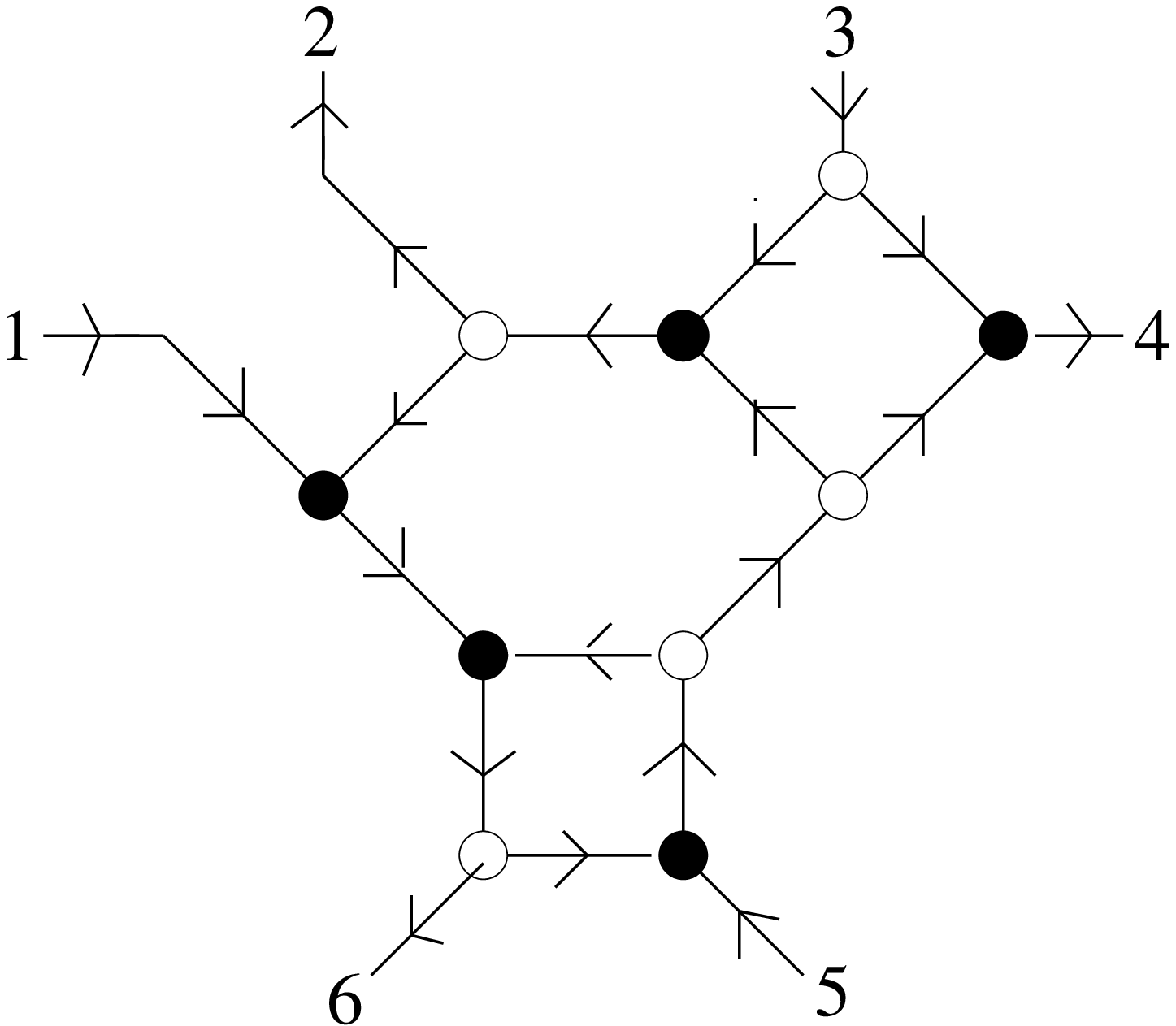}}}+\:\raisebox{-1.2cm}{\scalebox{.25}{\includegraphics{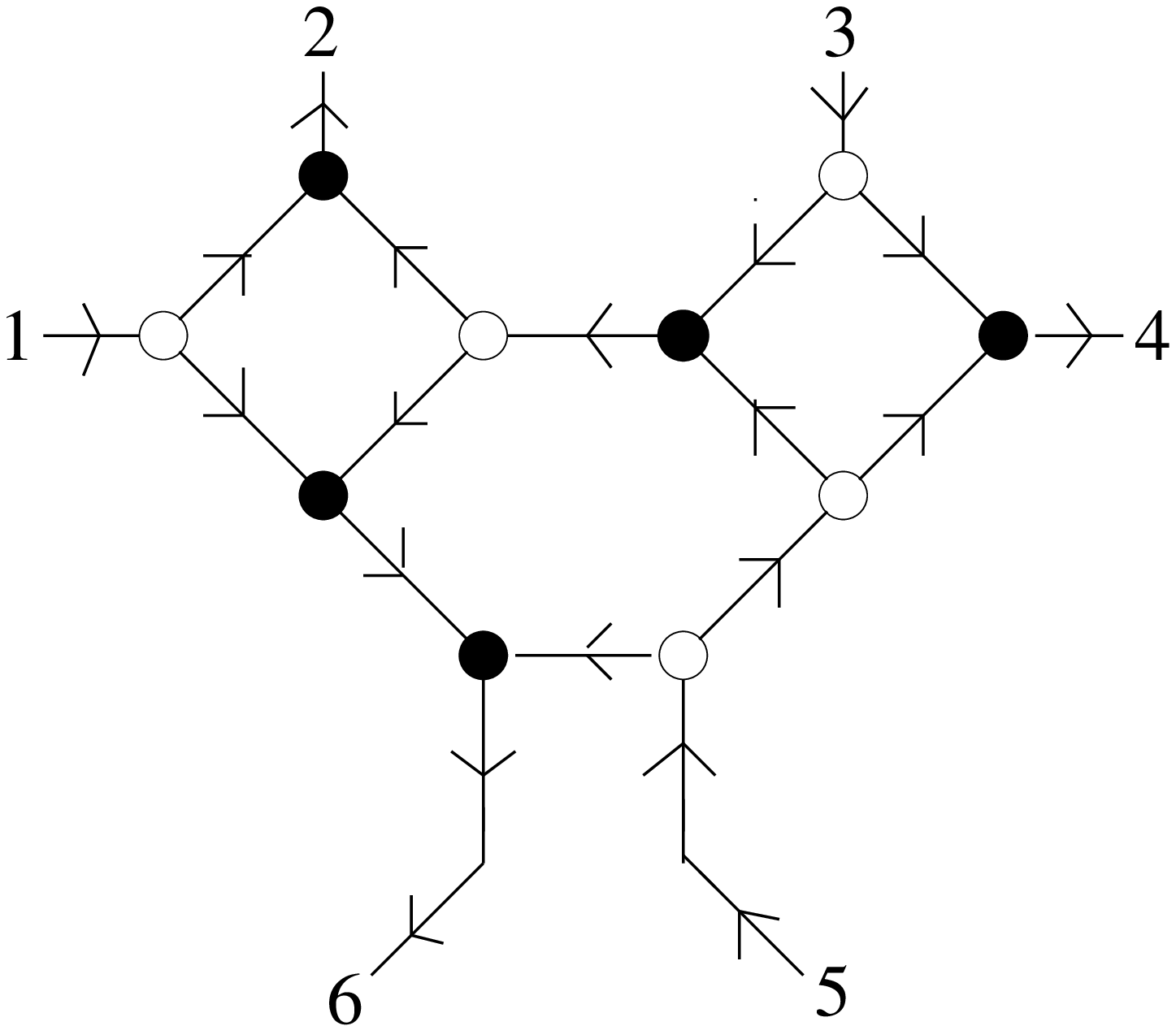}}}-
       \nonumber\\
     &\hspace{1.5cm}\mbox{\footnotesize $\{\Delta_{123}^{\mbox{\tiny $(0)$}}\,=\,0\}$}\hspace{2.6cm}\mbox{\footnotesize $\{\Delta_{234}^{\mbox{\tiny $(0)$}}\,=\,0\}$}
      \hspace{2.6cm}\mbox{\footnotesize $\{\Delta_{345}^{\mbox{\tiny $(0)$}}\,=\,0\}$}\\
     &-\raisebox{-1.2cm}{\scalebox{.25}{\includegraphics{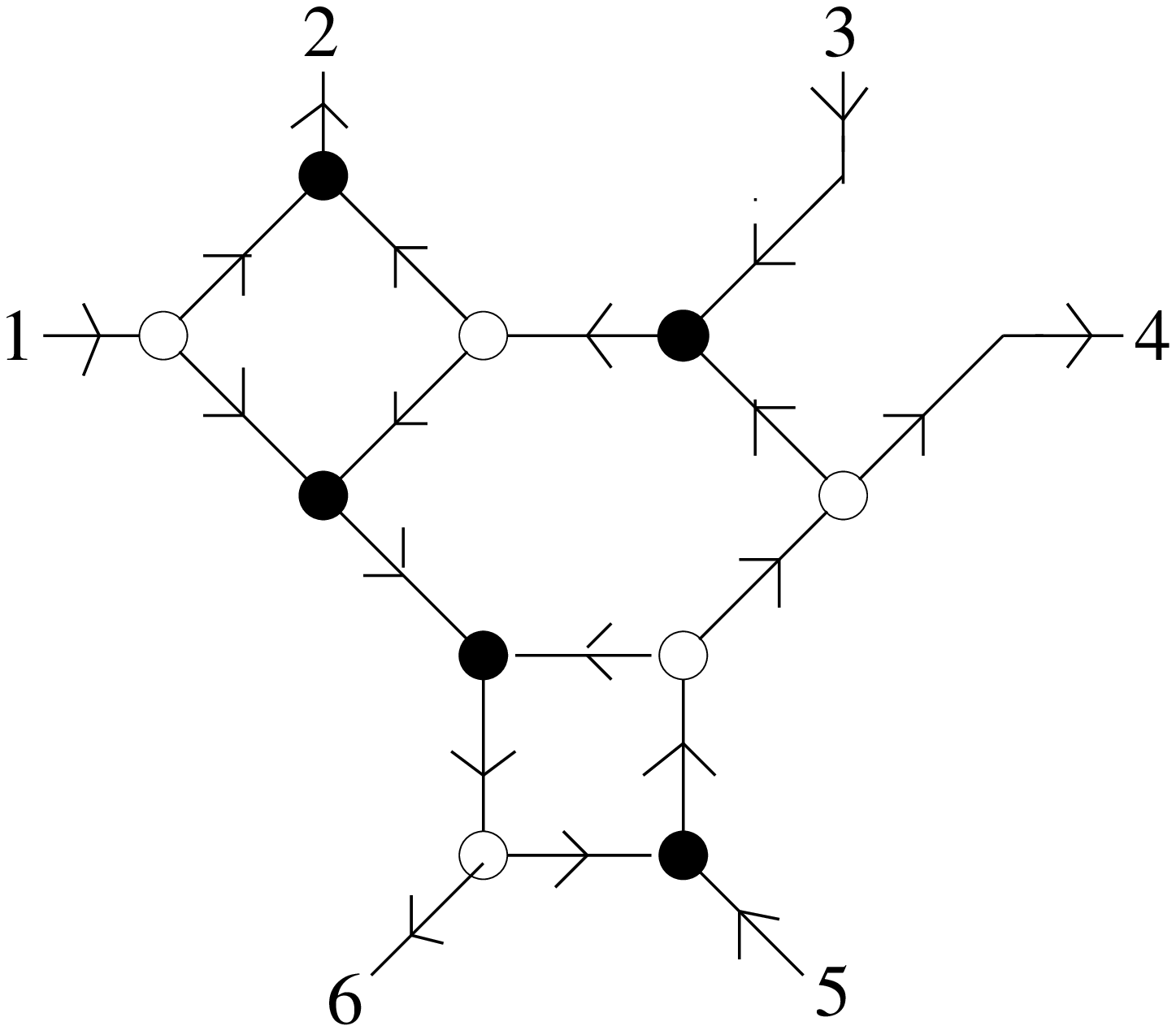}}}+\:\raisebox{-1.2cm}{\scalebox{.25}{\includegraphics{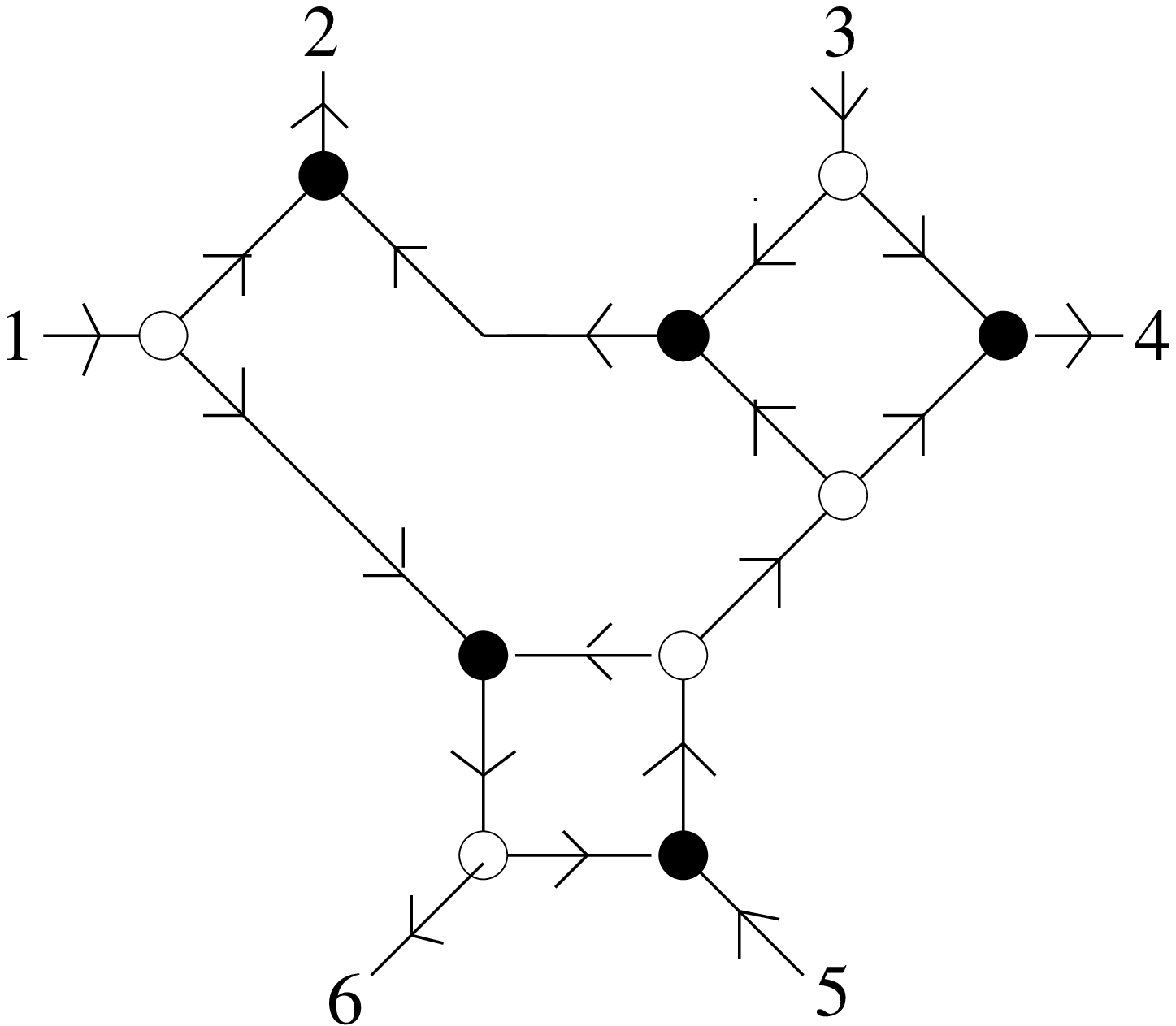}}}-\:\raisebox{-1.2cm}{\scalebox{.25}{\includegraphics{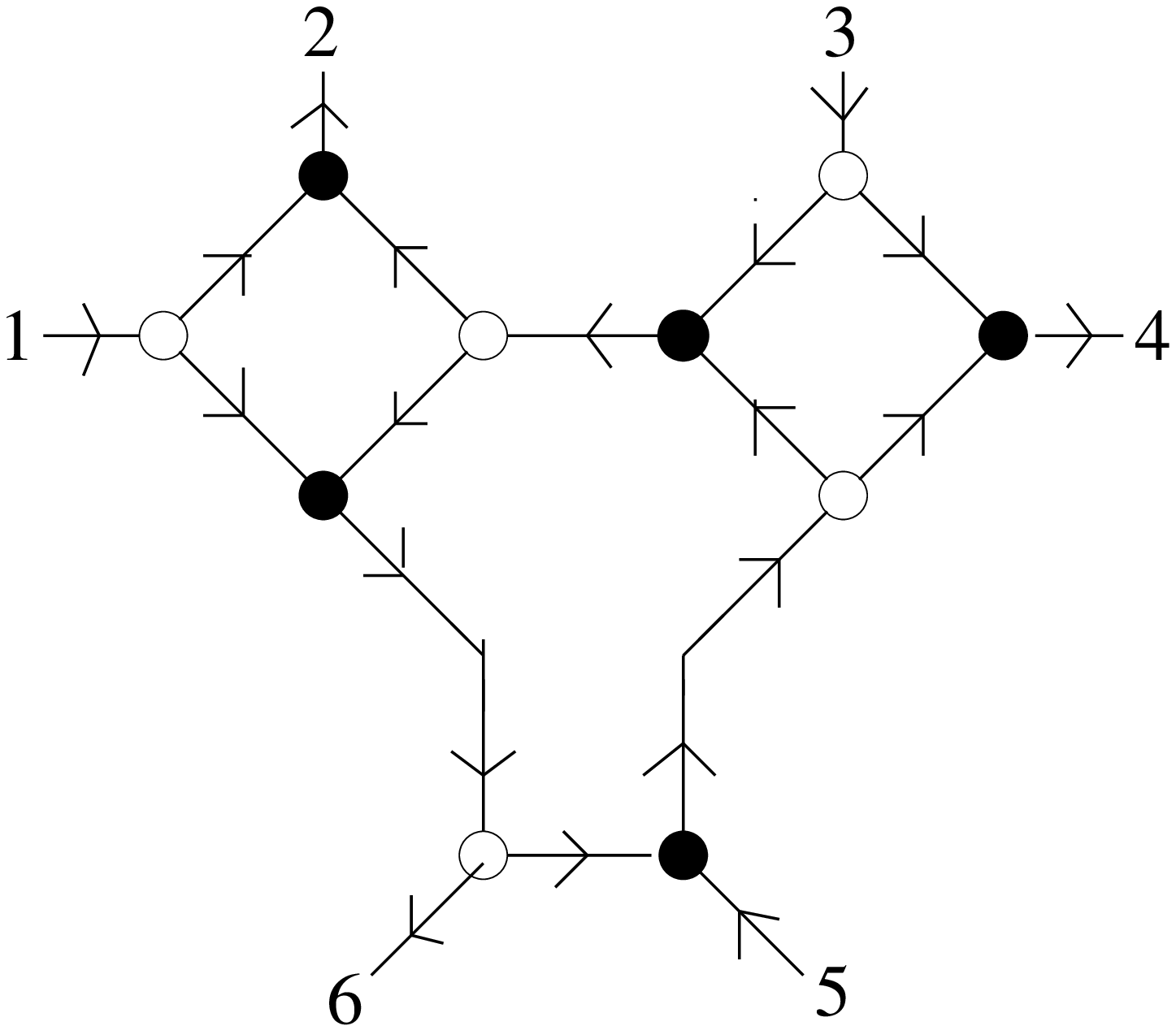}}}-
       \nonumber\\
     &\hspace{1.5cm}\mbox{\footnotesize $\{\Delta_{456}^{\mbox{\tiny $(0)$}}\,=\,0\}$}\hspace{2.5cm}\mbox{\footnotesize $\{\Delta_{561}^{\mbox{\tiny $(0)$}}\,=\,0\}$}
      \hspace{2.6cm}\mbox{\footnotesize $\{\Delta_{612}^{\mbox{\tiny $(0)$}}\,=\,0\}$}\displaybreak[0]\nonumber\\
     &-\raisebox{-1.2cm}{\scalebox{.25}{\includegraphics{Gr36cell346Dec4.eps}}}\nonumber\\
     &\hspace{1.2cm}\mbox{\footnotesize $\{\Delta_{346}^{\mbox{\tiny $(3-\mathcal{N})$}}\,=\,0\}$}.\nonumber
\end{align}
In the identity above, the label $\{\Delta_{ijk}^{\mbox{\tiny $(m)$}}\,=\,0\}$ below each decorated on-shell diagram identifies the $\delta^{\mbox{\tiny $(m)$}}$-support where each of them lives -- let us stress again that in all the 
cases these on-shell diagrams identify on-shell functions which live in the sub-cell $\Delta_{ijk}\,=\,0$ of $Gr(3,6)$.

Let us move on to the most interesting case given by the on-shell function in the first line of \eqref{eq:Gr36tc2}. The identity coming from the residue theorem is given by:
\begin{equation}\eqlabel{eq:Gr36Id2}
 \begin{split}
  0\:&=\raisebox{-1.2cm}{\scalebox{.25}{\includegraphics{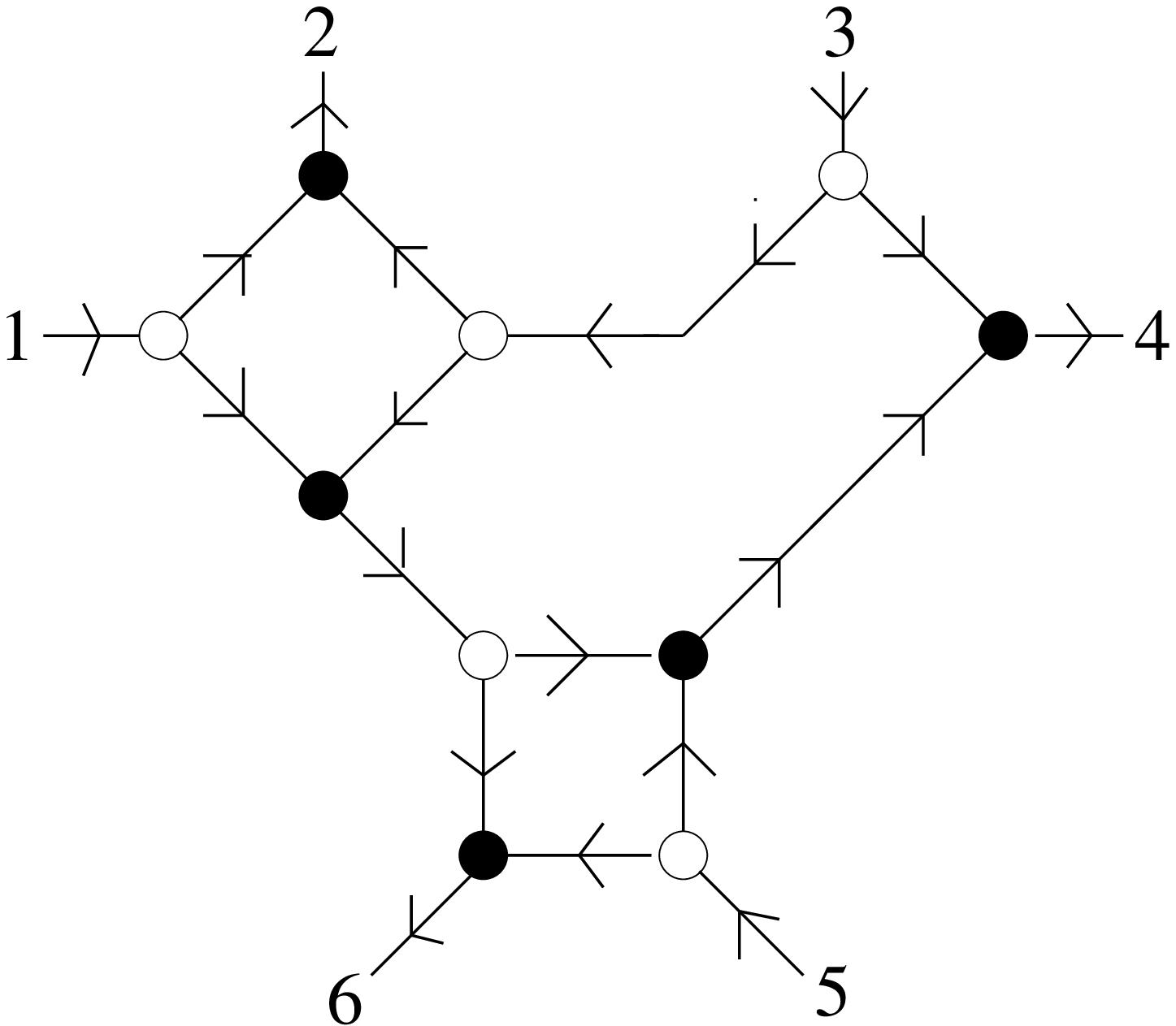}}}-\:\raisebox{-1.2cm}{\scalebox{.25}{\includegraphics{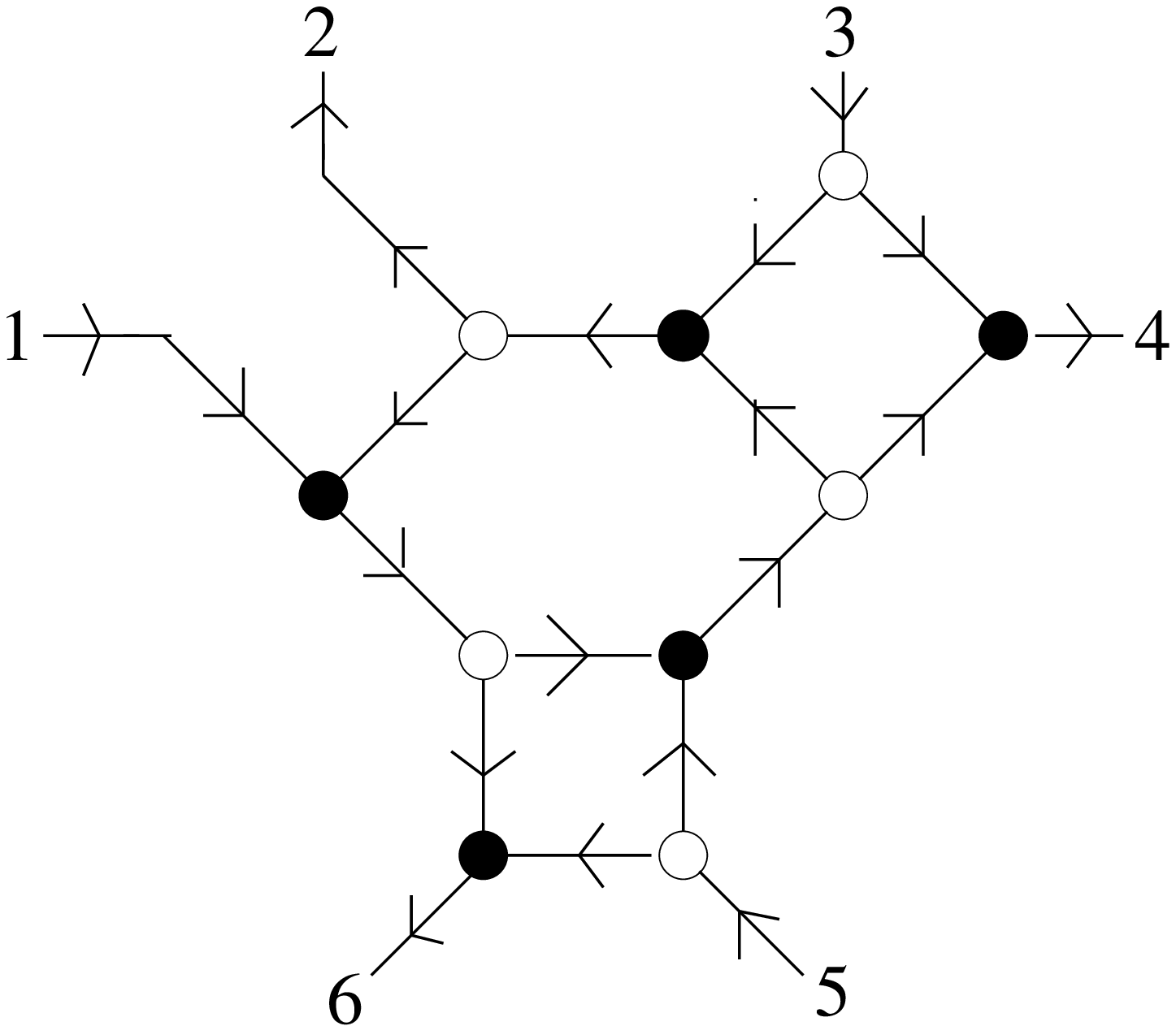}}}+\:\raisebox{-1.2cm}{\scalebox{.25}{\includegraphics{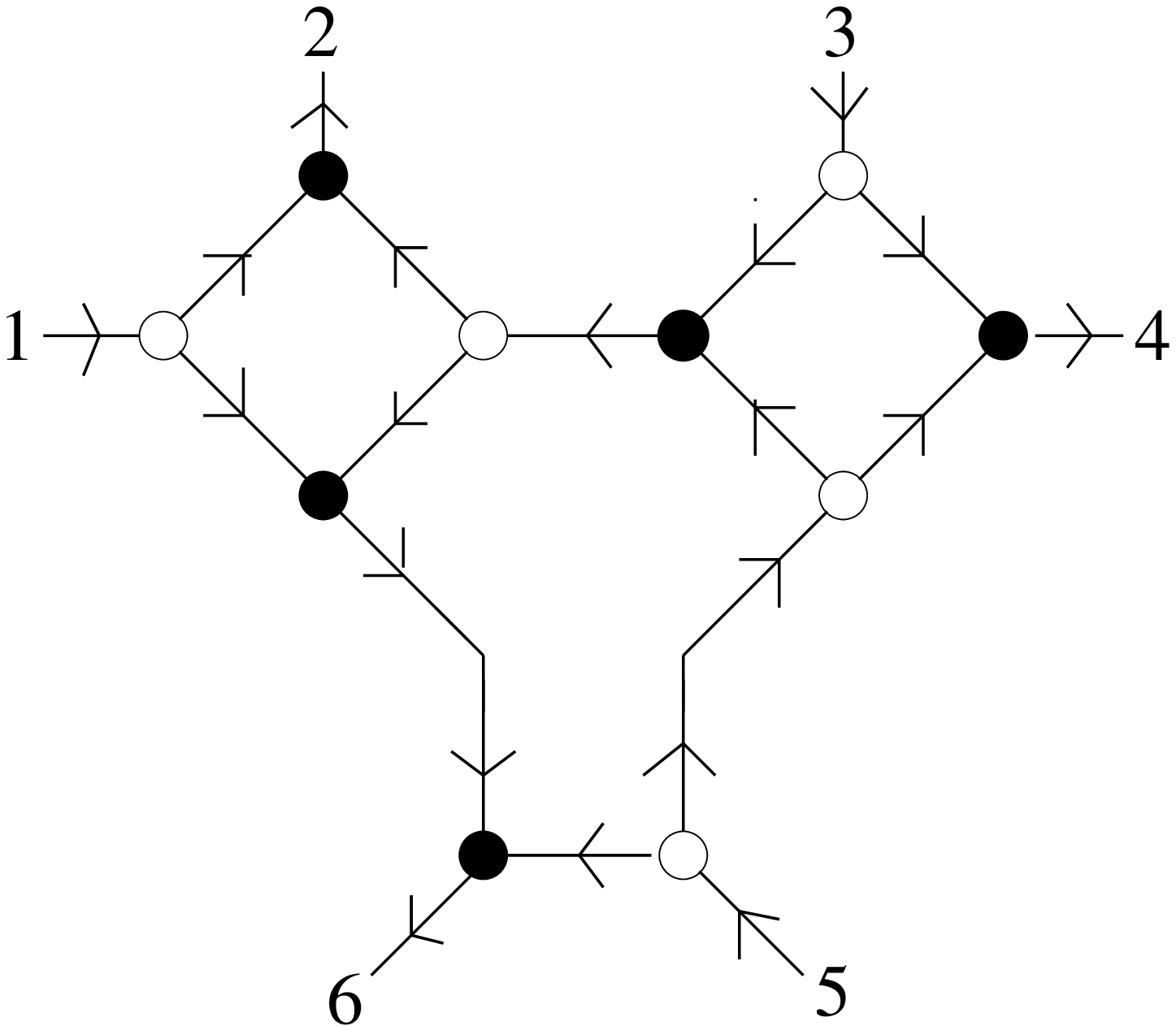}}}-\\
     &\hspace{1.5cm}\mbox{\footnotesize $\{\Delta_{123}^{\mbox{\tiny $(0)$}}\,=\,0\}$}\hspace{2.6cm}\mbox{\footnotesize $\{\Delta_{234}^{\mbox{\tiny $(0)$}}\,=\,0\}$}
      \hspace{2.6cm}\mbox{\footnotesize $\{\Delta_{345}^{\mbox{\tiny $(0)$}}\,=\,0\}$}\\
     &-\raisebox{-1.2cm}{\scalebox{.25}{\includegraphics{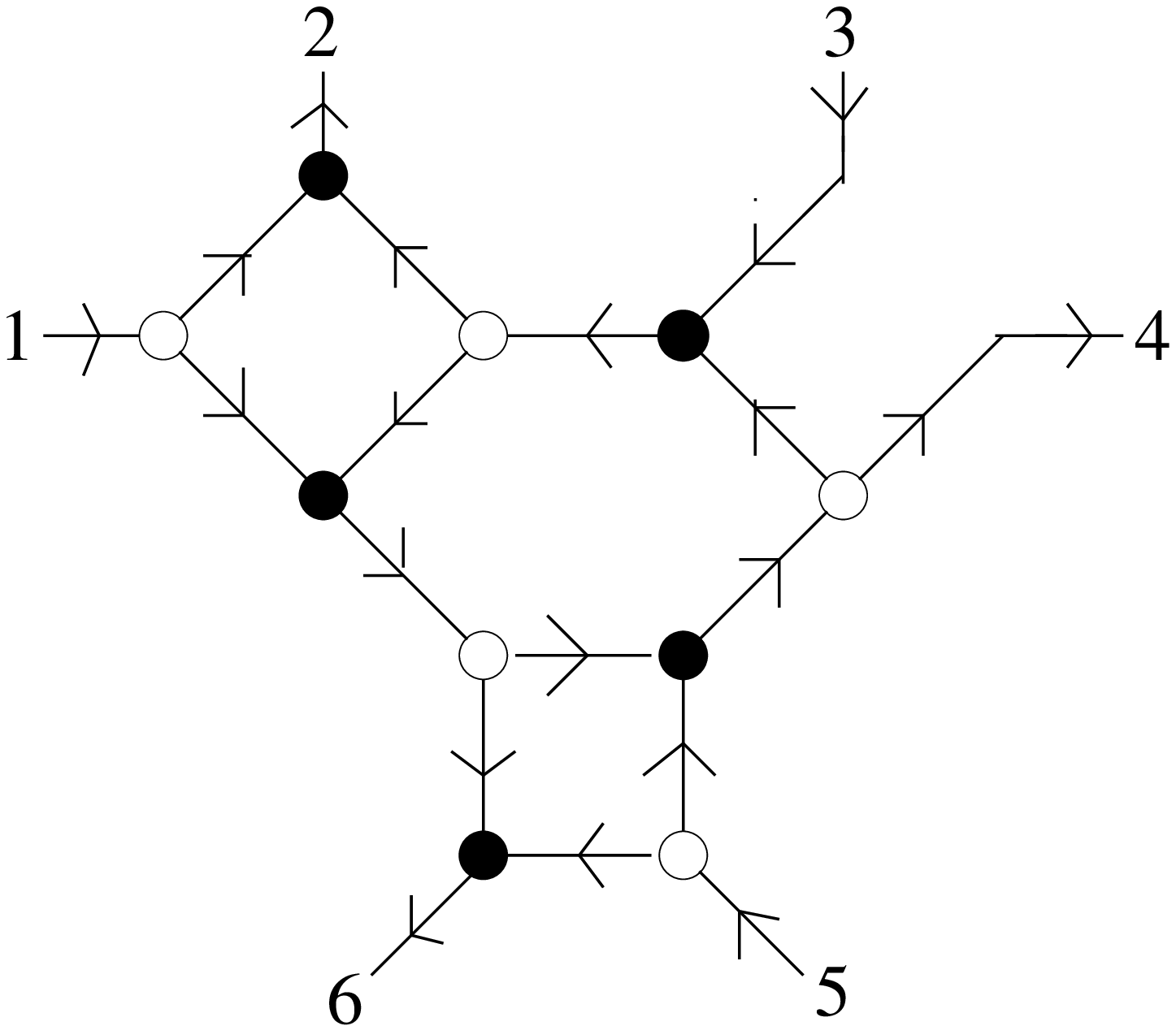}}}+\:\raisebox{-1.2cm}{\scalebox{.25}{\includegraphics{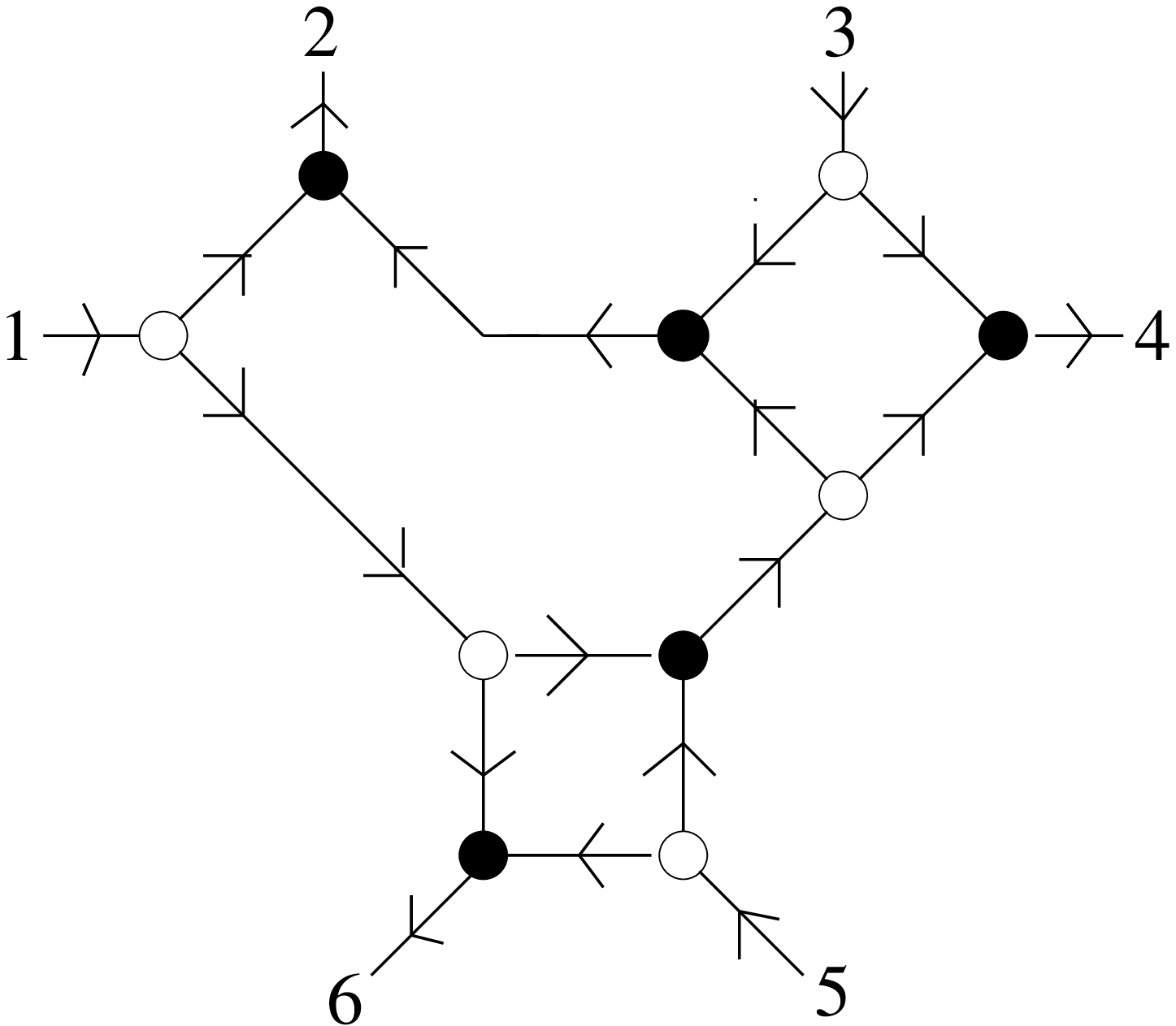}}}-\:\raisebox{-1.2cm}{\scalebox{.25}{\includegraphics{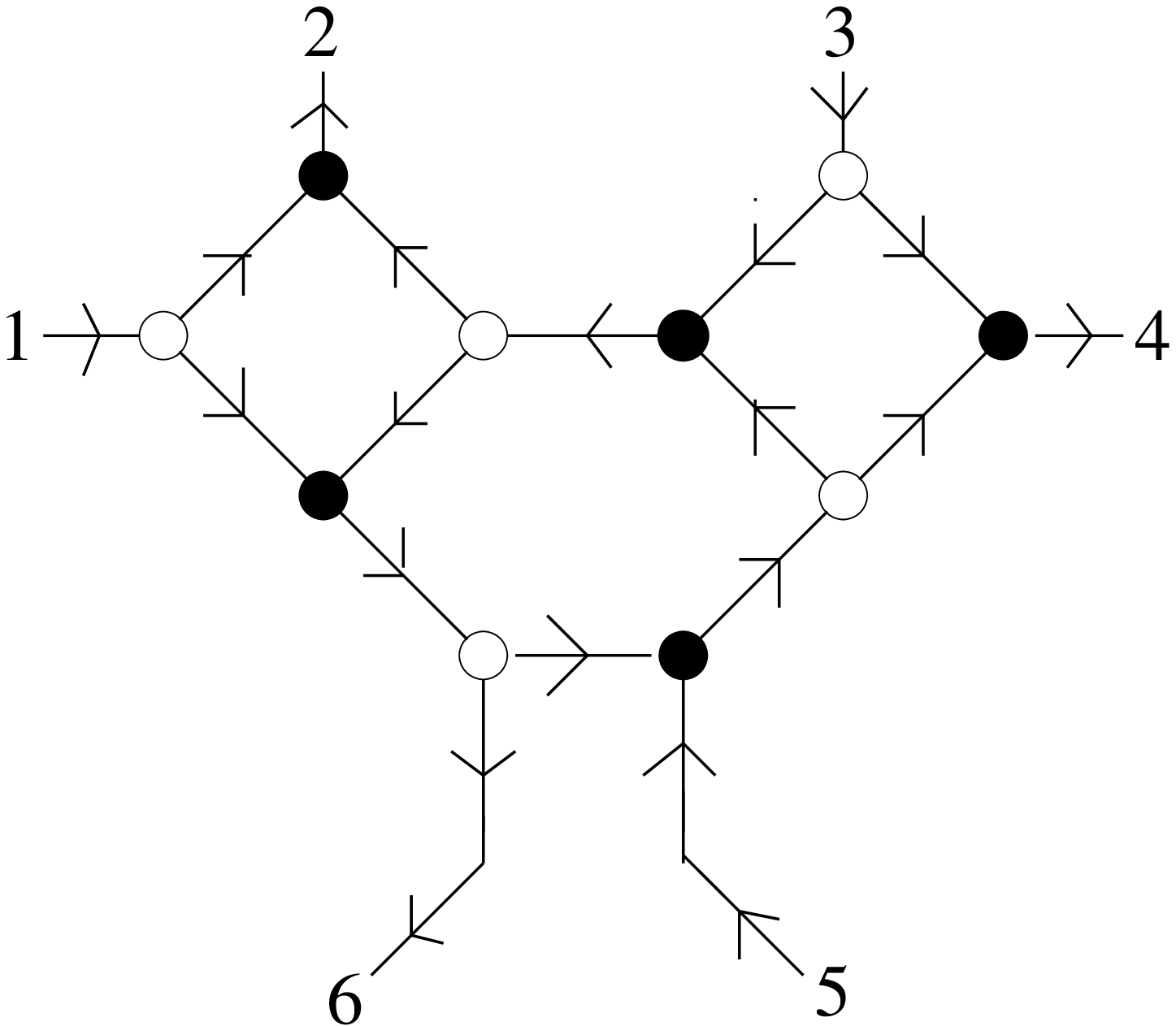}}}-\\
     &\hspace{1.5cm}\mbox{\footnotesize $\{\Delta_{456}^{\mbox{\tiny $(0)$}}\,=\,0\}$}\hspace{2.5cm}\mbox{\footnotesize $\{\Delta_{561}^{\mbox{\tiny $(0)$}}\,=\,0\}$}
      \hspace{2.6cm}\mbox{\footnotesize $\{\Delta_{612}^{\mbox{\tiny $(0)$}}\,=\,0\}$}\\
     &-\raisebox{-1.2cm}{\scalebox{.25}{\includegraphics{Gr36resNP.eps}}}\\
     &\hspace{1.2cm}\mbox{\footnotesize $\{\Delta_{34\mathfrak{c}}^{\mbox{\tiny $(3-\mathcal{N})$}}\,=\,0\}$}.
 \end{split}
\end{equation}
Very interestingly, notice that the sum of the diagrams corresponding to the residues in $\{\Delta_{123}\,=\,0,\,\Delta_{345}\,=\,0,\,\Delta_{561}\,=\,0\}$ is a representation of the tree-level $6$-particle amplitude with
helicity configuration $(-,+,-,+,-,+)$. The identity \eqref{eq:Gr36Id2} then provides another representation for such an amplitude, which involves diagrams with non-planar structures. Importantly, this representation does not come
from the recursion which a BCFW bridge on any of the pairs $\{(1,2),(3,\,4),\,(5,6)\}$ with helicity loop would generate.


\section{Conclusion}\label{sec:Concl}

The on-shell diagrammatics allows to define a theory, at least perturbatively, from first principles without any reference to a pre-existent Lagrangian and just in terms of observables, making manifest structures of theory that
are completely hidden in the most traditional approach of quantum field theory. This approach has been extensively explored in the context of maximally supersymmetric Yang-Mills theory, whose properties can be translated in
geometrical terms. What turns out to be very interesting is the relation between the on-shell diagrams and mathematical structures such as permutations, the positive Grassmannian and cluster algebras.

A general question that we have started addressing in the present paper as well as in a previous one \cite{Benincasa:2015zna} is whether much of the mathematical structure unveiled for $\mathcal{N}\,=\,4$ SYM survives for
less special theories. Indeed the procedure for constructing on-shell processes is not theory dependent: the three-particle amplitudes (the nodes of the diagrams) are fixed by Poincar{\'e} invariance for arbitrary
helicity configurations, and the gluing of these objects just amounts to integrate out the degrees of freedom along the edges which gets glued. Thus, the question is not really whether it is possible to build on-shell diagrams
for more general theories, rather which information about the theory they encode, given that, in these cases, it is not clear the direct relation between on-shell diagrams and the amplitudes.

The less/no-supersymmetric Yang-Mills theories offer a good arena for addressing these issues: They can be thought of as the next-to-simplest examples because, on one side, at least a class of diagrams have a direct physical 
interpretation, and on the other side they offer a richer structure ({\it e.g} UV divergences and rational terms at loop level). In the same direction, a discussion about on-shell diagrams in gravity has been recently pursued
\cite{Heslop:2016plj, Herrmann:2016qea}.

In this paper, we discuss the Grassmannian representation for on-shell processes. The richness of the structure appears immediately because of the presence of new poles which are typically a feature of non-planar diagrams. They are
multiple poles, except in $\mathcal{N}\,=\,3$ SYM whose individual diagrams show a very close structure with the non-planar diagrams in $\mathcal{N}\,=\,4$ SYM. These new singularities are associated to the helicity
loops which can appear in the on-shell diagrams, and can be identified either by the vanishing of a single Pl{\"u}cker coordinate involving at least two non-adjacent columns ({\it e.g} in $Gr(2,4)$), or as a non-Pl{\"u}cker relation
({\it e.g.} $Gr(3,6)$). The on-shell diagrams can be thought of as generalised functions on the Grassmannian, with the ones which live on the on-shell variety defined by a non-planar conditions having support on a derivative
delta-function.

We extensively discussed the on-shell functions on $Gr(2,4)$ and $Gr(3,6)$. In both cases, this formulation allows to obtain several identities among on-shell functions defined on codimension-$0$ and codimension-$1$ varieties
for $Gr(2,4)$ and $Gr(3,6)$ respectively. More precisely, on $Gr(2,4)$ it is possible to identify two classes of such identities: The first one provides the equivalence between two different BCFW representations of the four-particle
amplitude, with the explicit on-shell function for the contribution related to the multiple pole; The second one establishes an equivalence between the on-shell diagrams with helicity loops and non-planar functions.

On $Gr(3,6)$, the residue theorem allows to obtain a plethora of identities among the on-shell functions on a codimension-$1$ variety, the most interesting of which provides a new representation for the NMHV six-particle amplitude with 
helicity configuration $(-,+,-,+,-,+)$: Notice that such a representation, in terms of individual diagrams, is not the one that would be obtained by simply generating the amplitude recursively (with boundary term) using a BCFW bridge
with a helicity loop -- It can be indeed recast into it but applying several other identities among on-shell functions.

On the top-cell of $Gr(3,6)$, the delta-functions containing the kinematic data fix all the degrees of freedom of the Grassmannian but one. Analysing the non-Pl{\"u}cker relations in momentum space in terms of such a parameter,
one discovers that such relations are second-order algebraic equations. Thus the correct on-shell function living on the sheet identified by a non-Pl{\"u}cker relation, is obtained by summing over both the solutions.

This paper represents a first step towards a deeper understanding on the geometry of decorated on-shell diagrams for $\mathcal{N}\,<\,4$ SYM theories, and several questions remains open. The first one indeed is related to the
possibility of a systematic survey of these non-planar (higher-order) structures for $Gr(k,n)$ as well as of the identities among the on-shell functions on $Gr(k,n)$. Secondly, in the maximally supersymmetric case, the 
positive-preserving  diffeomorphisms on $Gr_{\mbox{\tiny $\ge\,0$}}(k,n)$ are related to the Yangian symmetry. In the present context, it is a fair question to ask whether the diffeomorphisms on $Gr(k,n)$ encode
some (so far) hidden symmetry. Furthermore, even if there is no real notion of positivity, it would be interesting to check whether there is a general criterium to fix the signs, as it occurs in the non-planar
sector of $\mathcal{N}\,=\,4$ SYM. Finally, it would be interesting to explore the possibility to have an amplituhedron formulation, along similar lines of what happens in the maximally supersymmetric cases
\cite{Arkani-Hamed:2013jha, Arkani-Hamed:2013kca, Bai:2014cna, Franco:2014csa, Lam:2014jda, Arkani-Hamed:2014dca, Bai:2015qoa, Ferro:2015grk, Bern:2015ple, Galloni:2016iuj}, 
and how the new structures are encoded in this geometry.

\section*{Acknowledgements}

P.B. would like to thank Nima Arkani-Hamed and the Institute for Advanced Studies in Princeton for hospitality while this work was in progress as well as Nima Arkani-Hamed, Jacob Bourjaily, Freddy Cachazo, Simon Caron-Huot and Yu-tin 
Huang for insightful discussions about the subject of the present paper and related topics. P.B. would also like to thank the developers of SAGE \cite{sage}, Maxima \cite{maxima} and Xfig \cite{xfig}. P.B. and D.G. are supported 
in part by Plan Nacional de Altas Energ{\'i}as (FPA2012-32828 and FPA2015-65480-P) and the Spanish MINECO's Centro de Excelencia Severo Ochoa Programme under grant SEV-2012-0249. D.G. is a La Caixa-Severo Ochoa Scholar and thanks La 
Caixa foundation for financial support.

\appendix

\section{On the structure on the $1$-loop integrand}\label{app:1Lint}

The identities discussed in Section \ref{subsec:IdGr24} relate on-shell functions defined on the top-cell of $Gr(2,4)$ with a given choice for the external sources/sinks to an on-shell function defined again on the top-cell of 
$Gr(2,4)$ but with a different ordering. They can be useful to rewrite the on-shell $4$-forms, which are related to the constructible part of the one-loop integrand \cite{Benincasa:2015zna}, in such a way that that the $d\log$ part and 
the part  which contains higher order poles get split:
\begin{equation}\eqlabel{eq:4formSplit}
 \begin{split}
  &\raisebox{-1.3cm}{\scalebox{.20}{\includegraphics{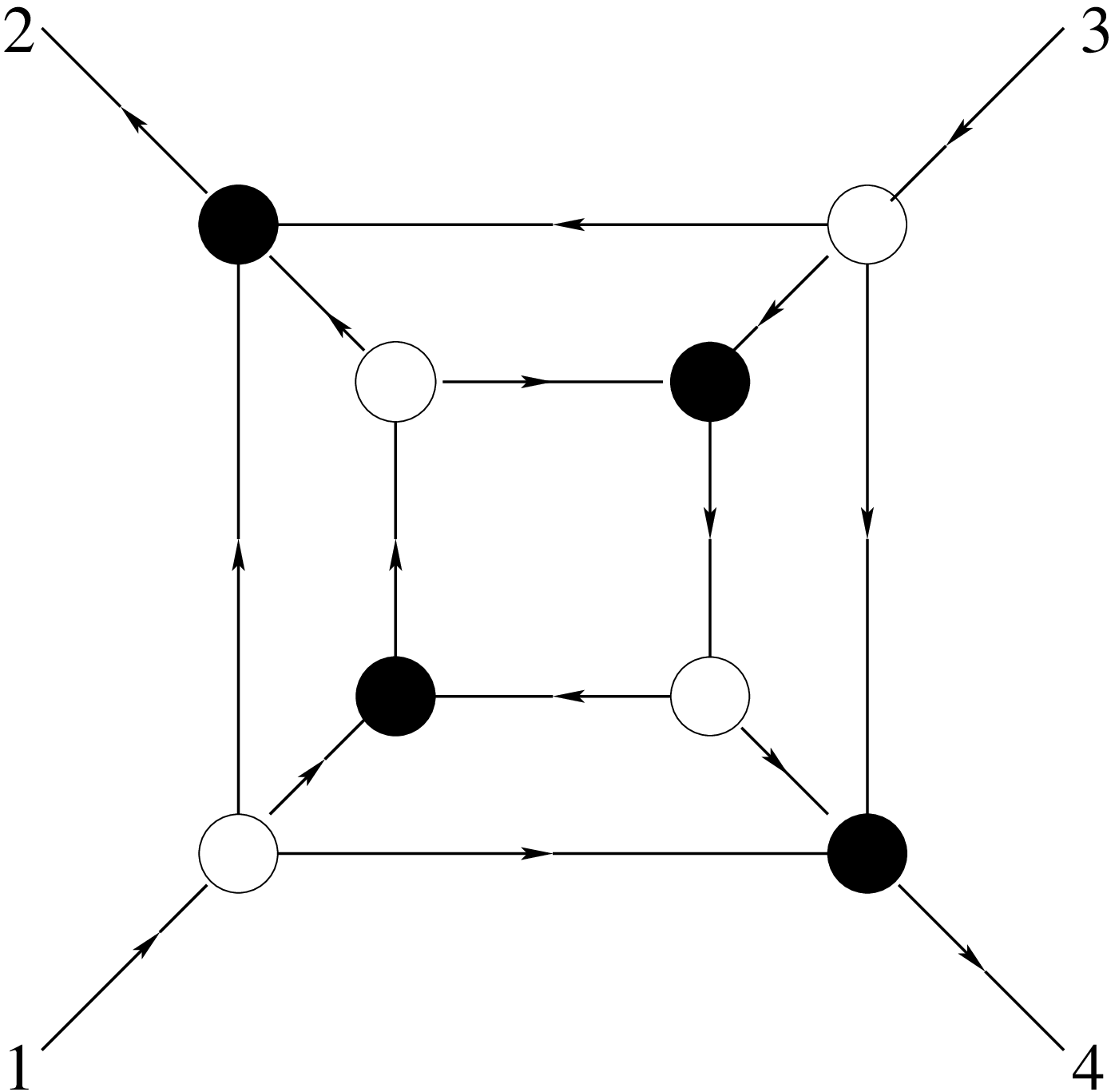}}}\:=\:\raisebox{-1.3cm}{\scalebox{.20}{\includegraphics{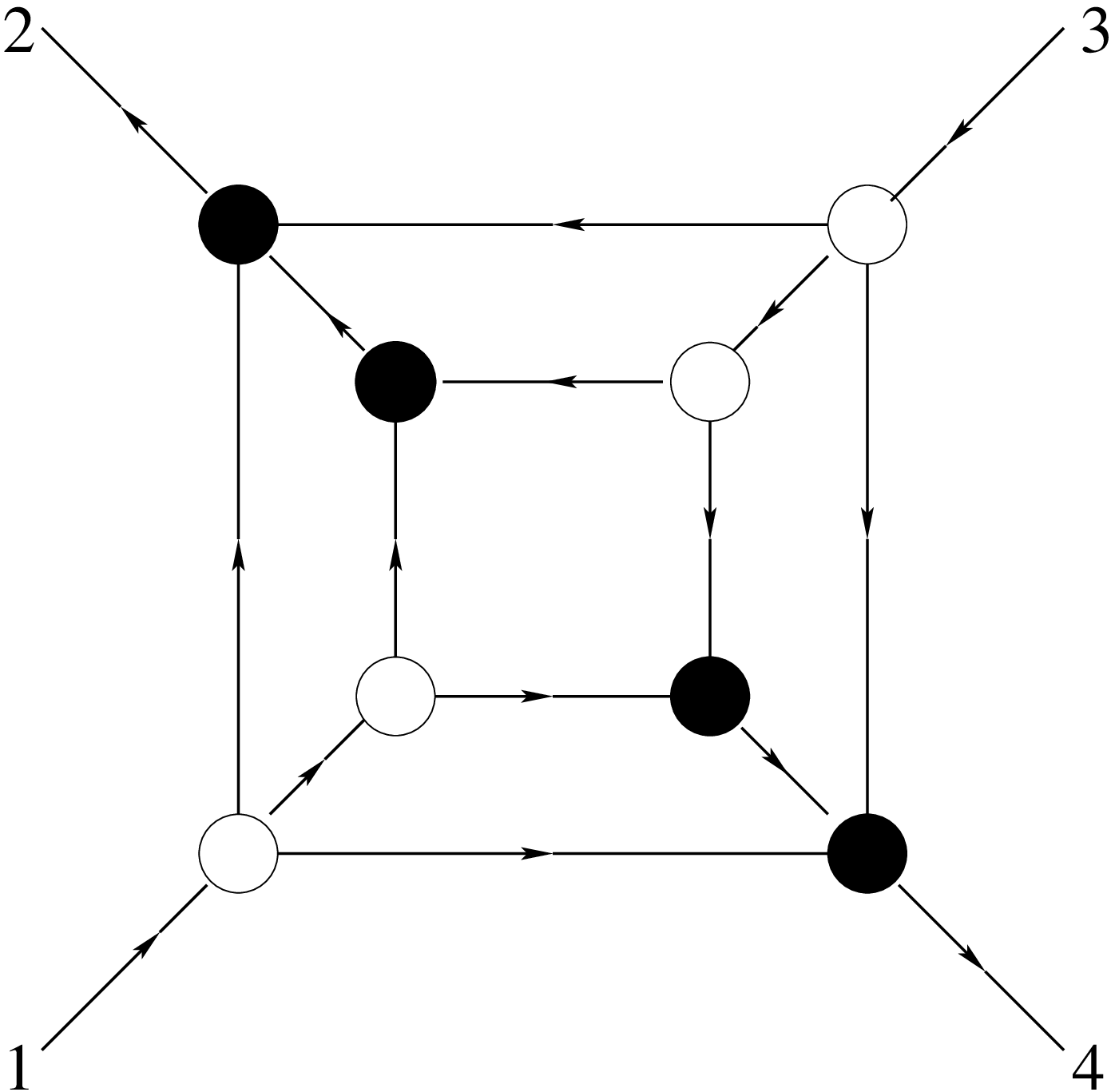}}}\:-\:\raisebox{-1.7cm}{\scalebox{.20}{\includegraphics{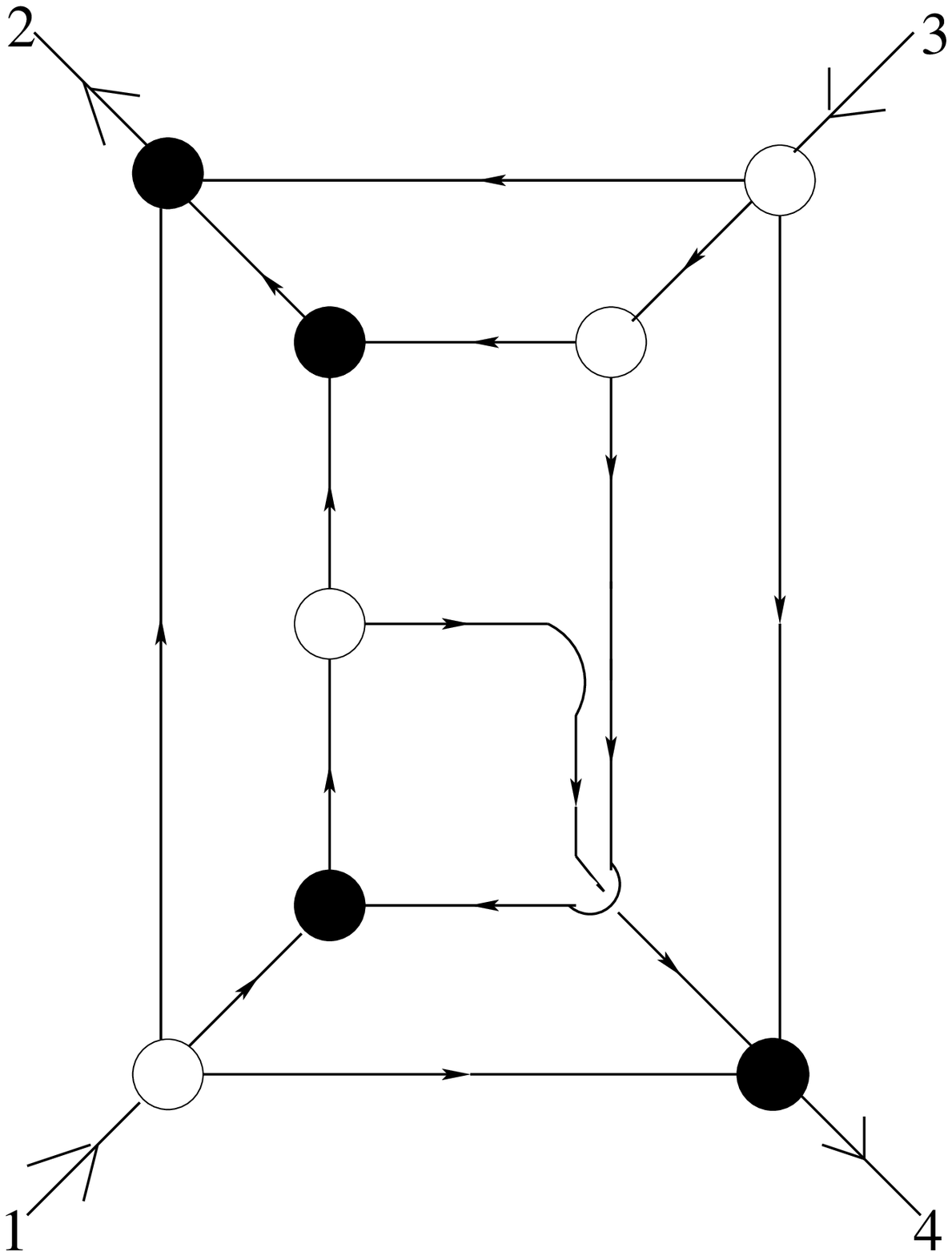}}}\\
  &\raisebox{-1.3cm}{\scalebox{.20}{\includegraphics{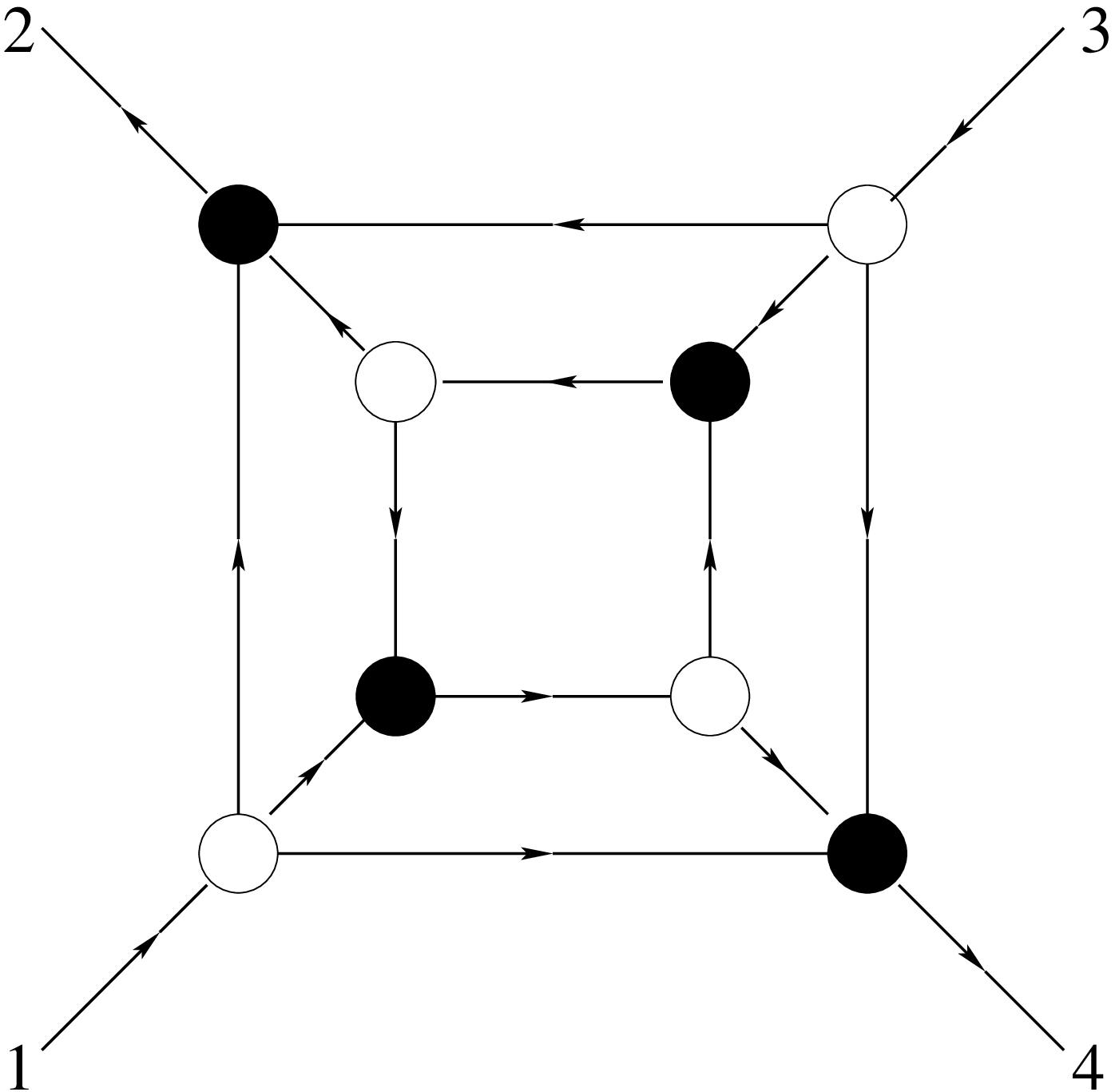}}}\:=\:\raisebox{-1.3cm}{\scalebox{.20}{\includegraphics{1l4ptLSlog.eps}}}\:-\:\raisebox{-1.3cm}{\scalebox{.20}{\includegraphics{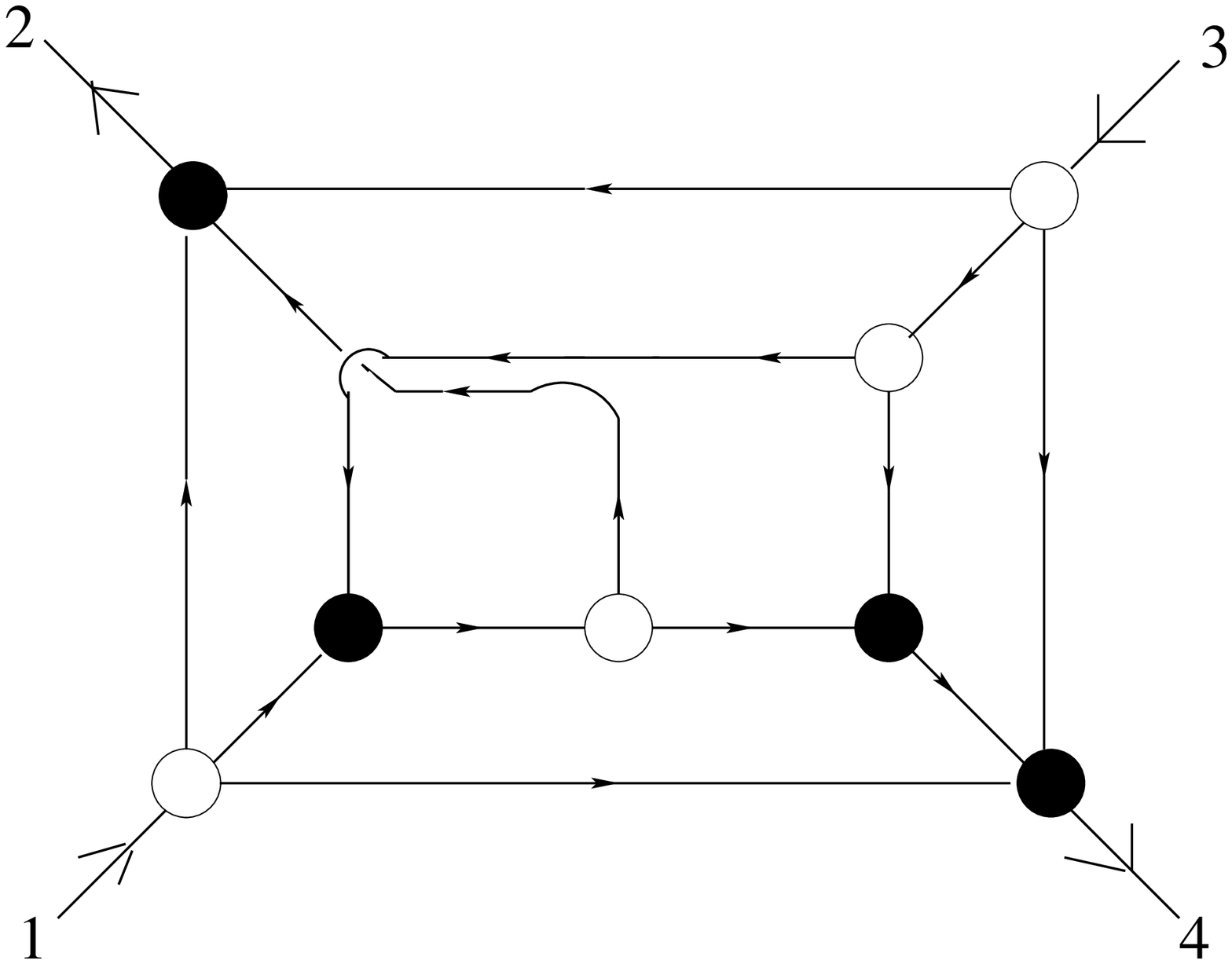}}}
 \end{split}
\end{equation}
The first diagram in the right-hand-side of both lines correspond to the purely $d\log$ contribution: using the merger equivalence relation and the bubble reduction, it can be written as the on-shell box which corresponds to the 
four-particle amplitude at tree level times four $d\log$'s. The other contributions contain a new higher order pole located at $\Delta_{24}(z)$. Actually, the expression in \eqref{eq:4ptHP} is a geometric sequence (truncated at order 
$4-\mathcal{N}$) in $\Delta_{24}(z)$. As shown in \cite{Benincasa:2015zna}, upon summation between the two contribution of this type in the two lines in \eqref{eq:4formSplit}, $\mathcal{N}\,=\,1,\,2$ SYM theories contain terms with 
single and double poles in $\Delta_{24}(z)$, while $\mathcal{N}\,=\,0$ theory shows poles up to order $4$. The terms with a simple pole correspond to triangle integrands, while the ones with double poles to bubble integrands 
\cite{Benincasa:2015zna}. The presence of higher order poles for $\mathcal{N}\,=\,0$ is a signature of the presence extra information, which is related to the rational contribution of the Passarino-Veltman reduction.

\section{Non-planar poles in momentum space}\label{app:NPpMomSp}

In Section \ref{sec:Gr36} we saw that on-shell functions on the top-cell of $Gr(3,6)$ can show (higher order) poles which are typical of non-planar diagrams. In particular, a class of such poles do not correspond to any Pl{\"u}cker
coordinate vanishing but rather it imposes relations among them. In this section we examine again the on-shell function of Section \ref{subsec:PnPr} with such a pole in momentum-space. Its expression on the twistor-space Grassmannian is
given by the first line of \eqref{eq:Gr36tc2}, and we will write it here for convenience in the momentum-space Grassmannian:
\begin{equation}\eqlabel{eq:Gr36tcMS}
 \begin{split}
  \raisebox{-1.2cm}{\scalebox{.25}{\includegraphics{Gr36topcellDec2.eps}}}\hspace{-.5cm}=\:
    \int\frac{d^{\mbox{\tiny $(3\times6)$}}C}{\mbox{Vol}\{GL(3)\}}\,&\frac{\delta^{\mbox{\tiny $(2\times3)$}}\left(\lambda\cdot C^{\perp}\right)\delta^{\mbox{\tiny $(2\times3$}}\left(C\cdot\tilde{\lambda}\right)
    \delta^{\mbox{\tiny $(3\times\mathcal{N})$}}\left(C\cdot\tilde{\eta}\right)
     }{
     \Delta_{123}\Delta_{234}\Delta_{345}\Delta_{456}\Delta_{561}\Delta_{612}}\times\\
   &\times
     \left[
      \frac{ 
             \Delta_{134}\Delta_{356}\Delta_{512}
           }{\Delta_{346}\Delta_{512}-\Delta_{345}\Delta_{612}}
     \right]^{4-\mathcal{N}}
 \end{split}
\end{equation}
The residue of the higher order pole is represented by a non-planar on-shell diagram:
\begin{equation}\eqlabel{eq:Gr36resnpMS}
 \begin{split}
  \raisebox{-1.2cm}{\scalebox{.25}{\includegraphics{Gr36resNP.eps}}}\hspace{-1cm}=\:
    \int&\frac{d^{\mbox{\tiny $(3\times6)$}}C}{\mbox{Vol}\{GL(3)\}}\,\frac{\delta^{\mbox{\tiny $(2\times3)$}}\left(\lambda\cdot C^{\perp}\right)\delta^{\mbox{\tiny $(2\times3)$}}\left(C\cdot\tilde{\lambda}\right)
     \delta^{\mbox{\tiny $(3\times\mathcal{N})$}}\left(C\cdot\tilde{\eta}\right)}{\Delta_{123}\Delta_{234}\Delta_{345}\Delta_{456}\Delta_{561}\Delta_{612}}\times\\
   &\times\left(\Delta_{134}\Delta_{356}\Delta_{512}\right)^{\mbox{\tiny $(4-\mathcal{N})$}}
    \delta^{\mbox{\tiny $(3-\mathcal{N})$}}\left(\Delta_{346}\Delta_{512}-\Delta_{345}\Delta_{612}\right).
 \end{split}
\end{equation}
As we already mentioned in the main text of the paper, the momentum-conserving delta-functions fix $8$ out of the $9$ parameter of $Gr(3,6)$. Let us choose the following representative for $C$:
\begin{equation}\eqlabel{eq:Gr36rep}
 C^{\mbox{$\star$}}\:=\:
 \begin{pmatrix}
  \lambda^{\mbox{\tiny $(1)$}}_1 & \lambda^{\mbox{\tiny $(2)$}}_1 & \lambda^{\mbox{\tiny $(3)$}}_1 & \lambda^{\mbox{\tiny $(4)$}}_1 & \lambda^{\mbox{\tiny $(5)$}}_1 & \lambda^{\mbox{\tiny $(6)$}}_1 \\
  \lambda^{\mbox{\tiny $(1)$}}_2 & \lambda^{\mbox{\tiny $(2)$}}_2 & \lambda^{\mbox{\tiny $(3)$}}_2 & \lambda^{\mbox{\tiny $(4)$}}_2 & \lambda^{\mbox{\tiny $(5)$}}_2 & \lambda^{\mbox{\tiny $(6)$}}_2 \\
  0                            & z[4,6]                       & 0                            & [5,6]-z[2,6]                 & [6,4]                        & [4,5]-z[4,6]
 \end{pmatrix}
\end{equation}
with $z$ being the unfixed parameter of $Gr(3,6)$. It is straightforward to see that the location of the higher-order pole is given by a second-order algebraic equation in the free parameter $z$:
\begin{equation}\eqlabel{eq:Gr36nploc}
 0\:=\:\Delta_{346}\Delta_{512}-\Delta_{345}\Delta_{612}\:=\:[6,5(z)]\langle1,2\rangle\langle3|5(z)+6|4]-z[6,4]\langle3|1+2(z)|5(z)+6|1\rangle,
\end{equation}
where $\lambda^{\mbox{\tiny $(2)$}}(z)\,\equiv\,\lambda^{\mbox{\tiny $(2)$}}+z\lambda^{\mbox{\tiny $(5)$}}$ and 
$\tilde{\lambda}^{\mbox{\tiny $(5)$}}(z)\,\equiv\,\tilde{\lambda}^{\mbox{\tiny $(5)$}}-z\tilde{\lambda}^{\mbox{\tiny $(2)$}}$. Therefore, in order to correctly compute the residue of this pole, one needs
to sum up over both the solutions of \eqref{eq:Gr36nploc}. Notice that the form of these solutions is $z_{\mbox{\tiny $\pm$}}\,=\,a\pm\sqrt{b}$, with $a$ and $b$ being rational functions of the Lorentz invariants.
At a generic point in momentum space, the function $b$ is not a perfect square and the individual residues at $z\,=\,z_{\mbox{\tiny $\pm$}}$ are no longer meromorphic functions. However, the square-roots cancel upon
summation of the two terms, returning a meromorphic function as it should, and which is represented by the decorated on-shell diagram on the left-hand-side of \eqref{eq:Gr36resnpMS}.

\bibliographystyle{utphys}
\bibliography{amplitudesrefs}	

\end{document}